\input epsf
\documentstyle[aps]{revtex}
\tighten
\begin{document}

\preprint{HUTP-03/A044,hep-th/0306208}
\title{Gravity in the dynamical approach to the cosmological constant}
\author{Shinji Mukohyama}
\address{
Department of Physics, Harvard University\\
Cambridge, MA, 02138, USA
}
\date{\today}

\maketitle

\begin{abstract} 

 One of the most disturbing difficulties in thinking about the
 cosmological constant is that it is not stable under radiative
 corrections. The feedback mechanism proposed in ref.~\cite{paper1} is
 a dynamical way to protect a zero or small cosmological constant
 against radiative corrections. Hence, while this by itself does not
 solve the cosmological constant problem, it can help solving the
 problem. In the present paper we investigate stability and gravity in
 this approach and show that the feedback mechanism is both classically
 and quantum mechanically stable and has self-consistent, stable
 dynamics. 

 \hfill\mbox{[HUTP-03/A044]}
\end{abstract}

\section{Introduction}

The cosmological constant is one of the most outstanding mysteries in
contemporary physics~\cite{reviews,Witten}. There are actually two 
cosmological constant problems. One is the ``old'' problem: ``why is the
cosmological constant vanishingly  small compared to the energy scale
of, say, the Planck scale?'' The ratio of the critical density of the
universe today, than which energy density of the cosmological constant
must be smaller, to the Planckian energy density is as small as
$10^{-120}$. Since the cosmological constant is not protected against 
quantum corrections, there is no manifest reason why it should be so
tiny. We now have the second problem: ``why is the dark energy roughly
of the same order as matter energy density today?'' The observation of
high-redshift, Type Ia supernovae is strongly inconsistent with a
universe with vanishing dark energy, whose typical form is a
cosmological constant~\cite{Perlmutter,Schmidt,Riess}. This implies that
there should be a small but non-vanishing dark energy and that it should
be comparable to matter energy density today. However, since dark energy
and matter scale differently as the universe expands, there seems to be
no manifest reason why they should be of the same order today. This
problem is often called the coincidence problem.

Although there are many approaches to the coincidence
problem~\cite{quintessence,k-essence,AHKM,Yokoyama}, all of them seem to 
have a common weakness: they can be spoiled by other contributions to
the cosmological constant. This is because it is a priori assumed that
the old problem can be solved. A tiny correction to the unknown solution
to the old problem can produce a relatively huge cosmological
constant. For example, in the quintessence scenario~\cite{quintessence}
it is a priori assumed that the asymptotic value of the quintessence
potential exactly vanishes. If we added a non-vanishing cosmological
constant to the potential then the scenario would not work. Moreover, if
there are many proposals to solve the coincidence problem and if they
can coexist then we will end up with a too much cosmological constant,
provided that each contribution is additive.  Hence, a good solution to
the coincidence problem must include a mechanism which almost completely
cancels other contributions to the cosmological constant. Unfortunately,
no phenomenologically acceptable solution with such a mechanism is
known.

There are also many approaches to the old problem. Among them are
quantum cosmology~\cite{Baum,Hawking,Coleman}, membrane
creation~\cite{Brown-Teitelboim,Bousso-Polchinski,Banks-Dine-Motl,FMSW},
a scalar field with a very small
mass~\cite{Garriga-Vilenkin,Weinberg2000}, 
backreaction models~\cite{Dolgov,Ford}, 
eternal inflation~\cite{eternal-inflation},  
extra dimensions~\cite{extra-dim}, and so
on~\cite{Rubakov,Hebecker-Wetterich}.

We do not consider the quantum cosmology approach based on the Euclidean
quantum gravity partly because it is not well understood and may be
ill-defined. (For example, see refs.~\cite{FKPS,Duff} for challenging
arguments.) There is another reason why we do not consider this
approach. Quantum cosmology is, in a sense, a theory of the beginning,
or the initial condition, of the universe. For example, the
Hartle-Hawking boundary condition~\cite{Hartle-Hawking} predicts a sharp
peak at a vanishing cosmological constant, but this value should be
considered as the effective cosmological constant at the beginning of
the universe, including all contributions from many fields and quantum
corrections. Hence, if this prediction is correct then the universe
cannot experience inflation nor any phase transitions after its
birth. If the effective cosmological constant were zero at the beginning
of the universe and if a phase transition occurred after that then the
cosmological constant now would be negative. If we try to avoid this
disagreeable conclusion by modifying or changing the proposal then the
peak must be moved to a non-vanishing cosmological constant. For
example, the proposals of Linde~\cite{Linde} and
Vilenkin~\cite{Vilenkin1986} prefer a large cosmological constant. This 
indicates the onset of a Planck scale inflation, but does not seem to
provide a solution to the cosmological constant problem without
anthropic consideration~\cite{Vilenkin1995}.

A proposal by Rubakov~\cite{Rubakov} might relax the cosmological 
constant in the Brans-Dicke frame to a small value if the cosmological
constant in the Einstein frame is exactly zero. However, for his
proposal to work, it is essential to include an inflaton sector in the
Einstein frame. Hence, we anyway need to explain why the cosmological
constant in the Einstein frame is exactly zero today. A similar 
proposal by Hebecker and Wetterich~\cite{Hebecker-Wetterich} not only
has a similar problem but also predicts a too much deviation from 
Einstein theory for weak gravity and, thus, is phenomenologically
unacceptable.

Dolgov's backreaction model~\cite{Dolgov} and its variants~\cite{Ford}
use the kinetic energy of a running scalar field to cancel the bare
cosmological constant. In these models the effective gravitational
coupling vanishes and, thus, the deviation from Einstein gravity 
is too large.

As far as the author knows, all other previously known approaches to the
old cosmological constant problem are either dependent of the anthropic 
principle or phenomenologically unacceptable (eg. a non-standard
Friedmann equation at low energy, a non-Einstein weak gravity, a naked
singularity, etc.). This situation has led to an increase in speculation
for the necessity of the anthropic principle~\cite{Anthropic}. Before
resorting to this, it is worthwhile to ask whether anything could
possibly do what the cosmological constant data
requires~\cite{Witten}. It seems likely that the correct way to
interpret the tiny value of the cosmological constant is that
conventional quantum field theory is not the whole story, so it is worth
seeking acceptable modifications.

In ref.~\cite{paper1} a new dynamical approach to the cosmological
constant problem was proposed. Although this idea by itself does not
explain why the cosmological constant is so small, it does provide a way
to protect a zero or small cosmological constant against radiative
corrections (without using supersymmetry) and thus can help solving the
cosmological constant problem. In this approach the true value of the
vacuum energy is negative, but the dynamics is such that the true
minimum is never attained, and the universe settles down to a near zero
energy state.

The purpose of the present paper is to investigate stability and gravity
in this dynamical approach. In particular, we show that (i) the
effective cosmological constant decreases in time and asymptotically
approaches zero from above; (ii) the evolution of a homogeneous,
isotropic universe is described by the standard Friedmann equation at
low energy; that (iii) classical, linearized gravity in Minkowski
background is described by Einstein gravity at distances longer than
$l_*=\sqrt{\alpha}\kappa$ and at energies lower than $l_*^{-1}$, where
$\kappa$ is the Planck length and $\alpha$ is a dimensionless, positive
parameter of the model; that (iv) the mechanism is stable under
radiative corrections and thus a zero or small cosmological constant is
protected against radiative corrections; and that (v) quantum
fluctuation at low energy is so small that the effective cosmological
constant at low energy does not overshoot zero even quantum
mechanically.

This paper is organized as follows. In Sec.~\ref{eqn:idea} we explain
the basic idea in analogy with a simple system of a charged particle. 
In Sec.~\ref{sec:mechanism} we review the feedback mechanism proposed in
ref.~\cite{paper1} in a more general setting. In
Sec.~\ref{sec:Friedmann} we derive equations of motion in detail and
show that the standard Friedmann equation is recovered at low
energy. The result in Sec.~\ref{sec:Friedmann} is confirmed numerically
in Sec.~\ref{sec:numerical}. In Sec.~\ref{sec:reheating} we consider
reheating in this model. In Sec.~\ref{sec:gravity} we investigate weak
gravity in Minkowski background and show the recovery of the linearized
Einstein theory. In Secs.~\ref{sec:quantum} we investigate stabilities
against radiative corrections and quantum
fluctuations. Sec.~\ref{sec:summary} is devoted to a summary of this
paper. 


\section{Basic idea}
\label{eqn:idea}

One of the most disturbing difficulties in thinking about the
cosmological constant is that it is not protected against radiative
corrections, which usually generate enormous vacuum energies compared to
what we observe. The feedback mechanism proposed in ref.~\cite{paper1}
can be considered as a dynamical way to protect a zero or small
cosmological constant against radiative corrections. Hence, although the
feedback mechanism by itself does not solve the cosmological constant
problem, it can help solving the problem. In this section we explain the
basic idea underlying the feedback mechanism.

Let us consider a simple system of a non-relativistic, charged particle
in a background electrostatic potential. The action of this simple
system is  
%
\begin{equation}
 I = \int dt 
  \left[\frac{M}{2}\dot{x}^2 - eV(x)\right], 
\end{equation}
where $V(x)$ is the electrostatic potential. Let us suppose that $eV(x)$
is an increasing function of $x$ so that the particle moves towards
the negative $x$ region. Suppose, however, that for some reason, we
would like to stop the particle just before its crossing $x=0$. How can
we stop it? One obvious solution is to control the electrostatic potential
so that it has a minimum near $x=0$, and introduce a friction. By
controlling the potential and the friction carefully, we can achieve our
purpose. A difficulty in this solution is that there might be forces
(eg. strong wind, gravity, someone who may kick the particle, etc.)
other than the electrostatic force and the friction. These forces make
the control of the potential extremely subtle and difficult.

A different, rather radical solution would be available if we could 
control the mass $M$ of the particle. In this case we make $M$ to
diverge when $x$ is sufficiently close to $x=0$. To make this
successful we of course need to watch the value of $x$ in real
time. This case does not involve a subtle control of anything: $x$
completely stops when $M$ diverges, irrespective of the value of
$eV'(x)$ and other forces. One might worry that the kinetic term in the
action might diverge. Actually, this is not the case, and the kinetic
term does not diverge but vanishes. Since the equation of motion of the
position $x$ of the charged particle is 
%
\begin{equation}
 \left(M\dot{x}\right)^{\cdot}+eV'(x)=0,
\end{equation}
the momentum $\pi\equiv M\dot{x}$ remains finite in finite time as far
as $eV'(x)$ is finite. (This is the reason why $x$ stops when $M$ 
diverges.) Hence, the kinetic term actually vanishes when $M$ diverges:
%
\begin{equation}
  \frac{M}{2}\dot{x}^2 = \frac{\pi^2}{2M} \to 0 \quad (M\to\infty). 
\end{equation}
Indeed, the more singular the mass is, the more quickly the kinetic term 
vanishes.

We apply this second idea to a scalar field, whose potential determines
the effective cosmological constant but is not fine-tuned. The reason
why we do not consider a fine-tuned potential, or the analogue of the
first idea above, is that it is not stable under radiative corrections,
which add vacuum energies to the potential and also change the form of
the potential. We still want to stop the scalar field at or near zero 
cosmological constant. Let us suppose that a scalar field $\phi$ is
described by the action  
%
\begin{equation}
 I = \int d^4x\sqrt{-g}
  \left[ -\frac{\partial^{\mu}\phi\partial_{\mu}\phi}{2f}
   -V(\phi)\right],
  \label{eqn:simple-action}
\end{equation}
where $f$ is chosen so that it vanishes at or near zero cosmological 
constant. A simple choice might be to make $f$ dependent of the value of 
$\phi$ itself so that $f$ vanishes at a root of the potential. In this
case, $\phi$ stops when the potential vanishes. However, this choice is
too naive and would not be stable under radiative corrections: radiative 
corrections add vacuum energies to the potential and, hence, change the
potential value at which $\phi$ stops. Indeed, if $f$ is a function of
$\phi$ only then we can always make a change of variable
$\phi\to\tilde{\phi}\equiv\int d\phi/\sqrt{f}$ so that the kinetic term
of the new variable $\tilde{\phi}$ is canonically normalized. In this
language, the choice of $f$ is nothing but a fine tuning of the
potential of $\tilde{\phi}$ and, thus, this is never stable against
radiative corrections. Hence, we cannot protect zero or small
cosmological constant against radiative corrections if $f$ depends only
on $\phi$. In other words, $\phi$ itself does not know where to stop.

On the other hand, the spacetime curvature seems to know where to
stop $\phi$ since it reflects the value of the effective cosmological
constant via the Friedmann equation. Hence, instead of considering a
$\phi$-dependent $f$, we postulate that $f$ depends on the spacetime
curvature so that the curvature can tell when $\phi$ should stop. For
simplicity, we restrict our discussion to the case where $f$ depends on
the Ricci scalar $R$ only and vanishes at $R=0$. In this case, $\phi$
stops rolling when $R=0$. We do not provide a reason why $f$ vanishes at
$R=0$. However, we shall show that this choice of $f$ is stable under 
radiative corrections and leads to a stable dynamics. In other words,
instead of providing a solution to the cosmological constant problem, we
propose a dynamical mechanism to protect a zero or small cosmological
constant against radiative corrections (without using supersymmetry) as
a first step. The singular kinetic term makes the scalar field stop at
zero curvature, even without a fine-tuned potential.

With this choice of $f$, we need to make it sure that $R=0$ really
corresponds to zero cosmological constant. If the kinetic term dominates
over the potential term then the value of the potential determined by
$R=0$ does not need to be zero and, thus, this model does not seem to
work. Hence, we are left with the following two possibilities. One
possibility is that there is a scaling solution to scalar dynamics in
which the kinetic and potential energies of the scalar field decrease in 
tandem, so that the field is slowed down when small potential energy is
achieved. Unfortunately, in our specific realizations, there always
seems an instability invalidating the scaling solutions we tried.

The second possibility, which we shall focus on in the rest of this
paper, is that the potential term dominates over the kinetic term 
so that $R$ is determined solely by the contribution from the
potential. In this case, the scalar field indeed stops at zero
cosmological constant automatically. We could make $f$ dependent of
$\phi$ as well, but such a dependence can be removed from the behavior
of the kinetic term near $R=0$ by redefinition of $\phi$ without loss
of generality. We suppose that the kinetic term in
(\ref{eqn:simple-action}) represents the term which is the most 
singular-looking at $R=0$ among many possible terms in the kinetic part
and we did not include less singular-looking terms since they are not
important at low energy. We also omitted all other dynamical fields
since, as we shall see in the following sections, the dynamics of $\phi$
is so slow that any dynamical fields other than $\phi$ will settle into
their ground state before the universe approaches a sufficiently low
energy state. The potential $V(\phi)$ includes the ground state energies
of all such fields.

Since $f$ is in the denominator and vanishes in the $R\to 0$ limit, the
kinetic term threatens to be singular at low energy. Moreover, since $f$
is a function of the Ricci scalar $R$, it introduces corrections to
Einstein equation and possibly destabilize the system. On the other
hand, from the charged-particle analogue above, we have learned that a
singular-looking kinetic term is actually regular. The same is true
here! The more singular $1/f$ is, the smaller the kinetic term and
corrections to Einstein equation are! This rather counter-intuitive but
very simple fact means that by making $1/f$ (for the most
singular-looking term) sufficiently singular at $R=0$, we can make
corrections to Einstein equation as small as we like. In particular, it
is indeed possible to make the corrections due to the kinetic term much
smaller than those from usual higher derivative terms like $R^2$ at low
energy so that higher derivative corrections and stability are
completely controlled by the usual higher derivative terms. Note that if
there are several (or many) terms in the kinetic part, we do not need to
require the singular behavior for all coefficients: just one term is
enough.


\section{Feedback mechanism}
\label{sec:mechanism}

In the previous section we have learned from a simple analogous system
that a scalar field stops in a completely regular way when the
coefficient of a kinetic term diverges. In this section we consider a
scalar field with such a singular-looking kinetic term.

The field Lagrangian we consider is
%
\begin{eqnarray}
 I & = & \int d^4x\sqrt{-g}
  \left[ \frac{R}{2\kappa^2} + \alpha R^2 + L_{kin} - V(\phi) \right],
  \nonumber\\
 L_{kin} & = & \frac{\kappa^{-4}K^q}{2qf^{2q-1}},
  \label{eqn:action}
\end{eqnarray}
where $f$ is a function of the Ricci scalar $R$, $\kappa$ is the
Planck length, $\alpha$ and $q$ are constants, and 
$K \equiv -\kappa^4\partial^{\mu}\phi\partial_{\mu}\phi$. Our sign 
convention for the metric is $(-+++)$. In a word, the idea of this
model is to assume a non-standard kinetic term. The dependence on
curvature is what stalls the field at small cosmological constant.

We can absorb any nonzero cosmological term into $V(\phi)$ without
loss of generality. We assume that (i) the minimum of the potential
$V(\phi)$ is negative; that (ii) $q>1/2$; that (iii) $\alpha>0$; and
that (iv) the function $f(R)$ is non-negative and behaves near $R=0$ as 
%
\begin{equation}
 f(R) \sim (\kappa^2R)^{2m},
  \label{eqn:fine-tuning}
\end{equation}
where $m$ is an integer satisfying
%
\begin{equation}
 2(m-1) > \frac{q}{2q-1}.
  \label{eqn:stability-cond}
\end{equation}
In the $m\to\infty$ limit, this choice of $f(R)$ may be replaced
by $f(R)\sim\exp[-\kappa^{-4}R^{-2}]$ or similar functions. We suppose
that the kinetic term in (\ref{eqn:action}) is the most singular-looking
one among many possible terms in the kinetic part. We assume that the 
most singular-looking term behaves like (\ref{eqn:fine-tuning}) with the
condition (\ref{eqn:stability-cond}), but other terms do not need to
behave like that. If there are several terms in the kinetic part then
the most singular-looking term is dominant and other terms are not
important at low energy. Thus, eg. adding the standard canonical kinetic
term does not affect anything.

Since $f$ is in the denominator of the kinetic term and vanishes in the 
$\kappa^2R\to 0$ limit, the kinetic term threatens to be singular
at low energy. We shall see below that the large $m$ makes $\phi$
evolve slowly, that the numerator $K^{q}$ vanishes more quickly
than the denominator and that the kinetic term is actually
regular.

Now let us explain the motivations for the assumptions (i)-(iv). (i) We
would like to propose a mechanism in which the scalar field stops
rolling at or near zero vacuum energy. For this purpose, the minimum of
the potential $V(\phi)$ should be negative so that $V(\phi)$ has a root.
(ii) For the stability of inhomogeneous perturbations, it is necessary
that the sound velocity squared
$c_s^2=L_{kin,K}/(2KL_{kin,KK}+L_{kin,K})$ is
positive~\cite{Garriga-Mukhanov}. In our model this condition is reduced
to $q>1/2$.
(iii) The $R$-dependence of the kinetic term $L_{kin}$ produces higher
derivative corrections to Einstein equation which might destabilize
gravity. As we shall explain below, the term $\alpha R^2$ can stabilize
gravity at low energy if $\alpha$ is positive.
(iv) For the term  $\alpha R^2$ to control the stability, we need to
make it sure that the term $\alpha R^2$ is dominant over $L_{kin}$ at
low energy. As shown below, this is  the case if and only if the
condition (\ref{eqn:stability-cond}) is satisfied.

We do not know a parent theory that will provide our Lagrangian as the
low-energy effective theory. In particular, we do not give a reason why
$f$ vanishes at $R=0$. In this sense, this model by itself does not
solve the cosmological constant problem. However, we shall show below
that this choice of $f$ is radiatively stable, leads to a stable
dynamics and could help solve the cosmological constant problem. We
treat this model as a purely phenomenological suggestion that might
motivate further research into the possible parent theory, which is
presumably not based entirely on conventional four-dimensional field
theory.

We could make $f$ dependent of $\phi$, but such a dependence
can be removed from the behavior of $L_{kin}$ near $R=0$ by redefinition
of $\phi$ without loss of generality. Note that $L_{kin}$ above
represents the term which is the most singular-looking at $R=0$ among
many possible terms in the kinetic part and that we did not include less
singular-looking terms since they are not important at low energy. We
also omitted all other dynamical fields since, as we shall see below,
the dynamics of $\phi$ is so slow that any dynamical fields other than
$\phi$ will settle into their ground state before the universe
approaches a sufficiently low energy state.

We now argue that the Lagrangian (\ref{eqn:action}) gives a feedback
mechanism that makes the field stall at or near zero vacuum energy. 
The argument for the $q=1$ case was already given in
ref.~\cite{paper1}. Here, let us simply generalize it to a general $q$
($>1/2$).

In the flat Friedmann-Robertson-Walker (FRW) background
%
\begin{equation}
 ds^2 = -dt^2 + a(t)^2(dx^2+dy^2+dz^2),\label{eqn:FRW-metric}
\end{equation}
the equation of motion for a homogeneous $\phi$ is
%
\begin{equation}
 \dot{\pi} + 3H\pi + V'(\phi) = 0, \quad
 \pi \equiv \frac{\dot{\phi}}{f}\cdot
  \left(\frac{\kappa^4\dot{\phi}^2}{f^2}\right)^{q-1}, 
  \label{eqn:eom-homogeneous-phi}
\end{equation}
where $H=\dot{a}/a$, a dot denotes the time derivative and a prime
applied to $V(\phi)$ denotes the derivative with respect to $\phi$. Near
$V=0$, $V$ can be approximated by a linear function as 
%
\begin{equation}
 V \simeq c\kappa^{-3}(\phi-\phi_0), 
\end{equation}
where $c$ and $\phi_0$ are constants. This approximation is extremely
good since, as we shall see, $\phi$ rolls very slowly near $V=0$ and
does not probe the global shape of the potential.

Without any fine-tuning, the
dimensionless constant $c$ should be of order unity. The equation of
motion (\ref{eqn:eom-homogeneous-phi}) can be rewritten as
%
\begin{equation}
 \frac{d{\cal B}}{dx} 
  + \left(3-\frac{\dot{H}}{H^2}\right){\cal B} + 1 = 0, 
  \label{eqn:eom-B}
\end{equation}
where ${\cal B}\equiv\kappa^3H\pi/c$ and $x\equiv\ln a(t)$. This implies
that ${\cal B}$ approaches $-(3-\dot{H}/H^2)^{-1}$, if $\dot{H}/H^2$
changes slowly compared to $a$ and is smaller than $3$. See
Fig.~\ref{fig:B} for confirmation of this statement by numerical
calculation. Hence, the asymptotic behavior of $\pi$ is  
%
\begin{equation}
 \kappa^2\pi \sim -c\kappa^{-1}H^{-1}.
  \label{eqn:asymptotic-pi}
\end{equation}

If the kinetic term is small compared to the potential term then the
Friedmann equation implies that 
%
\begin{equation}
 3H^2 \simeq \kappa^2 V, 
\end{equation}
and (\ref{eqn:asymptotic-pi}) can be rewritten as
%
\begin{eqnarray}
 \kappa \partial_t(\kappa^4V) &\sim  &
  c\kappa^2\dot{\phi} = cf\cdot\kappa^2\pi\cdot
  (\kappa^4\pi^2)^{\frac{1}{2(2q-1)}-\frac{1}{2}} \nonumber\\
 & \sim & -(c^2)^{\frac{q}{2q-1}}(\kappa^2H^2)^{2m-\frac{1}{2(2q-1)}}
  \nonumber\\
  & \sim & -(c^2)^{\frac{q}{2q-1}}(\kappa^4V)^{2m-\frac{1}{2(2q-1)}}.
\end{eqnarray}
Hence, we obtain 
%
\begin{equation}
 (\kappa^4V)^{-\gamma/2} \sim 
  (c^2)^{\frac{q}{2q-1}}\frac{t-t_0}{\kappa}, \quad
  \gamma = 4m-\frac{4q-1}{2q-1},
  \label{eqn:V-sol}
\end{equation}
where $t_0$ is a constant. From the condition
(\ref{eqn:stability-cond}), $\gamma>3$ and 
%
\begin{equation}
 \kappa^4V   \to +0 \quad (t/\kappa\to\infty).
\end{equation}
This result means that the field stalls near $V=0$. See
Figs.~\ref{fig:Lambda} and \ref{fig:Lambda_long} for confirmation of
this behavior by numerical calculation.

We have assumed that the kinetic energy is small compared to the
potential energy. This assumption is easily verified. At low energy
($\kappa H\ll 1$),
%
\begin{eqnarray}
 L_{kin} & = &
  \frac{\kappa^{-4}f}{2q}(\kappa^4\pi^2)^{\frac{q}{2q-1}} 
  \sim \kappa^{-4}(\kappa H)^{4m}
  \left(\frac{c^2}{\kappa^2H^2}\right)^{\frac{q}{2q-1}}  \nonumber\\
  & \ll & \kappa^{-4}(\kappa H)^4 \ll \kappa^{-4}(\kappa H)^2 \sim V,
   \label{eqn:Lkin-is-small}
\end{eqnarray}
provided that $c=O(1)$. Here, we have used the condition
(\ref{eqn:stability-cond}) to obtain the first inequality. We have
implicitly assumed that the standard Friedmann equation is valid at low
energy. Let us justify this assumption in the next section.


\section{Classical stability I - Recovery of Friedmann equation}
\label{sec:Friedmann}

We can also show the essential, and somewhat surprising result,
that the standard Friedmann equation is recovered at low energy,
starting from the action (\ref{eqn:action}). There are higher
derivative corrections to the Einstein equation due to the
$R$-dependence of $L_{kin}$ and $\alpha R^2$.

Since there are higher-derivative terms, the stability of the system is
a non-trivial question. In order to see the non-triviality, let us
consider the Klein-Gordon equation $(\Box-M^2)\varphi=0$ as a standard
equation and add $\epsilon(\varphi)\Box^2\varphi/M^2$ to the right hand
side. One might expect that the standard equation should be recovered
whenever $\epsilon\to 0$. Actually, this is not true. The standard
equation is certainly recovered in the $\epsilon\to 0$ limit if
$\epsilon>0$. On the other hand, if $\epsilon<0$ then the system has a
tachyonic degree and is unstable. Moreover, if $\varphi$ crosses a root
of $\epsilon$ then the system experiences a singularity ($\Box^2\varphi$
diverges) and the low energy effective theory cannot be trusted unless a
miracle cancellation occurs.

In our system, from the estimate of $L_{kin}$ in
(\ref{eqn:Lkin-is-small}) with (\ref{eqn:stability-cond}), $L_{kin}$ is
much smaller than $\alpha R^2$ ($\sim H^4$) at low energy if 
$\alpha\ne 0$. Hence, $\alpha R^2$ should dominate the higher derivative
corrections and control the stability. For the theory
$R/2\kappa^2+\alpha R^2$, roughly speaking, the parameter $\alpha$ plays
the role of $\epsilon$ above (including the sign) and it is known that
the low energy dynamics is stable if and only if  
$\alpha\ge 0$~\cite{Muller-Schmidt}. Here, stability means that as the 
universe expands, the system keeps away from unphysical spurious
solutions and approaches the standard low energy evolution
asymptotically. If we did not include the term $\alpha R^2$ then, as we
shall see below, $L_{kin}$ would make the quantity corresponding to
$\epsilon$ above to be negative essentially because $f(R)$ is in the
denominator. Therefore our system should be stable and the standard
Friedmann equation should be recovered at low energy if and only if
$\alpha>0$ and (\ref{eqn:stability-cond}) are satisfied.

The equation of motion derived from the action (\ref{eqn:action})
includes up to fourth order derivatives of the metric. However, since
the Friedmann equation is a constraint equation, it does not include the
highest order derivatives of the metric. This means that the generalized
Friedmann equation describing a homogeneous, isotropic universe in our
model includes derivatives of the metric only up to third order. 
Hence, it should be of the following form. 
%
\begin{equation}
 \frac{\epsilon(t)}{H}\partial_t\left(\frac{\dot{H}}{H^2}\right) = 
  \frac{\kappa^2(\rho_{other}+\rho_{\phi,0}+\Delta\rho_{\phi})-3H^2}
  {H^2},
  \label{eqn:generalized-Friedmann-eq}
\end{equation}
where the dimensionless coefficient $\epsilon(t)$ is written in terms of
($\dot{H}/H^2$, $\kappa H$, $\kappa^2\pi$), $\rho_{other}$ is the energy
density of fields other than $\phi$, 
$\rho_{\phi,0}\equiv 2KL_{kin,K}-L_{kin}+V$ is a part of the energy 
density of $\phi$ which would be obtained by neglecting the
$R$-dependence of $L_{kin}$, $\Delta\rho_{\phi}$ is corrections to
$\rho_{\phi,0}$ which depends on $H$ and $\dot{H}$ but not on
$\ddot{H}$. We shall see the detailed form of the generalized Friedmann
equation soon.

In this form of the generalized Friedmann equation, $\epsilon$ and
$\Delta\rho_{\phi}$ characterize corrections to the standard Friedmann
equation. Indeed, if $\epsilon\equiv 0$ and if 
$\Delta\rho_{\phi}\equiv 0$ then this equation reduces to the standard
Friedmann equation $3H^2=\kappa^2(\rho_{other}+\rho_{\phi,0})$. However,
this does NOT necessarily guarantees the recovery of the standard
Friedmann equation in the limit $\epsilon\to 0$, 
$\Delta\rho_{\phi}\to 0$. As in the above example of a Klein-Gordon
equation with a higher derivative correction, what makes the limit
rather subtle is the sign of the coefficient $\epsilon$ of the highest 
order derivative term. As we shall see soon, if $\epsilon$ approaches
zero from the positive side then the standard Friedmann equation is
indeed recovered in the limit. This situation is somehow similar to
having an infinitely massive extra degree of freedom. On the other hand,
if $\epsilon$ approaches zero from the negative side then the situation
is similar to having an infinitely unstable tachyonic degree and, thus,
the system is completely unstable. Moreover, if $\epsilon$ reaches zero
in finite time then the system encounters a singularity where
$\partial_t(\dot{H}/H^2)$ diverges. Hence, both the stability and the
recovery of the standard Friedmann equation require $\epsilon$ to be
positive.

In our model there are two contributions to $\epsilon$: one from
$L_{kin}$ and the other from $\alpha R^2$. We shall see below that the
first contribution is negative. This implies that the system would be
unstable in the absence of the term $\alpha R^2$ and that we really need
this term (or other ordinary higher derivative terms). On the other
hand, we shall see below that the sign of the second contribution is the
same as the sign of $\alpha$. Moreover, as we have already seen in 
the previous section, $L_{kin}$ is much smaller than $H^4$ and, thus,
than $\alpha R^2$ at low energy. This means that the contribution from
the $\alpha R^2$ term determines the sign of $\epsilon$ at low energy
and that $\epsilon$ is positive if and only if $\alpha$ is positive. 
Therefore, if and only if $\alpha$ is positive, the system is stable 
and the standard Friedmann equation is recovered at low energy. In the
following, we shall show this statement more explicitly.

For the stability analysis and numerical works, it is sometimes more
convenient to work with a set of first order differential equations than
a set of higher order differential equations. In the end of this
section, we shall use the set of first-order equations to show the
recovery of the standard Friedmann equation analytically. Some numerical
results are shown in Sec.~\ref{sec:numerical}.

A set of first order differential equations is obtained in
Appendix~\ref{app:derivation} for a general action of the form
(\ref{eqn:more-general-action}) with (\ref{eqn:special-L}). Restricting
it to our model action (\ref{eqn:action}), we obtain
%
\begin{eqnarray}
 \dot{\phi} & = & f\pi\cdot
  (\kappa^4\pi^2)^{\frac{1}{2(2q-1)}-\frac{1}{2}}, \nonumber\\
 \dot{\pi} & = & -3H\pi-V'(\phi), \nonumber\\
 \dot{H} & = & H^2\Omega, \nonumber\\
 \epsilon\dot{\Omega} & = & H{\cal F},  \nonumber\\
 \kappa^2\dot{\varphi} & = & -H^2({\cal G}_0 +\Delta{\cal G}), \nonumber\\ 
 \dot{\rho}_{other} & = & -3H(\rho_{other}+p_{other}), 
  \label{eqn:dphi_i}
\end{eqnarray}
where 
%
\begin{eqnarray}
 \Omega & \equiv & \frac{\dot{H}}{H^2}, \nonumber\\
 \varphi & \equiv & 2(L_R)^{\cdot} - 2HL_R, \nonumber\\
 L_R & = & 2\alpha R + L_{kin,R},
\end{eqnarray}
and
%
\begin{eqnarray}
 \epsilon & \equiv & 18\kappa^2 H^2
  \left[ 4\alpha- \frac{(2q-1)f''}{q\kappa^4}
   (\kappa^4\pi^2)^{\frac{q}{2q-1}}\right], \label{eqn:epsilon-t}\\
 {\cal F} & \equiv & \frac{3\kappa^2\varphi}{H} 
  + \frac{3(2q-1)}{q\kappa^2}
  \left[12\Omega(\Omega+2)H^2f''-f'\right]
  (\kappa^4\pi^2)^{\frac{q}{2q-1}} \nonumber\\
 & &  - \frac{6(3H\pi+V')f'}{H}\kappa^2\pi\cdot
  (\kappa^4\pi^2)^{-\frac{q-1}{2q-1}}
  + 72\alpha\kappa^2H^2(1-2\Omega)(\Omega+2), \nonumber\\
 {\cal G}_0 & \equiv & 2\Omega 
  + \frac{1}{\kappa^2H^2}\left[\kappa^4(\rho_{other}+p_{other})
  + f\cdot(\kappa^4\pi^2)^{\frac{q}{2q-1}}\right], \nonumber\\
 \Delta{\cal G} & \equiv & 
  -\frac{3(2q-1)\Omega f'}{q\kappa^2}(\kappa^4\pi^2)^{\frac{q}{2q-1}}
  +72\alpha\kappa^2H^2\Omega(\Omega+2),  \nonumber\\
  R & = & 6H^2(\Omega+2). \nonumber
\end{eqnarray}
Here, $\rho_{other}$ and $p_{other}$ are energy density and pressure of
fields and/or matter other than $\phi$. Besides the above dynamical
equations, there is a constraint equation of the form 
%
\begin{equation}
 \frac{3\kappa^2\varphi}{H} = {\cal F}_0+\Delta{\cal G}
		-36\alpha\kappa^2H^2(\Omega+2)^2,
  \label{eqn:constraint-eq}
\end{equation}
where
%
\begin{eqnarray}
 {\cal F}_0 & \equiv &
  \frac{\kappa^2(\rho_{other} + \rho_{\phi,0}) -3H^2}{H^2},
  \nonumber\\
 \rho_{\phi,0} & \equiv & 2KL_{kin,K}-L_{kin}+V 
  = \frac{2q-1}{2q\kappa^4}f\cdot(\kappa^4\pi^2)^{\frac{q}{2q-1}}
  + V. 
  \label{eqn:calF0}
\end{eqnarray}
Note that the standard Friedmann equation and the standard dynamical
equation are ${\cal F}_0=0$ and ${\cal G}_0=0$, respectively. When the 
system is analyzed numerically, we can use the constraint equation to
set the initial condition for $\varphi$ and also to check numerical
accuracy.

It is also possible to use the constraint equation to eliminate
$\varphi$ from the set of first-order equations: ${\cal F}$
in the forth equation in (\ref{eqn:dphi_i}) is rewritten as
%
\begin{equation}
 {\cal F} = {\cal F}_0 + \Delta{\cal F},
	 \label{eqn:alt-F}
\end{equation}
where
%
\begin{eqnarray}
 \Delta{\cal F} & = &
  \frac{3(2q-1)}{q\kappa^2}
  \left[12\Omega(\Omega+2)H^2f''-(\Omega+1)f'\right]
  (\kappa^4\pi^2)^{\frac{q}{2q-1}} \nonumber\\
  & & - \frac{6(3H\pi+V')f'}{H}\kappa^2\pi\cdot
  (\kappa^4\pi^2)^{-\frac{q-1}{2q-1}}
  -108\alpha\kappa^2H^2\Omega(\Omega+2). 
\end{eqnarray}
Hence, the fourth equation in (\ref{eqn:dphi_i}) is reduced to
(\ref{eqn:generalized-Friedmann-eq}) with 
$\Delta\rho_{\phi}=\kappa^{-2}H^2\Delta{\cal F}$.

Now let us show the recovery of the standard Friedmann equation at low
energy by using the above set of first-order equations.

From the inequalities in (\ref{eqn:Lkin-is-small}), both $L_{kin}$ and 
$\alpha R^2$ ($\sim H^4$) are small compared to the Einstein-Hilbert
term $R/2\kappa^2$ ($\sim H^2/\kappa^2$) at low energy 
($\kappa H\ll 1$). Thus, even without any calculations we can conclude
that 
%
\begin{equation}
 \epsilon(t) \to 0, \quad \frac{\kappa^2}{H^2}(\rho_{\phi,0}-V) \to 0, 
  \quad \frac{\kappa^2\Delta\rho_{\phi}}{H^2}=\Delta{\cal F} \to 0, 
  \label{eqn:low-energy}
\end{equation}
in the low energy limit $\kappa H\to 0$. Of course, it is
straightforward to confirm this by using the explicit expressions
presented here. More specifically, the second term in the expression
(\ref{eqn:epsilon-t}) of $\epsilon(t)$ is small compared to the first
term at low energy ($\kappa H\ll 1$): 
%
\begin{equation}
 \kappa^{-4}f''(R)\cdot(\kappa^4\pi^2)^{\frac{q}{2q-1}}
  \sim (c^2)^{\frac{q}{2q-1}}(\kappa^2H^2)^{2(m-1)-\frac{q}{2q-1}}
  \ll \alpha, 
\end{equation}
where we have used the condition (\ref{eqn:stability-cond}). Note that
the second term in (\ref{eqn:epsilon-t}) is negative if $f''(R)$ is
positive. Therefore, if and only if $\alpha$ is positive and the
condition (\ref{eqn:stability-cond}) is satisfied, $\epsilon(t)$ 
is positive at low energy. In the next paragraph we shall see that the
positivity of $\epsilon(t)$ and, thus, the positivity of $\alpha$ and
the condition (\ref{eqn:stability-cond}) are essential for the recovery
of the standard Friedmann equation.

Having the behavior (\ref{eqn:low-energy}) and the positivity of
$\epsilon(t)$, it is easy to see that
%
\begin{equation}
 [\kappa^2(\rho_{other}+V)-3H^2]/H^2
  \to 0 \quad (t \to \infty), \label{eqn:recovery}
\end{equation}
provided that $H>0$ and $\dot{H}/H^2<0$. This is equivalent to 
${\cal F}_0\to 0$ ($t\to\infty$) since the kinetic part in
$\rho_{\phi,0}$ in (\ref{eqn:calF0}) goes to zero much faster than the
potential $V$. Actually, if the right hand side of
(\ref{eqn:generalized-Friedmann-eq}) is positive (or negative) then
$\dot{H}/H^2$ will increase (or decrease, respectively). Hence, $H$ will
decrease less (or more, respectively) rapidly. Moreover, $\rho_{other}$
decreases more (or less, respectively) rapidly if it satisfies the weak
energy condition. Thus, as a result, the right hand side of
(\ref{eqn:generalized-Friedmann-eq}) will decrease (or increase,
respectively). The decrease (or increase, respectively) should continue
until the right hand side of (\ref{eqn:generalized-Friedmann-eq})
becomes almost zero, and we expect the behavior (\ref{eqn:recovery}),
which is nothing but the recovery of the standard Friedmann equation at
low energy. This behavior, namely the recovery of the standard Friedmann
equation, can be confirmed numerically. See
Figs.~\ref{fig:F0}-\ref{fig:F0_long} in the next section for numerical
confirmation of the recovery of the standard Friedmann equation 
(${\cal F}_0\to 0$) and the standard dynamical equation 
(${\cal G}_0\to 0$). In the low energy Friedmann equation, $\kappa^2 V$
plays the role of the cosmological constant $\Lambda_{eff}$.

\section{Numerical results}
\label{sec:numerical}

It is straightforward to integrate the set of first order equations
(\ref{eqn:dphi_i}) numerically. In this section we show some results. In
the following we set 
%
\begin{equation}
 q = 1, \quad m = 2, \quad c=1, \quad \alpha=1; \quad 
  \rho_{other} \equiv 0, \quad p_{other} \equiv 0,
\end{equation}
for simplicity. In order to set initial values we need four independent
conditions at $t=0$ since we have five variables and one constraint 
equation. In the following we show numerical results for four different
sets of initial values. We shall see that the system exhibits an
attractor behavior. Because of the attractor behavior, in figures
showing results of long numerical integration, plots for those four
different initial values are degenerate.

\begin{figure}
 \centering\leavevmode\epsfysize=8cm \epsfbox{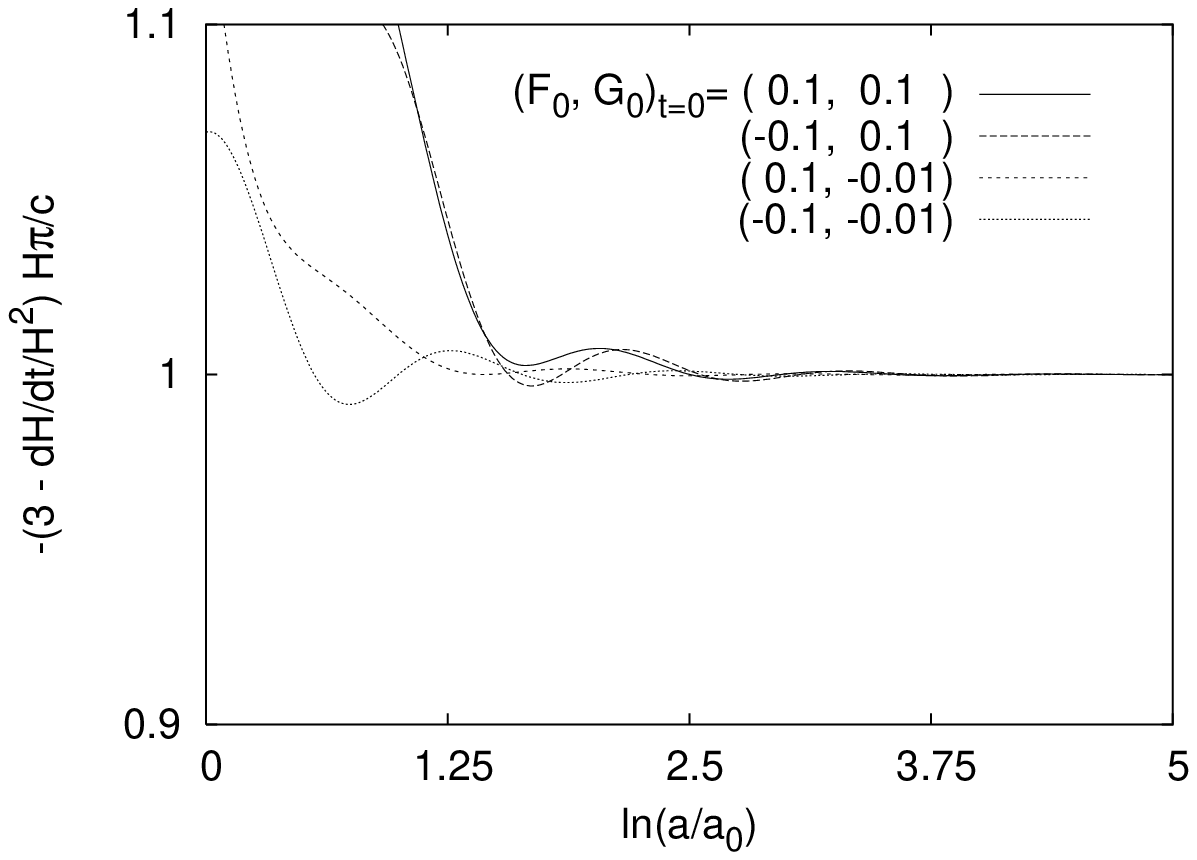}
 \caption{\label{fig:B}
 The behavior of ${\cal B}\equiv\kappa^3H\pi/c$ for four different sets
 of initial values. In the figure, $a$ is the scale factor and
 $a_0=a(t=0)$. The initial values of $\phi$ and $\dot{H}/H^2$ are 
 $c\kappa(\phi-\phi_0)|_{t=0}=0.01$ and $\dot{H}/H^2|_{t=0}=-0.02$. 
 Two more independent conditions at $t=0$ are given by setting 
 ${\cal F}_0$ and ${\cal G}_0$ in four different ways as
 $({\cal F}_0,{\cal G}_0)_{t=0}=(0.1,0.1)$, $(-0.1,0.1)$, 
 $(0.1,-0.01)$, $(-0.1,-0.01)$. 
 This figure indeed shows that $-(3-\dot{H}/H^2){\cal B}$ quickly
 converges to $1$.  
 }
\end{figure}

\begin{figure}
 \centering\leavevmode\epsfysize=8cm \epsfbox{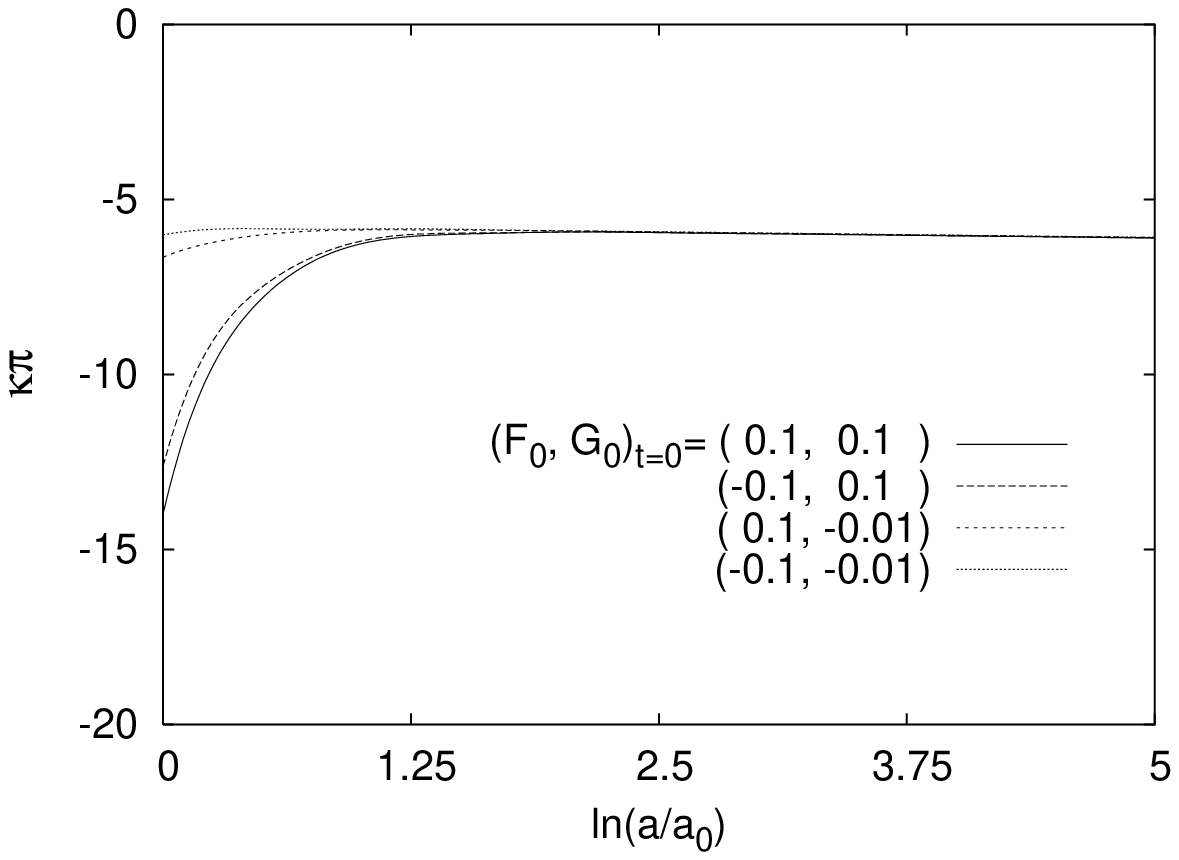}
 \caption{\label{fig:pi}
 The evolution of $\pi$ for four different sets
 of initial values. In the figure, $a$ is the scale factor and
 $a_0=a(t=0)$. The initial values of $\phi$ and $\dot{H}/H^2$ are 
 $c\kappa(\phi-\phi_0)|_{t=0}=0.01$ and $\dot{H}/H^2|_{t=0}=-0.02$. 
 Two more independent conditions at $t=0$ are given by setting 
 ${\cal F}_0$ and ${\cal G}_0$ in four different ways as
 $({\cal F}_0,{\cal G}_0)_{t=0}=(0.1,0.1)$, $(-0.1,0.1)$, 
 $(0.1,-0.01)$, $(-0.1,-0.01)$. 
 }
\end{figure}

\begin{figure}
 \centering\leavevmode\epsfysize=8cm \epsfbox{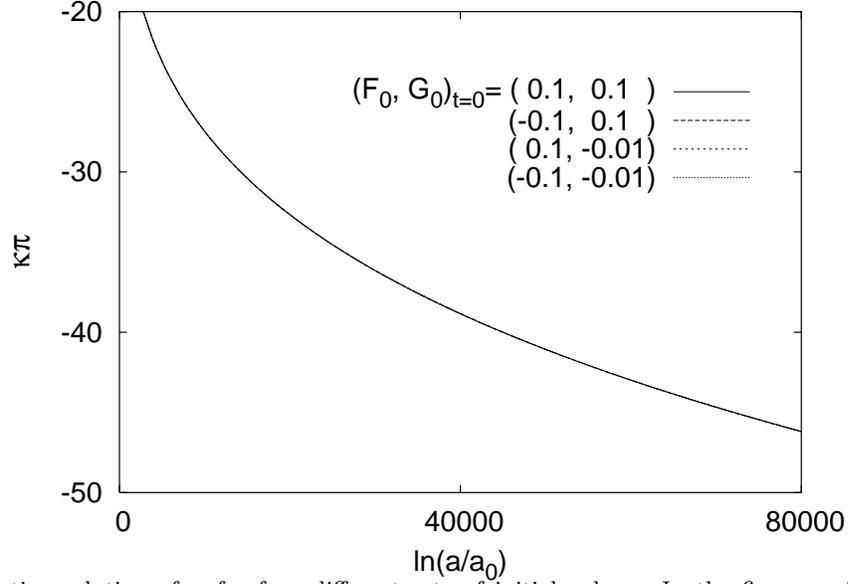}
 \caption{\label{fig:pi_long}
 The asymptotic evolution of $\pi$ for four different sets
 of initial values. In the figure, $a$ is the scale factor and
 $a_0=a(t=0)$. The initial values of $\phi$ and $\dot{H}/H^2$ are 
 $c\kappa(\phi-\phi_0)|_{t=0}=0.01$ and $\dot{H}/H^2|_{t=0}=-0.02$. 
 Two more independent conditions at $t=0$ are given by setting 
 ${\cal F}_0$ and ${\cal G}_0$ in four different ways as
 $({\cal F}_0,{\cal G}_0)_{t=0}=(0.1,0.1)$, $(-0.1,0.1)$, 
 $(0.1,-0.01)$, $(-0.1,-0.01)$. Four lines are degenerate because of
 the attractor behavior. 
 }
\end{figure}

First, let us confirm the behavior 
${\cal B}\equiv\kappa^3H\pi/c$ $\to -(3-\dot{H}/H^2)^{-1}$ stated just
after (\ref{eqn:eom-B}). Note that the feedback mechanism and the
stability analysis are based on this important
behavior. Fig.~\ref{fig:B} shows that $-(3-\dot{H}/H^2){\cal B}$ indeed 
converges to $1$ rather quickly. Thus, the evolution of $\pi$ is
determined by this attractor behavior indecently of the initial
condition. Figs~\ref{fig:pi} and \ref{fig:pi_long} show the evolution
of $\pi$.


\begin{figure}
 \centering\leavevmode\epsfysize=8cm \epsfbox{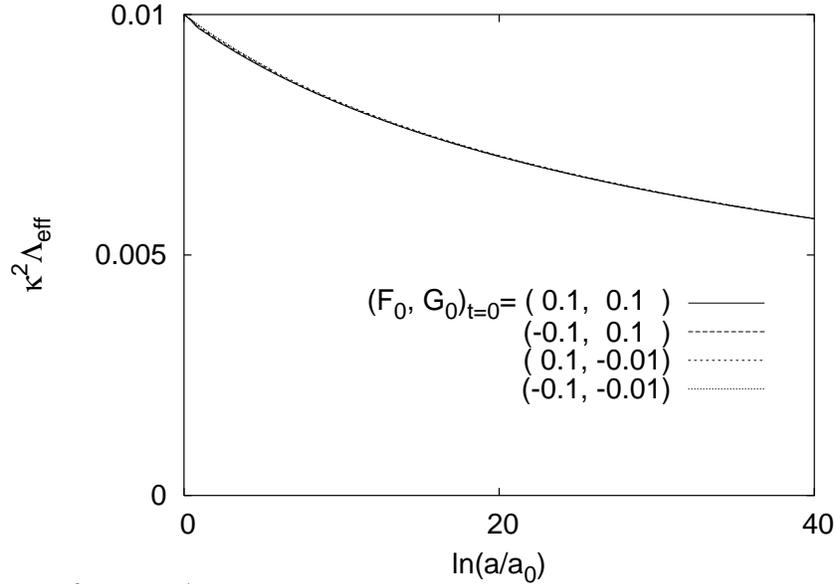}
 \caption{\label{fig:Lambda}
 The evolution of
 $\kappa^2\Lambda_{eff}=\kappa^4V(\phi)=c\kappa(\phi-\phi_0)$ for four
 different sets of initial values. In the figure, $a$ is the scale
 factor and $a_0=a(t=0)$. The initial values of $\phi$ and $\dot{H}/H^2$
 are $c\kappa(\phi-\phi_0)|_{t=0}=0.01$ and $\dot{H}/H^2|_{t=0}=-0.02$.
 Two more independent conditions at $t=0$ are given by setting 
 ${\cal F}_0$ and ${\cal G}_0$ in four different ways as
 $({\cal F}_0,{\cal G}_0)_{t=0}=(0.1,0.1)$, $(-0.1,0.1)$, 
 $(0.1,-0.01)$, $(-0.1,-0.01)$. 
 }
\end{figure}

\begin{figure}
 \centering\leavevmode\epsfysize=8cm \epsfbox{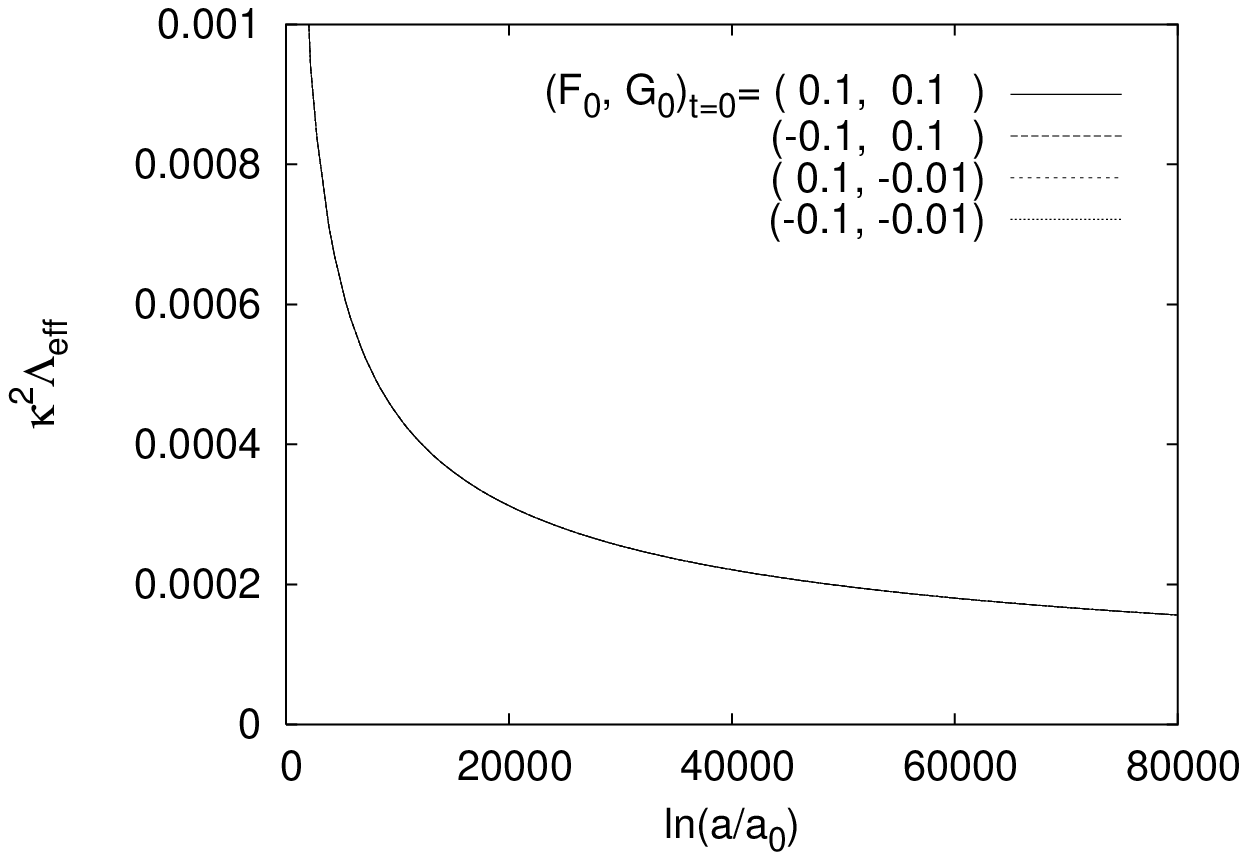}
 \caption{\label{fig:Lambda_long}
 The asymptotic evolution of
 $\kappa^2\Lambda_{eff}=\kappa^4V(\phi)=c\kappa(\phi-\phi_0)$ for four
 different sets of initial values. In the figure, $a$ is the scale
 factor and $a_0=a(t=0)$. The initial values of $\phi$ and $\dot{H}/H^2$
 are $c\kappa(\phi-\phi_0)|_{t=0}=0.01$ and $\dot{H}/H^2|_{t=0}=-0.02$.
 Two more independent conditions at $t=0$ are given by setting 
 ${\cal F}_0$ and ${\cal G}_0$ in four different ways as
 $({\cal F}_0,{\cal G}_0)_{t=0}=(0.1,0.1)$, $(-0.1,0.1)$, 
 $(0.1,-0.01)$, $(-0.1,-0.01)$. Four lines are degenerate because of the
 attractor behavior. 
 }
\end{figure}

Second, let us see that the feedback mechanism indeed works and the 
effective cosmological constant stalls near zero. Figs.~\ref{fig:Lambda}
and \ref{fig:Lambda_long} show that. It is worth while stressing again
that $V(\phi)$ includes the ground state energies of all fields and that
adding finite terms to the potential does never spoil the feedback
mechanism as far as the minimum of $V(\phi)$ remains negative. Although
the additional finite terms change the form of $V(\phi)$ but the
behavior of $V(\phi)$ near its root is always characterized by two
parameters $c$ and $\phi_0$ (i.e. the first order Taylor
expansion). The feedback mechanism works irrespective of the values of
$c$ and $\phi_0$. 

\begin{figure}
 \centering\leavevmode\epsfysize=8cm \epsfbox{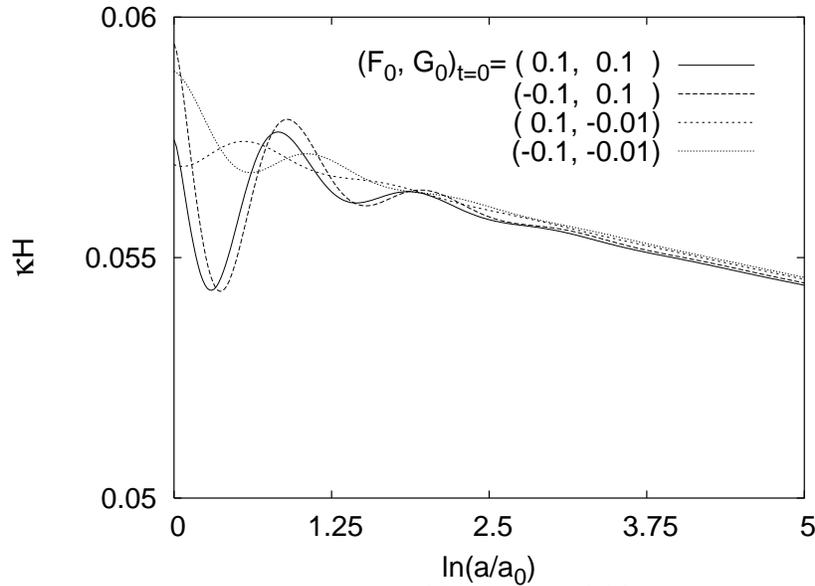}
 \caption{\label{fig:H}
 The evolution of the Hubble parameter $H$ for four
 different sets of initial values. In the figure, $a$ is the scale
 factor and $a_0=a(t=0)$. The initial values of $\phi$ and $\dot{H}/H^2$
 are $c\kappa(\phi-\phi_0)|_{t=0}=0.01$ and $\dot{H}/H^2|_{t=0}=-0.02$.
 Two more independent conditions at $t=0$ are given by setting 
 ${\cal F}_0$ and ${\cal G}_0$ in four different ways as
 $({\cal F}_0,{\cal G}_0)_{t=0}=(0.1,0.1)$, $(-0.1,0.1)$, 
 $(0.1,-0.01)$, $(-0.1,-0.01)$. 
 }
\end{figure}

\begin{figure}
 \centering\leavevmode\epsfysize=8cm \epsfbox{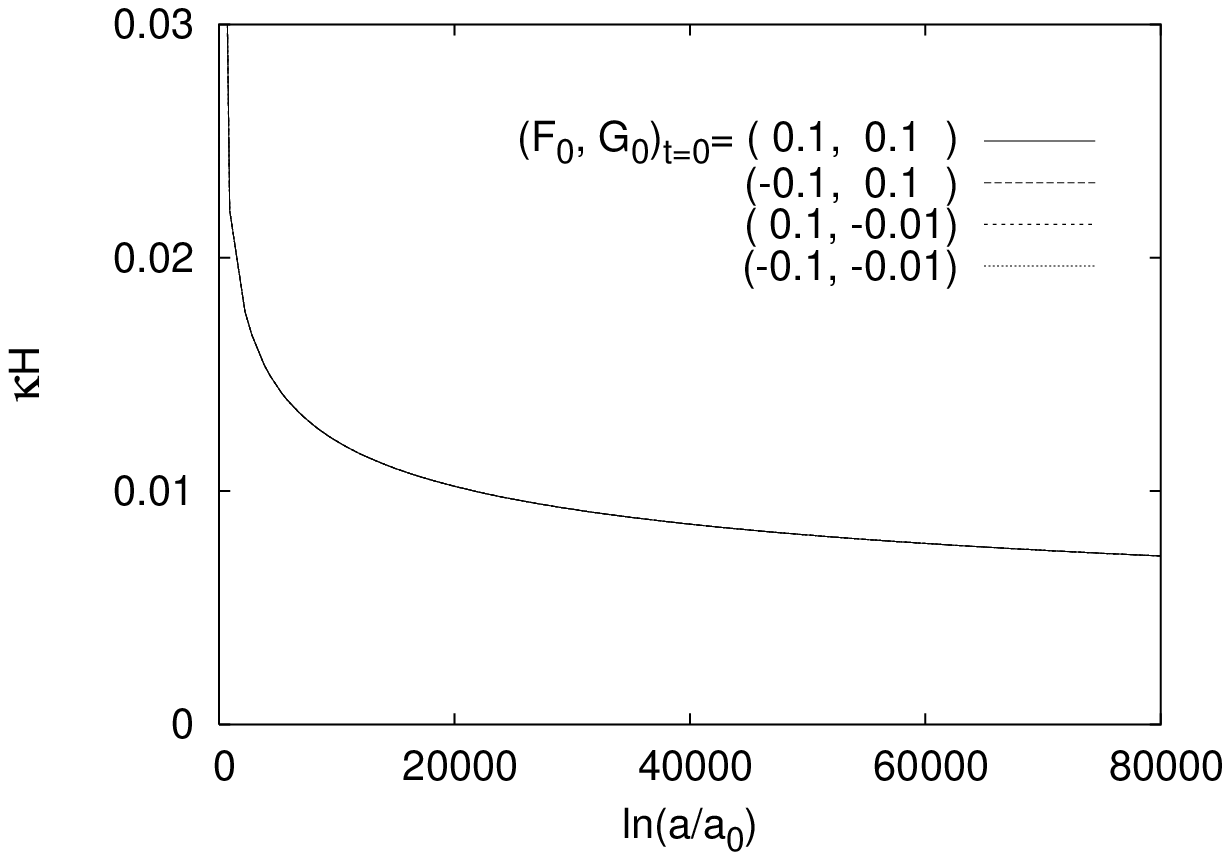}
 \caption{\label{fig:H_long}
 The asymptotic evolution of the Hubble parameter $H$ for four
 different sets of initial values. In the figure, $a$ is the scale
 factor and $a_0=a(t=0)$. The initial values of $\phi$ and $\dot{H}/H^2$
 are $c\kappa(\phi-\phi_0)|_{t=0}=0.01$ and $\dot{H}/H^2|_{t=0}=-0.02$.
 Two more independent conditions at $t=0$ are given by setting 
 ${\cal F}_0$ and ${\cal G}_0$ in four different ways as
 $({\cal F}_0,{\cal G}_0)_{t=0}=(0.1,0.1)$, $(-0.1,0.1)$, 
 $(0.1,-0.01)$, $(-0.1,-0.01)$. Four lines are degenerate because of the
 attractor behavior. 
 }
\end{figure}

\begin{figure}
 \centering\leavevmode\epsfysize=8cm \epsfbox{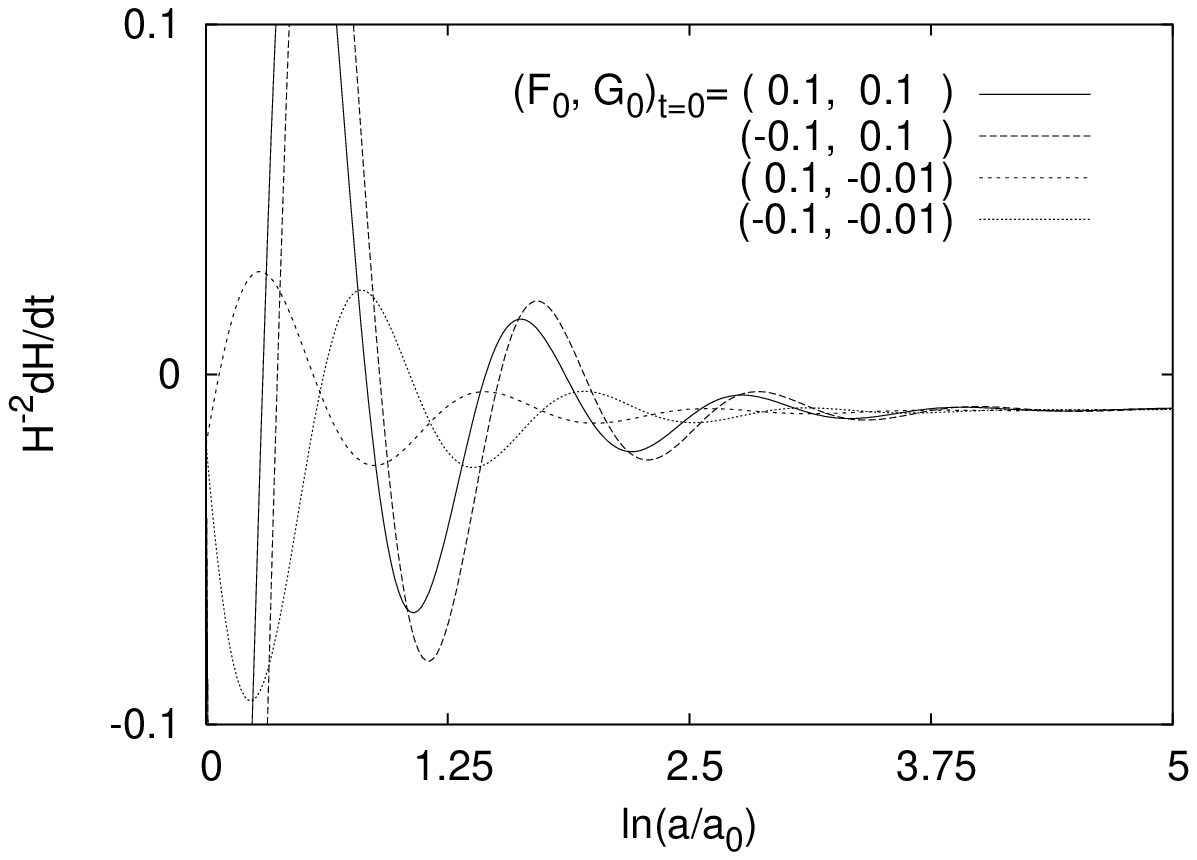}
 \caption{\label{fig:HdotoverH2}
 The evolution of the dimensionless evolution rate $\dot{H}/H^2$ of
 $H$ for four different sets of initial values. In the figure, $a$
 is the scale factor and $a_0=a(t=0)$. The initial values of $\phi$ and
 $\dot{H}/H^2$ are $c\kappa(\phi-\phi_0)|_{t=0}=0.01$ and
 $\dot{H}/H^2|_{t=0}=-0.02$. Two more independent conditions at $t=0$
 are given by setting ${\cal F}_0$ and ${\cal G}_0$ in four different
 ways as $({\cal F}_0,{\cal G}_0)_{t=0}=(0.1,0.1)$, $(-0.1,0.1)$, 
 $(0.1,-0.01)$, $(-0.1,-0.01)$. 
 }
\end{figure}

\begin{figure}
 \centering\leavevmode\epsfysize=8cm \epsfbox{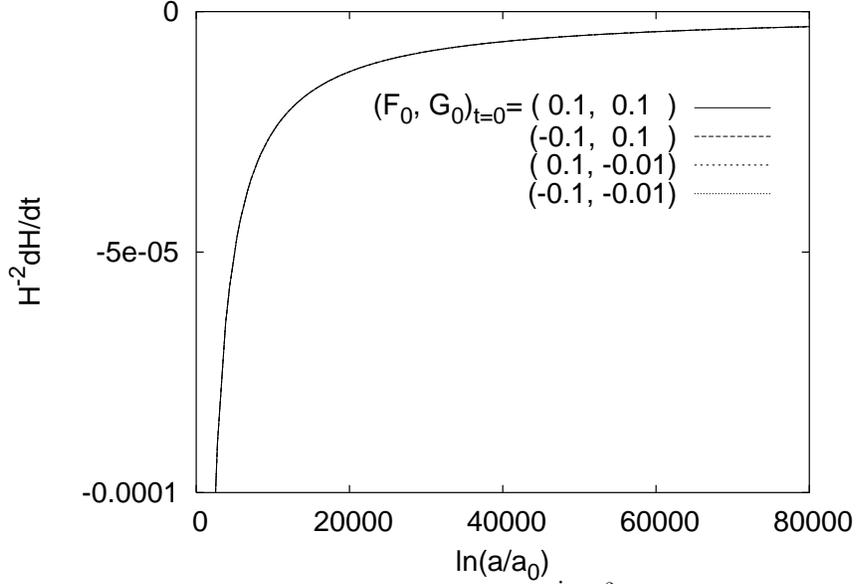}
 \caption{\label{fig:HdotoverH2_long}
 The asymptotic evolution of the dimensionless evolution rate
 $\dot{H}/H^2$ of $H$ for four different sets of initial values. In
 the figure, $a$ is the scale factor and $a_0=a(t=0)$. The initial
 values of $\phi$ and $\dot{H}/H^2$ are
 $c\kappa(\phi-\phi_0)|_{t=0}=0.01$ and $\dot{H}/H^2|_{t=0}=-0.02$. Two
 more independent conditions at $t=0$ are given by setting ${\cal F}_0$
 and ${\cal G}_0$ in four different ways as $({\cal F}_0,{\cal
 G}_0)_{t=0}=(0.1,0.1)$, $(-0.1,0.1)$, $(0.1,-0.01)$,
 $(-0.1,-0.01)$. Four lines are degenerate because of the attractor
 behavior. 
 }
\end{figure}

Third, because of the behavior of $\Lambda_{eff}$ governed by the
feedback mechanism, the Hubble parameter $H$ also approaches zero very 
slowly. Figs~\ref{fig:H} and \ref{fig:H_long} show this behavior of
$H$. Figs~\ref{fig:HdotoverH2} and \ref{fig:HdotoverH2_long} show that 
the dimensionless evolution rate $\dot{H}/H^2$ of $H$ indeed approaches
zero. Thus, asymptotically $H$ does not change in cosmological time
scale. These again confirm the statement that $\Lambda_{eff}$ stalls
near zero and that it does more slowly than matter or radiation. 

\begin{figure}
 \centering\leavevmode\epsfysize=8cm \epsfbox{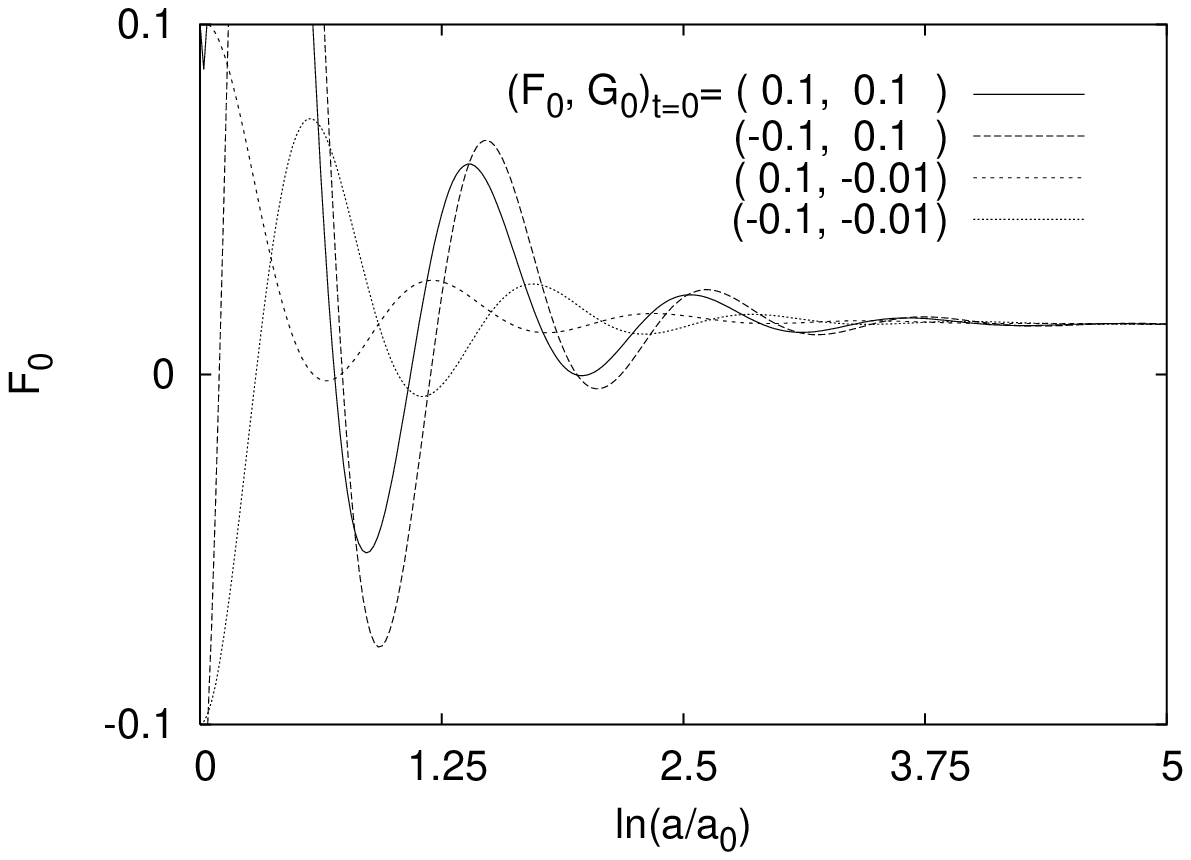}
 \caption{\label{fig:F0}
 The early stage of the recovery of the standard Friedmann equation
 (${\cal F}_0\to 0$) for four different sets of initial values. In
 the figure, $a$ is the scale factor and $a_0=a(t=0)$. The initial
 values of $\phi$ and $\dot{H}/H^2$ are
 $c\kappa(\phi-\phi_0)|_{t=0}=0.01$ and $\dot{H}/H^2|_{t=0}=-0.02$. Two
 more independent conditions at $t=0$ are given by setting ${\cal F}_0$
 and ${\cal G}_0$ in four different ways as $({\cal F}_0,{\cal
 G}_0)_{t=0}=(0.1,0.1)$, $(-0.1,0.1)$, $(0.1,-0.01)$,
 $(-0.1,-0.01)$. 
 }
\end{figure}

\begin{figure}
 \centering\leavevmode\epsfysize=8cm \epsfbox{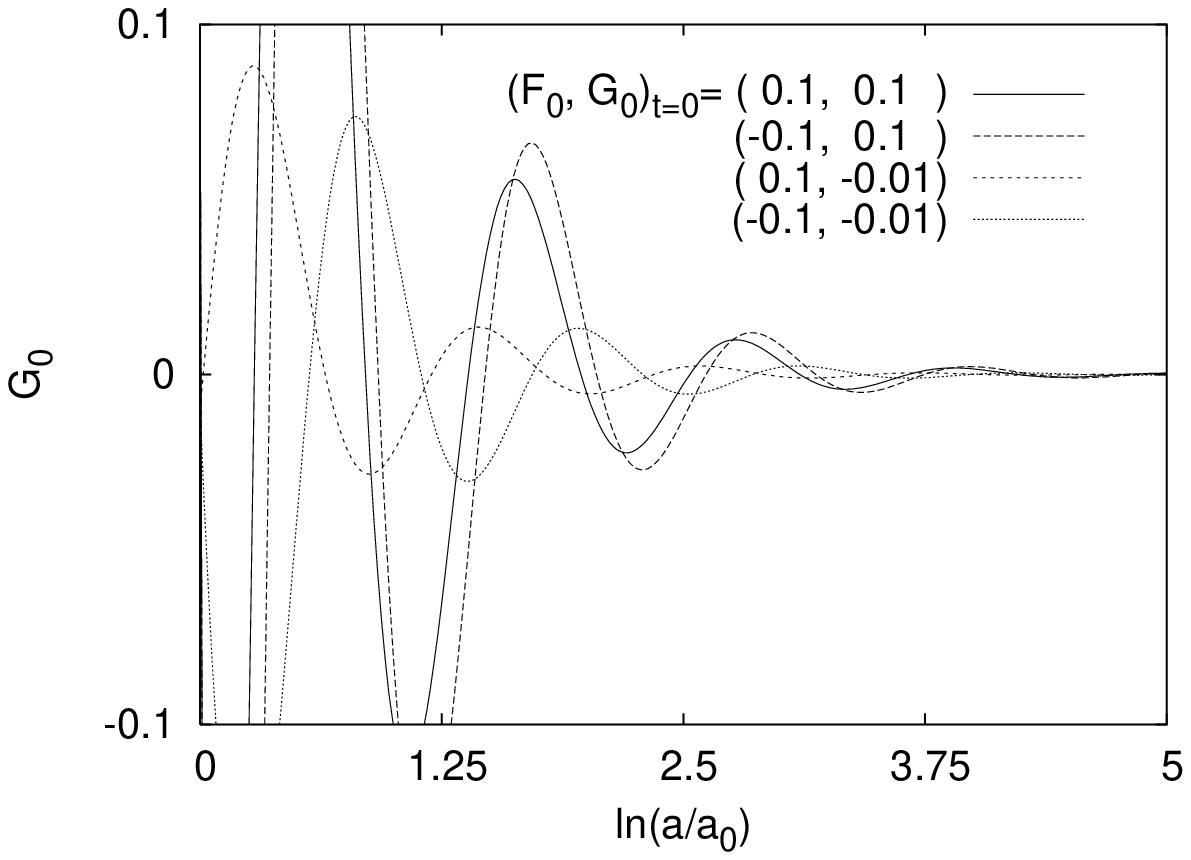}
 \caption{\label{fig:G0}
 Recovery of the standard dynamical equation (${\cal G}_0\to 0$) for
 four different sets of initial values. In the figure, $a$ is the
 scale factor and $a_0=a(t=0)$. The initial values of $\phi$ and
 $\dot{H}/H^2$ are $c\kappa(\phi-\phi_0)|_{t=0}=0.01$ and
 $\dot{H}/H^2|_{t=0}=-0.02$. Two more independent conditions at $t=0$
 are given by setting ${\cal F}_0$ and ${\cal G}_0$ in four different
 ways as $({\cal F}_0,{\cal G}_0)_{t=0}=(0.1,0.1)$, $(-0.1,0.1)$,
 $(0.1,-0.01)$, $(-0.1,-0.01)$. 
 }
\end{figure}

\begin{figure}
 \centering\leavevmode\epsfysize=8cm \epsfbox{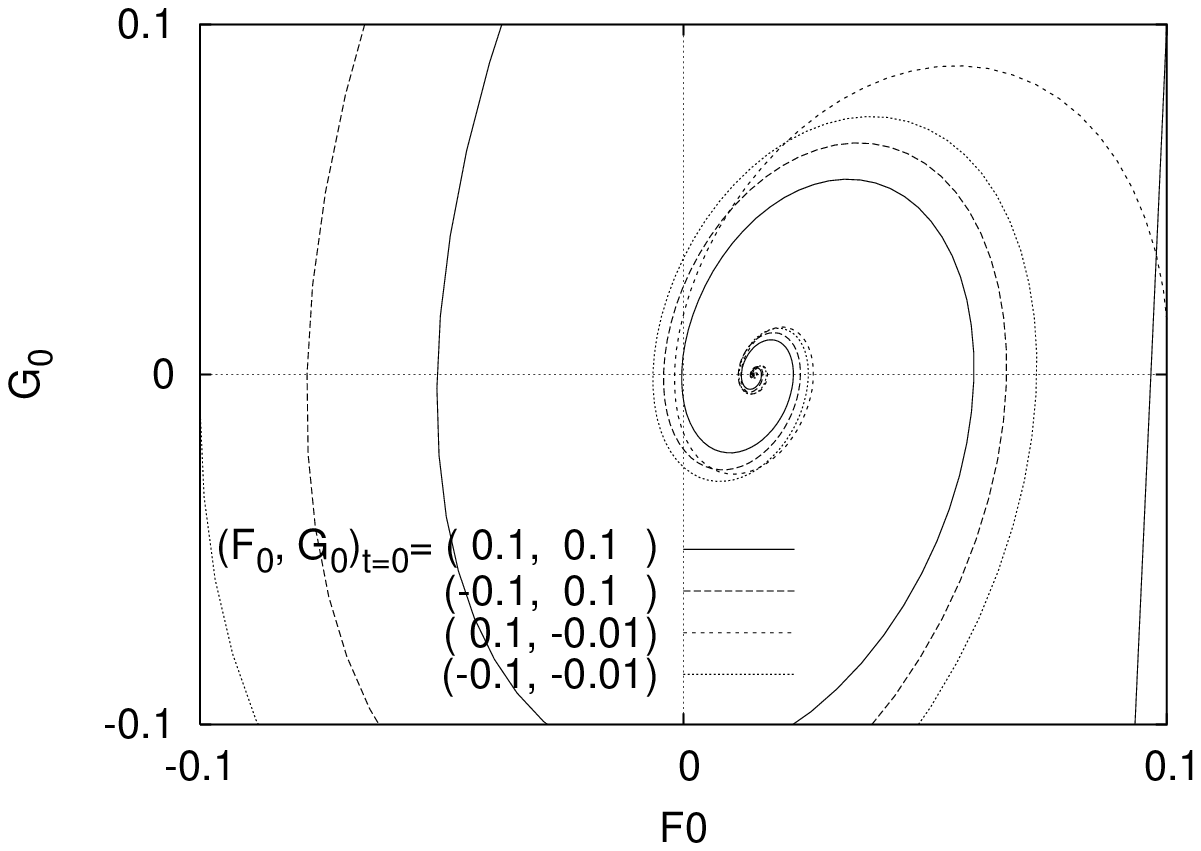}
 \caption{\label{fig:F0G0}
 The early stage of the recovery process of the standard Friedmann and
 dynamical equations (${\cal F}_0\to 0$, ${\cal G}_0\to 0$) for four
 different sets of initial values. In the figure, $a$ is the scale
 factor and $a_0=a(t=0)$. The initial values of $\phi$ and $\dot{H}/H^2$
 are $c\kappa(\phi-\phi_0)|_{t=0}=0.01$ and
 $\dot{H}/H^2|_{t=0}=-0.02$. Two more independent conditions at $t=0$ 
 are given by setting ${\cal F}_0$ and ${\cal G}_0$ in four different
 ways as $({\cal F}_0,{\cal G}_0)_{t=0}=(0.1,0.1)$, $(-0.1,0.1)$,
 $(0.1,-0.01)$, $(-0.1,-0.01)$. 
 }
\end{figure}

\begin{figure}
 \centering\leavevmode\epsfysize=8cm \epsfbox{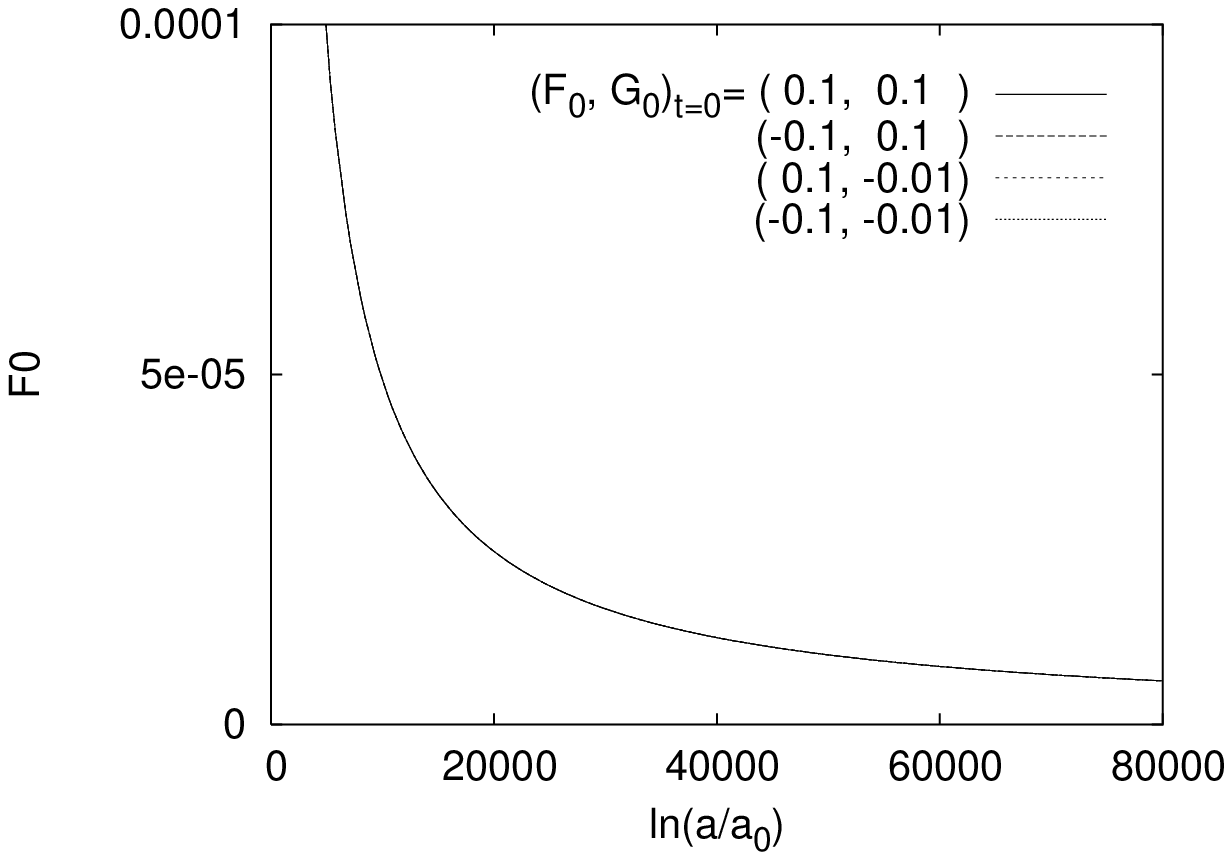}
 \caption{\label{fig:F0_long}
 The late stage of the recovery of the standard Friedmann equation
 (${\cal F}_0\to 0$) for four different sets of initial values. In
 the figure, $a$ is the scale factor and $a_0=a(t=0)$. The initial
 values of $\phi$ and $\dot{H}/H^2$ are
 $c\kappa(\phi-\phi_0)|_{t=0}=0.01$ and $\dot{H}/H^2|_{t=0}=-0.02$. Two
 more independent conditions at $t=0$ are given by setting ${\cal F}_0$
 and ${\cal G}_0$ in four different ways as $({\cal F}_0,{\cal
 G}_0)_{t=0}=(0.1,0.1)$, $(-0.1,0.1)$, $(0.1,-0.01)$,
 $(-0.1,-0.01)$. Four lines are degenerate because of the attractor
 behavior. 
 }
\end{figure}

Fourth, we can also see that the standard Friedmann equation 
is recovered (${\cal F}_0\to 0$) and that the standard dynamical
equation is also recovered (${\cal G}_0\to 0$). What actually happens is
that the standard dynamical equation is recovered earlier than the
standard Friedmann equation and that the latter is gradually recovered
after that. Figs.~\ref{fig:F0}, \ref{fig:G0} and \ref{fig:F0G0} show the
early stage of the recovery process. Fig.~\ref{fig:F0_long} shows the
late stage the recovery of the standard Friedmann equation.

\begin{figure}
 \centering\leavevmode\epsfysize=8cm \epsfbox{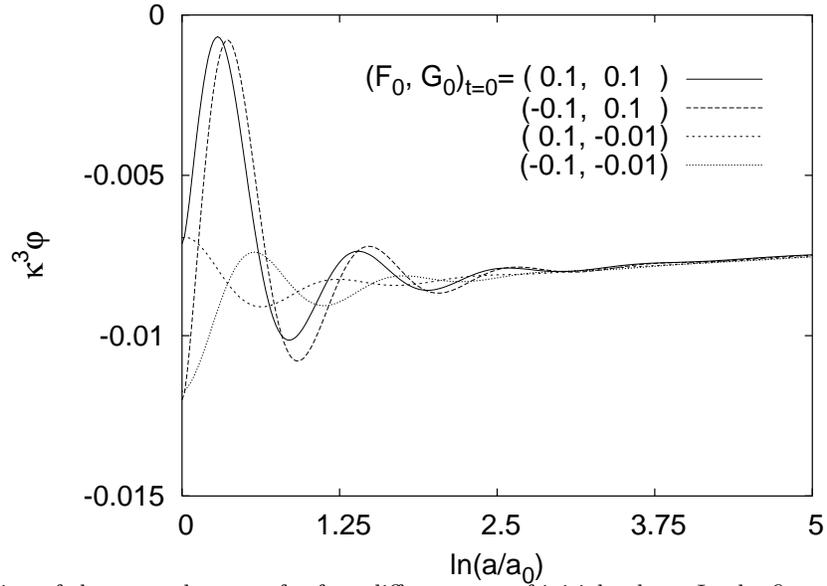}
 \caption{\label{fig:varphi}
 The evolution of the extra degree $\varphi$ for four different sets of
 initial values. In the figure, $a$ is the scale factor and
 $a_0=a(t=0)$. The initial values of $\phi$ and $\dot{H}/H^2$ are 
 $c\kappa(\phi-\phi_0)|_{t=0}=0.01$ and $\dot{H}/H^2|_{t=0}=-0.02$. Two
 more independent conditions at $t=0$ are given by setting ${\cal F}_0$
 and ${\cal G}_0$ in four different ways as $({\cal F}_0,{\cal
 G}_0)_{t=0}=(0.1,0.1)$, $(-0.1,0.1)$, $(0.1,-0.01)$,
 $(-0.1,-0.01)$. 
 }
\end{figure}

\begin{figure}
 \centering\leavevmode\epsfysize=8cm \epsfbox{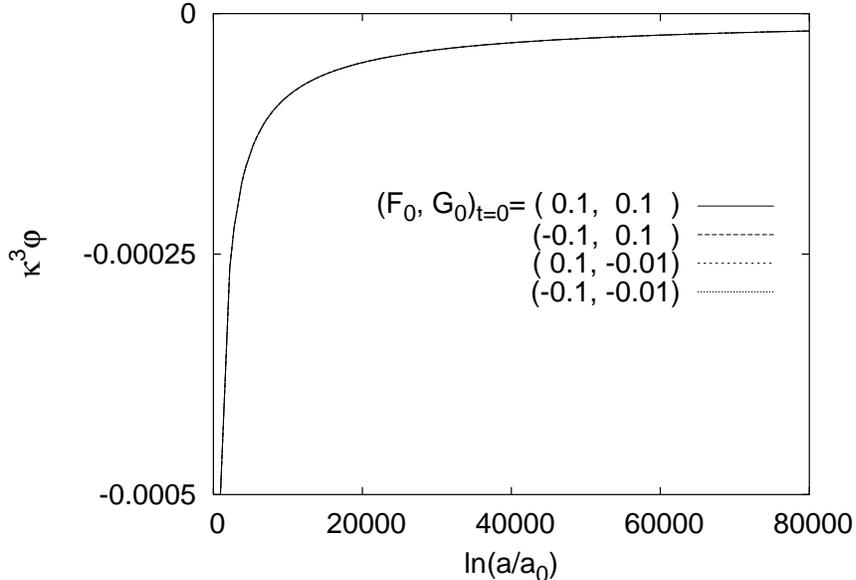}
 \caption{\label{fig:varphi_long}
 The asymptotic evolution of the extra degree $\varphi$ for four
 different sets of initial values. In the figure, $a$ is the scale
 factor and $a_0=a(t=0)$. The initial values of $\phi$ and $\dot{H}/H^2$
 are $c\kappa(\phi-\phi_0)|_{t=0}=0.01$ and
 $\dot{H}/H^2|_{t=0}=-0.02$. Two more independent conditions at $t=0$
 are given by setting ${\cal F}_0$ and ${\cal G}_0$ in four different
 ways as $({\cal F}_0,{\cal G}_0)_{t=0}=(0.1,0.1)$, $(-0.1,0.1)$,
 $(0.1,-0.01)$, $(-0.1,-0.01)$. Four lines are degenerate because of the
 attractor behavior. 
 }
\end{figure}

Finally, we can check that the extra degree of freedom $\varphi$ goes to
zero. Figs~\ref{fig:varphi} and \ref{fig:varphi_long} confirm that
behavior. 


\section{Reheating}
\label{sec:reheating}

We have achieved the vanishing cosmological constant in a way
that is stable under radiative corrections and that has self-consistent,
stable dynamics. However, although the cosmological constant approaches
zero, it does so more slowly than matter or radiation so that without 
additional structure, the universe would be empty. It is not entirely
clear that a dynamical model where this is not the case could be
successful since it is this property that makes it possible for all
fields other than $\phi$ to settle into their ground state before $\phi$
stalls at zero curvature so that the zero curvature really corresponds
to the vanishing cosmological constant. Moreover, the singular behavior
of a kinetic term coefficient and, thus, the slow evolution of $\phi$
are required for stability. This does imply, however, that should this
mechanism be responsible for a low cosmological constant, reheating
would be required to thermally populate the universe after the
cosmological constant has decreased to a small value. Nonetheless, as we
shall see below, this model can at the very least reduce fine-tuning by
$60$ orders of magnitude or provide a new mechanism for sampling
possible cosmological constants and implementing the anthropic
principle.

The reheat process requires further speculation, and is a subject for
future research. A couple of possibilities are:

(I) Low-energy inflation. One can consider an extra scalar field $\chi$
with mass $m_{\chi}\sim 10^{-3}eV$ and a term like $-R\chi^2$. When
$R\sim m_{\chi}^2$, a phase transition would occur (as in hybrid
inflation \cite{Lindebook}) and the universe would be reheated up to
temperature $\sim TeV$.
This phase transition happens when the energy stored in $\chi$ plus
the energy stored in $\phi$  yields a Hubble constant of approximately
$m_{\chi}$. The energy in $\chi$ will decrease during the phase
transition; the energy after the phase transition
must be very small. In this case the cosmological constant problem is
reduced from $(M_{Pl}/10^{-3}eV)^4\sim 10^{120}$ to
$(TeV/10^{-3}eV)^4\sim 10^{60}$. For smaller $m_{\chi}$, the reheat
temperature would be lower and the tuning of the cosmological constant
would presumably be smaller.

(II) Energy inflow from extra dimensions. For example, in a
non-elastic scattering of branes, a part of the kinetic energy due
to the relative motion can be converted to radiation on our brane
without changing the brane tension and the cosmological constant.
For this to work, branes should be sufficiently flat and parallel.

After reheating, the conventional standard cosmology with a vanishingly
small cosmological constant can be restored. This is because the scalar
field $\phi$ is just frozen at low energy. Namely,
%
\begin{equation}
 |\kappa^2\dot{\phi}| \sim
  (\kappa H)^{4m}\times\left|\kappa^2\pi\right|^{\frac{1}{2q-1}}
  \label{eqn:Vdot}
\end{equation}
can be made arbitrarily small at low energy ($\kappa H\ll 1$) by
considering a sufficiently large $m$, where $\kappa^2\pi$ is estimated
by (\ref{eqn:asymptotic-pi}). In the scenario (I), both $\pi$ and $H$
evolve continuously through the reheating epoch. Hence we obtain
%
\begin{equation}
 |\kappa^5\dot{V}| \sim (\kappa H)^{4m-\frac{1}{2q-1}},
  \label{eqn:dphi-small-I}
\end{equation}
and the $4$d cosmological constant does not overshoot zero within the
present age of the universe if $m$ is large enough to ensure that
$(\kappa^2R_{reh})^{2m-\frac{1}{2(2q-1)}}<\kappa H_{today}
[\kappa^2R_{reh}/4-\kappa^4(U(\chi_+)|_{R=R_{reh}}-U(\chi_-))]$, 
where $R_{reh}$ is the value of the Ricci scalar at reheating, $\chi_+$
and $\chi_-$ are the expectation value of $\chi$ before and after the
phase transition, respectively. Hereafter, we set $c=O(1)$.

On the other hand, in the scenario (II), while $\pi$ evolves
continuously, $H$ increases suddenly by energy flow from the extra
dimension. Hence, $H$ in (\ref{eqn:asymptotic-pi}) and (\ref{eqn:Vdot})
should be estimated before and after reheating, respectively. Therefore,
we obtain from (\ref{eqn:Vdot}) the following estimate for the period
soon after reheating.
%
\begin{equation}
 \kappa|\dot{V}|/V  \sim
  (\kappa H_{after})^{4m}/
  (\kappa H_{before})^{\frac{1}{2q-1}+2},
 \label{eqn:recovery-cond}
\end{equation}
where we have used the Friedmann equation
$3H_{before}^2\sim \kappa^2V$. Here, $H_{before}$ and $H_{after}$ are
the Hubble parameter before and after the energy inflow from extra
dimensions, respectively. By using (\ref{eqn:eom-B}) it is easily
shown that $\kappa^3\pi H+c/5\propto a^{-5}$ during the
radiation-dominated epoch ($\dot{H}/H^2=-2$). Hence, $\kappa^3\pi H$
decays from $\sim H_{after}/H_{before}$ to $O(1)$ in the time scale
$\Delta t_{relax}\sim (H_{after}/H_{before})^{2/5}H_{after}^{-1}$.
Thus, if $m$ is large enough to ensure that
$(\kappa H_{after})^{4m-\frac{3}{5}}
<(\kappa H_{before})^{\frac{1}{2q-1}+\frac{12}{5}}$
then $V$ does not overshoot zero
in the time scale $\Delta t_{relax}$, provided that
$\kappa H_{after}\ll 1$. After that, we can use the estimate
(\ref{eqn:dphi-small-I}) with $H$ being the realtime value, and $V$ does
not overshoot zero if $m$ is sufficiently large. Let us recall that a
choice like $f(R)\sim\exp[-\kappa^{-4}R^{-2}]$ corresponds to
$m\to\infty$.

One might worry about symmetry restoration and phase transitions
that occur  after reheating. This is not a problem in both
examples if $m$ is sufficiently large. The reason   is that
$R_{reh}/4-\kappa^2(U(\chi_+)|_{R=R_{reh}}-U(\chi_-))$ (or $\kappa^2 V$,
respectively) is the cosmological constant at zero temperature since the
temperature before the reheating is zero. The large $m$ ensures that
$\Lambda_{today}$ is still positive and small since $\phi$ continues to
be almost frozen all the way down to the present epoch including the
time when the symmetry is restored and during the time the phase
transition takes place.

Even if all else fails, although not our initial subjective, the
existence of $\phi$ can at the very least provide a natural framework in
which to implement the anthropic principle. If we assume eternal
inflation, there would be many inflationary universes. In each universe,
the cosmological constant is determined by how much $\phi$ has rolled
when inflation ends. That in turn depends on the number of $e$-foldings
that occurred before inflation stopped. In an eternal inflation
scenario, different numbers of $e$-foldings would occur in different
domains and therefore different $\phi$ values, and hence different
values of the cosmological constant would occur in different regions.

Our model in general predicts
$w_{\phi}\equiv p_{\phi}/\rho_{\phi}\simeq -1$ today because of
(\ref{eqn:dphi-small-I}) since pre- and post-reheating behavior requires
a large $m$. In the scenario (I), $|\dot{V}|$ continuously decreases. In
the scenario (II), $|\dot{V}|$ increases suddenly at reheating but is
adjusted to the behavior (\ref{eqn:dphi-small-I}) in the time scale
$\Delta t_{relax}$. Note that
$H_{today}\Delta t_{relax}<(H_{before}/H_{after})^{3/5}\ll 1$ since
$H_{after}^2\gg H_{before}^2\simeq \Lambda_{before}/3>\Lambda_{today}/3
\simeq H_{today}^2$.
This means that the behavior (\ref{eqn:dphi-small-I}) and
$w_{\phi}\simeq -1$ are realized before the present stage of the
universe.


\section{Classical stability II - Linearized gravity}
\label{sec:gravity}

In this section, restricting to the $q=1$ case for simplicity, we show
that linearized Einstein gravity in a Minkowski background is recovered
at distances longer than the length scale $l_*=\sqrt{\alpha}\kappa$ and
at energies lower than $l_*^{-1}$. The extension to general $q$ ($>1/2$)
should be straightforward.

A possible source of instability which would disturb weak gravity in a
Minkowski background could be an inhomogeneous fluctuation of $\phi$,
which could possibly generate violent breakdown of linearized Einstein
gravity. Quite surprisingly, this is not the case and the seemingly
most dangerous part, the kinetic term, is not as dangerous as it
looks. Essential to this conclusion is the constraint equation, which
prevents $\phi$ from fluctuating freely and forces the denominator and
the numerator to fluctuate in a strongly correlated way so that the
contributions of the kinetic term to the equation of motion are regular
and much smaller than those of the $\alpha R^2$ term. Namely, the scalar
field and metric cannot fluctuate independently. In fact, it is a
well-known fact in the standard scalar field cosmology that there is
only one physical degree of freedom out of scalar field and metric
degrees for scalar-type perturbations. In the longitudinal gauge (see
(\ref{eqn:longitudinal-gauge}) or
(\ref{eqn:longitudinal-gauge-appendix})), one of the constraint
equations is the ($0i$)-component of the perturbed Einstein equation,
which is roughly of the form
%
\begin{equation}
 \pi \partial_i(\delta\phi) \simeq \frac{\delta G_{0i}}{\kappa^2} -
  \delta T_{0i}^{(other)},
\end{equation}
where $\pi$ is the momentum conjugate to the homogeneous background
$\phi$ defined in (\ref{eqn:eom-homogeneous-phi}), $\delta\phi$ is the
inhomogeneous perturbation of $\phi$, $\delta T_{0i}^{(other)}$ is the 
($0i$)-component of the stress-energy tensor of fields and matter other
than $\phi$. Note that homogeneous perturbations can be absorbed into
the background and that, without loss of generality, we can restrict our
analysis to inhomogeneous perturbations, for which
$\partial_i(\delta\phi)\ne 0$. The right hand side of this equation is
manifestly regular and we already know that $\pi$ behaves like
(\ref{eqn:asymptotic-pi}). Thus, we conclude that 
%
\begin{equation}
 \delta\phi \simeq \kappa H \times (\mbox{regular expression}). 
\end{equation}
This means that $\delta\phi$ is indeed suppressed at low energy 
($\kappa H\ll 1$). Similarly, the other constraint equation, namely the
traceless part of the ($ij$)-components of the perturbed Einstein
equation, leads to 
%
\begin{equation}
 (\delta\phi)^{\cdot} = 
  (\kappa H)^3 \times (\mbox{regular expression}). 
\end{equation}
This again means that inhomogeneous perturbation of $\phi$ is
suppressed at low energy. Hence, from the constraint equations, we
expect that the stress energy tensor of the scalar field should be
small enough. More rigorous treatment of the constraint equations is 
included in the complete analysis shown in
Appendix~\ref{app:linear-gravity}.

Since contributions of the singular-looking kinetic term are expected to
be small enough, the only possibly important correction to Einstein
gravity is again due to the $\alpha R^2$ term. This tells us that the
linearized gravity in our model should be similar to that in the
theory $R/2\kappa^2+\alpha R^2$. Hence, we should be able to recover the
linearized Einstein gravity in Minkowski background at distances longer 
than the length scale $l_*=\sqrt{\alpha}\kappa$ and at energies lower
than $l_*^{-1}$~\cite{Stelle}.

A complete analysis of the weak gravity in Minkowski background requires
much more carefulness. One of the reasons is that the action for the
scalar field is not manifestly well defined in Minkowski background
since the denominator of the kinetic term vanishes at $R=0$. Moreover,
it is not totally clear how to treat the perturbations of the
denominator. This situation requires a sort of regularization for the
field equation. In the end of the calculation we of course need to take
the limit where the regularization is turned off. Unlike calculations in
quantum field theories, we must not renormalize anything since we are
dealing with classical dynamics. Namely, before turning off the
regularization, we must not subtract anything from the regularized field
equation and must treat it as it is.

For the purpose of the rigorous treatment of weak gravity, we
investigate in detail perturbations around the flat FRW background in 
the longitudinal gauge and take the $\kappa H\to +0$ limit in the end of
the calculation. This strategy makes it possible to analyze linear
perturbations around Minkowski background, on which the action of the
scalar field perturbation is not apparently well-defined. For
simplicity, we assume that the FRW background is driven by the scalar
field $\phi$. Of course, we include a general matter stress-energy
tensor into perturbations and carefully investigate how they couple to
gravity. A critical point for this analysis is that constraint equations
(($0i$)-components and the traceless part of ($ij$)-components of
Einstein equation) prevent $\phi$ from fluctuating freely, and  it turns
out that the $\phi$ perturbation is of sufficiently high order in the
$\kappa H$ expansion that the perturbative analysis is under control.

The result of the analysis is simple. The linearized gravity in our
model is similar to that in the theory $R/2\kappa^2+\alpha R^2$ in the
sense that the only difference is the existence of an extra scalar-type
massless mode in our model. The extra massless mode is decoupled from
the matter stress-energy tensor at the linearized level. Hence, as far
as we are concerned with classical, linear perturbations generated by
matter sources, these two theories give exactly the same
prediction. Therefore, in our model the linearized Einstein gravity in
Minkowski background is recovered at distances longer than the length
scale $l_*=\sqrt{\alpha}\kappa$ and at energies lower than $l_*^{-1}$.

Let us begin by defining the metric perturbation $\delta g_{\mu\nu}$ by
%
\begin{equation}
 g_{\mu\nu} = g^{(0)}_{\mu\nu} + \delta g_{\mu\nu}, 
\end{equation}
where
%
\begin{equation}
 g^{(0)}_{\mu\nu}dx^{\mu}dx^{\nu} = -dt^2 + a(t)^2\delta_{ij}dx^idx^j
\end{equation}
($i=1,2,3$). We shall investigate the system of coupled equations for
the scalar field perturbation $\delta\phi$, the metric perturbation
$\delta g_{\mu\nu}$ and the stress energy tensor contributions from
other fields and matter, taking into account all higher derivative
corrections. In order to investigate linearized gravity in Minkowski 
background, in the end of calculations, we shall take the limit  
%
\begin{equation}
 \kappa H\to +0,
\end{equation}
keeping $\partial_t^nH/H^{n+1}$ ($n=1,2,3$) finite, where
$H=\partial_ta/a$. In the following, we shall denote $\partial_t$ by an 
overdot.

In order to take the full advantage of the symmetry of the background
spacetime (with $H\ne 0$), we expand the scalar field perturbation, the
metric perturbation and the stress energy tensor contributions from
other matter and fields by harmonics on the $3$-plane with each 
expansion coefficient being a function of the time $t$ only. Now it is
well-known and easily shown that there are three distinct types of
perturbations and that each type is decoupled from the others in the
linearized level. They are called scalar-, vector- and tensor-type
perturbations and each of them is an irreducible representation of the
symmetry of the background FRW spacetime: spin-$0$, spin-$1$ and
spin-$2$ representations, respectively. Since scalar-, vector-, and
tensor-type perturbations are decoupled from each other in the 
linearized level, we can analyze perturbations of each type separately.

The vector- and tensor-type perturbations are decoupled from scalar-type
perturbations and, thus, from the perturbation of the scalar field
$\phi$. Moreover, the linear perturbation of the Ricci scalar vanishes
for vector- and tensor-type perturbations. Thus, the $\kappa H\to +0$
limit of perturbation equations is manifestly well-defined. Actually, in
the $\kappa H\to +0$ limit, the linear perturbation of the stress energy
tensor $T_{\phi\mu\nu}$ of $\phi$ vanishes. Therefore, we obtain 
%
\begin{equation}
 \delta G_{\mu\nu} = \kappa^2T^{other}_{\mu\nu}. 
\end{equation}
This is nothing but the linearized Einstein equation. In other words,
possible corrections to the linearized Einstein equation appear only for
the scalar-type perturbations.

So, let us concentrate on the scalar-type perturbations. As is
well-known, in linear perturbations we have not only physical degrees of
freedom but also gauge degrees of freedom so we need to fix gauge. (See 
subsection~\ref{subsec:gauge-choice} of
Appendix~\ref{app:linear-gravity} for the formula of infinitesimal gauge
transformation and a gauge choice.) In the longitudinal gauge, the
perturbed metric is written as 
%
\begin{equation}
 g_{\mu\nu}dx^{\mu}dx^{\nu} = 
  -(1+2\Phi Y)dt^2 + (1-2\Psi Y)a^2\delta_{ij}dx^idx^j,
  \label{eqn:longitudinal-gauge}
\end{equation}
where $\Phi$ and $\Psi$ are functions of $t$, and $Y$ is the scalar
harmonics on the $3$-plane with $3$-momentum $k_i$. Here, we omit to
write $k_i$ and the integration over all possible values of $k_i$.

As already stated, we shall analyze the system of coupled equations for
the scalar field perturbation $\delta\phi$, the metric perturbation
$\delta g_{\mu\nu}$ and the stress energy tensor contributions from
other fields and matter in the FRW background, taking into account all
higher derivative corrections. In the end of calculations, we shall take
the limit $\kappa H\to +0$, keeping $\partial_t^nH/H^{n+1}$ ($n=1,2,3$)
finite. For this purpose, we introduce a small dimensionless parameter
$\epsilon$ so that 
%
\begin{eqnarray}
 \kappa H & = & O(\epsilon), \nonumber\\
 \frac{\kappa^2k^2}{a^2} & = & O(\epsilon^0).
\end{eqnarray}
and expand all quantities and equations. We shall keep the expanded
equations up to the order $O(\epsilon^2)$ to obtain equations governing
the quantities of order $O(\epsilon^0)$. In the end of the calculation,
we shall take the limit $\epsilon\to 0$.

Complete analysis of the scalar-type perturbations is given in
subsection~\ref{subsec:lineargravity-scalar} of
Appendix~\ref{app:linear-gravity}. The equations governing the
$O(\epsilon^0)$ order metric perturbations are obtained after going
through the procedure described above and eliminating the scalar field
perturbation. The result is summarized as the following equations for 
linearized gravity in Minkowski background. 
%
\begin{eqnarray}
 \kappa^2k^2\Psi_+  +\kappa^4\left(\tau_{00}
	    +2k^2\tau_{(LL)}\right) & = & 0, \nonumber\\ 
 \Box\left[\left(1-12\alpha\kappa^2\Box\right)\Psi_- 
      + 4\alpha\kappa^2
      \left(3\ddot{\Psi}_+ + k^2\Psi_+\right)
      -2\kappa^2\tau_{(LL)}
     \right] & = & 0,\label{eqn:result-scalar}
\end{eqnarray}
where $\tau$'s are scalar-type perturbations of the stress energy tensor
contributions from other fields and matter defined by
(\ref{eqn:T-harmonics}). Here, we have redefined the spatial coordinates
as $a_0x^i\to x^i$ and 
%
\begin{equation}
 \Box \equiv -\partial_t^2 -k^2.
\end{equation}
Before taking the $\kappa H\to 0$ limit, $a_0$ is defined as the value
of the scale factor at the time when a weak gravity experiment is
performed, assuming that the duration of the experiment is much shorter
than the cosmological time scale $H^{-1}$. After taking the 
$\kappa H\to 0$ limit, the scale factor is a constant and, thus, the
above rescaling of the spatial coordinates makes perfect sense at any
time. The perturbed metric, after the $\kappa H\to +0$ limit and the
above redefinition of the spatial coordinates, is 
%
\begin{equation}
 ds^2 = 
  -(1+2\Phi Y)dt^2 + (1-2\Psi Y)\delta_{ij}dx^idx^j,
  \label{eqn:scalar-pert-flat}
\end{equation}
and $\Psi_{\pm}=\Psi\pm\Phi$. The conservation equation is 
%
\begin{eqnarray}
 \dot{\tau}_{00}+k^2\tau_{(L)0} & = & 0,\nonumber\\
 \dot{\tau}_{(L)0} + \frac{4}{3}k^2\tau_{(LL)}
  - \tau_{(Y)} & = & 0. 
\end{eqnarray}

Now it is time to show that the linearized Einstein gravity in 
Minkowski background is recovered at distances longer than the length
scale $l_*=\sqrt{\alpha}\kappa$ and at energies lower than
$l_*^{-1}$. (Note that we have assumed that $\alpha$ is positive.) For
this purpose and to make the arguments qualitative, let us compare the
result (\ref{eqn:result-scalar}) for scalar-type perturbations in our
model with the corresponding equations in the higher-curvature theory
whose gravitational action is 
%
\begin{equation}
 I_{HD} = \int d^4x\sqrt{-g}
  \left(\frac{1}{2\kappa^2}R + \alpha R^2\right).
  \label{eqn:ordinary-HD}
\end{equation}
For scalar perturbations given by (\ref{eqn:scalar-pert-flat}) and 
(\ref{eqn:T-harmonics}), linearized gravity equations in the theory 
(\ref{eqn:ordinary-HD}) are 
%
\begin{eqnarray}
 \kappa^2k^2\Psi_+  +\kappa^4\left(\tau_{00}
	    +2k^2\tau_{(LL)}\right) & = & 0, \nonumber\\ 
 \left(1-12\alpha\kappa^2\Box\right)\Psi_- 
      + 4\alpha\kappa^2
      \left(3\ddot{\Psi}_+ + k^2\Psi_+\right)
      -2\kappa^2\tau_{(LL)} & = & 0. 
      \label{eqn:result-ordinary-HD}
\end{eqnarray}
Therefore, for linearized gravity the only difference between our model 
and the higher-curvature theory (\ref{eqn:ordinary-HD}) is that there is
an extra massless mode for $\Psi_-$ in our model. (Notice the extra
$\Box$ in the second equation in (\ref{eqn:result-scalar}).) However,
this zero mode does not couple to the matter stress energy 
$T^{other}_{\mu\nu}$ directly since the extra $\Box$ in the second
equation in (\ref{eqn:result-scalar}) is applied to the whole
expression. Hence, as far as we are concerned with linear perturbations
generated by matter sources, these two theories give exactly the same 
prediction. Thus, it is concluded that the scalar-type linearized
gravity in Minkowski background in our model is effectively described by
the theory (\ref{eqn:ordinary-HD}).

In the theory (\ref{eqn:ordinary-HD}), the linearized Einstein gravity
is recovered at distances longer than the length scale
$l_*=\sqrt{\alpha}\kappa$ and at energies lower than $l_*^{-1}$. To see
this, let us quote the equations governing scalar perturbations in
Einstein gravity: 
%
\begin{eqnarray}
 \kappa^2k^2\Psi_+  +\kappa^4\left(\tau_{00}
	    +2k^2\tau_{(LL)}\right) & = & 0, \nonumber\\ 
 \Psi_- -2\kappa^2\tau_{(LL)} & = & 0. 
\end{eqnarray}
It is easy to see that these equations in Einstein theory are recovered
from the corresponding equations (\ref{eqn:result-ordinary-HD}) in the
theory (\ref{eqn:ordinary-HD}) if
$\alpha\kappa^2\ddot{\Psi}_{\pm}/\Psi_{\pm}$ and $\alpha\kappa^2k^2$ are
sufficiently small. For the recovery, it is fairy important that we had
assumed that $\alpha$ is positive. Actually, the higher derivative terms
in the second equation in (\ref{eqn:result-ordinary-HD}) do not
introduce extra instabilities if and only if $\alpha$ is
non-negative. Note that the stability of homogeneous, isotropic
evolution of the universe has also required $\alpha>0$.


\section{Quantum mechanical stability}
\label{sec:quantum}

Now let us think about quantum mechanical stabilities.

The stability under radiative corrections is an important feature of our
model. Radiative corrections will produce additional regular terms in
the action, but the field will stall whether or not these are
present. Adding finite terms to the potential part does not change
anything since the curvature $R$ feels the corrected potential via the
gravity equation. As for the kinetic part, we would like to stress again
that the singular behavior of the kinetic term coefficient is imposed
only on the most singular-looking term among many possible terms in the
kinetic part and, thus, adding any kinetic terms which are less
singular-looking at $R=0$ does not change anything. Of course, adding a
more singular-looking kinetic term just makes the condition more
robust. The more singular a kinetic term looks, the more stable it is 
under radiative corrections.

Since $f(R)$ is in the denominator of a kinetic term and vanishes at
$R=0$, one might also worry about additional singular potential terms
like $1/f(R)$ being generated through radiative corrections. However,
this does not happen. In order to see this, it is convenient to
normalize quantum fluctuation of the scalar field $\phi$ around the
homogeneous classical background. For simplicity, we restrict our
discussion to the $q=1$ case and work in the unit where $\kappa=1$. The 
kinetic term is expanded as 
%
\begin{eqnarray}
 -\frac{\partial^{\mu}\phi\partial_{\mu}\phi}{2f} & = & 
  -\frac{1}{2}
  \left[\frac{1}{f_0}+\left(\frac{1}{f}\right)'_0(\delta R + \delta_2 R)
   + \left(\frac{1}{f}\right)''_0\frac{(\delta R)^2}{2}\right]
  \nonumber\\
 & & \times
  \left(g_0^{\mu\nu}-h^{\mu\nu}+h^{\mu\lambda}h_{\lambda}^{\nu}\right)
  \partial_{\mu}\left(\phi_0+\sqrt{f_0}\delta\phi_c\right)
  \partial_{\nu}\left(\phi_0+\sqrt{f_0}\delta\phi_c\right)
\end{eqnarray}
up to the quadratic order in perturbations, where a quantity with the
subscript $0$ represents the value on the FRW background considered in
Sec.~\ref{sec:Friedmann}, $h_{\mu\nu}$ is
metric perturbation, $\delta R$ and $\delta_2 R$ are the $O(h)$ and
$O(h^2)$ parts of the Ricci scalar perturbation, respectively, and 
%
\begin{equation}
 \delta\phi_c \equiv \frac{\delta\phi}{\sqrt{f_0}}
\end{equation}
is the normalized quantum fluctuation of the scalar field. Hence, the
contributions of the kinetic term to the full quadratic action
multiplied by $-2$ are 
%
\begin{eqnarray}
&& g_0^{\mu\nu}\partial_{\mu}(\delta\phi_c)\partial_{\nu}(\delta\phi_c) 
 \nonumber\\
&& - \frac{R_0f'_0}{f_0}\cdot\frac{\dot{R}_0}{R_0}
  \delta\phi_c(\delta\phi_c)^{\cdot}
  - \frac{1}{4}\left(\frac{R_0f'_0}{f_0}\cdot
		\frac{\dot{R}_0}{R_0}\right)^2(\delta\phi_c)^2 
  - 2\frac{\sqrt{f_0}\pi}{R_0}\cdot
  R_0f_0\left(\frac{1}{f}\right)'_0\cdot
  \delta R (\delta\phi_c)^{\cdot} 
  -\frac{\sqrt{f_0}\pi}{R_0}\cdot R_0\cdot
  h^{t\mu}\partial_{\mu}(\delta\phi_c)
  \nonumber\\
 && 
  - \frac{\sqrt{f_0}\pi}{R_0}\cdot R_0f_0\left(\frac{1}{f}\right)'_0\cdot
  \frac{R_0f'_0}{f_0}\cdot \frac{\dot{R}_0}{R_0}\delta R\delta\phi_c
  -\frac{\sqrt{f_0}\pi}{R_0}\cdot 
  \frac{R_0f'_0}{f_0}\cdot\frac{\dot{R}_0}{R_0}
  \cdot R_0\cdot h^{tt}\delta\phi_c
  -\left(\frac{\sqrt{f_0}\pi}{R_0}\right)^2\cdot R_0^2\cdot
  h^{t\lambda}h^t_{\lambda}
  \nonumber\\
 &&  
  - \frac{1}{2}\left(\frac{\sqrt{f_0}\pi}{R_0}\right)^2\cdot
  R_0^2f_0\left(\frac{1}{f}\right)''_0\cdot (\delta R)^2 
  -\left(\frac{\sqrt{f_0}\pi}{R_0}\right)^2\cdot
  R_0f_0\left(\frac{1}{f}\right)'_0\cdot R_0\cdot\delta_2 R
  -\left(\frac{\sqrt{f_0}\pi}{R_0}\right)^2\cdot 
  R_0f_0\left(\frac{1}{f}\right)'_0\cdot
  R_0\cdot h^{tt}\delta R .
\end{eqnarray}
Actually, all except for the first term vanish in the low energy limit
since
%
\begin{eqnarray}
 R_0 & \propto & H^2 \to 0, \nonumber\\
 \frac{\dot{R}_0}{R_0} & \propto & H\cdot\frac{\dot{H}}{H^2} \to 0, 
  \nonumber\\
 \frac{\sqrt{f_0}\pi}{R_0} & \propto & H^{2m-3} \to 0
\end{eqnarray}
in the low energy limit $H\to 0$ and 
%
\begin{equation}
 \frac{R_0f'_0}{f_0} \sim R_0f_0\left(\frac{1}{f}\right)'_0
  \sim R_0^2f_0\left(\frac{1}{f}\right)''_0 \sim 1. 
\end{equation}
This means that $\delta\phi_c$ asymptotically approaches a canonically
normalized fluctuation in the low energy limit. In terms of
$\delta\phi_c$, each term in the potential part can include a positive
power of $\sqrt{f_0}$ but not negative power. Therefore, loop
contributions of $\delta\phi_c$ to the radiatively corrected potential
part can include positive powers of $\sqrt{f_0}$ but not negative
power.

More rigorous treatment requires careful consideration of constraint
equations among the scalar field perturbation and the metric
perturbation. Namely, we need to introduce a variable analogous to the
Mukhanov variable~\cite{Mukhanov} in the standard field theory
cosmology. Hence, rigorous treatment seems much more complicated than
the above. Nonetheless, the above argument is convincing enough and we 
do not expect terms with negative power of $\sqrt{f_0}$ to appear in the
potential part of the proper perturbation variable.

In the above, it has been shown that the feedback mechanism is stable
under radiative corrections and thus a zero or small cosmological
constant is protected against radiative corrections. In the following,
we show that quantum fluctuation at low energy is so small that the
effective cosmological constant at low energy does not overshoot zero
even quantum mechanically.

In the $\kappa^2R\to 0$ limit the scalar field becomes completely weekly
coupled. Hence, one might think that the scalar field could overshoot
zero curvature by quantum fluctuation. This does not happen, as far as
the (effective) cosmological constant is a substantial component
of the energy density of the universe. The reason is as follows. (For 
simplicity we consider the $q=1$ case only, but it is easy to generalize 
it to a general $q>1/2$. We shall again work in the unit where
$\kappa=1$, and assume that the stability condition $m>3/2$ is
satisfied.)

The fluctuation $\delta\phi$ of the scalar field $\phi$ around a
homogeneous background does not have a canonically normalized kinetic 
term. As we showed in the above, the normalized fluctuation 
$\delta\phi_c \equiv \delta\phi/\sqrt{f_0}$ has a canonically normalized
kinetic term plus additional terms which vanish in the low energy limit
$H\to 0$. Hence, by dimensionality, amplitude of the quantum fluctuation
$\delta\phi_c$ should be $|\delta\phi_c| \sim H$. Hence, the quantum
fluctuation of the cosmological constant is estimated as  
%
\begin{equation}
 |\delta\Lambda_{eff}| \sim |c\delta\phi| \propto H^{2m+1},
  \label{eqn:dLambda}
\end{equation}
where we have used the behavior $f_0\propto H^{4m}$ near $H=0$.
On the other hand, if the (effective) cosmological constant is a
substantial component of the cosmological energy density then
%
\begin{equation}
 \Lambda_{eff} \sim H^2. 
  \label{eqn:FeqLambda}
\end{equation}
Note that under the stability condition
$m>3/2$ (see (\ref{eqn:stability-cond}) with $q=1$), the cosmological
constant asymptotically dominates the cosmological energy density. Since
$2m+1>2$ from the stability condition, the estimates (\ref{eqn:dLambda})
and (\ref{eqn:FeqLambda}) imply that 
%
\begin{equation}
 |\delta\Lambda_{eff}| \ll \Lambda_{eff}
\end{equation}
at low energy ($H\ll 1$). This means that $\Lambda_{eff}$ does not jump
to a negative value by quantum fluctuation.


\section{Summary}
\label{sec:summary}

In the present paper, we have investigated gravity in the recently
proposed dynamical approach to the cosmological constant. We have shown
that (i) the effective cosmological constant decreases in time and
asymptotically approaches zero from above; (ii) the evolution of a
homogeneous, isotropic universe is described by the standard Friedmann
equation at low energy; that (iii) classical, linearized gravity in
Minkowski background is described by Einstein gravity at distances
longer than $l_*=\sqrt{\alpha}\kappa$ and at energies lower than
$l_*^{-1}$, where $\kappa$ is the Planck length and $\alpha$ is a
dimensionless, positive parameter of the model; that (iv) the mechanism
is stable under radiative corrections and thus a zero or small
cosmological constant is protected against radiative corrections; and
that (v) quantum fluctuation at low energy is so small that the
effective cosmological constant at low energy does not overshoot zero
even quantum mechanically.

One of the most disturbing difficulties in thinking about the
cosmological constant is that it is not protected against radiative
corrections, which usually generate enormous vacuum energies compared to
what we observe. Because of the above properties (i)-(v), the feedback
mechanism can be considered as a dynamical way to protect a zero or
small cosmological constant against radiative corrections. Hence,
although the feedback mechanism by itself does not solve the
cosmological constant problem, it can help solving the problem.

\begin{acknowledgments}
 The author would like to thank Lisa Randall for collaborating on this
 subject and Nima Arkani-Hamed for many stimulating discussions. He is
 thankful to Raphael Bousso, Hsin-Chia Cheng, Paolo Creminelli, Eanna
 Flanagan, Debashis Ghoshal, Dileep Jatkar, Lubos Motl, Simon Ross,
 Howard Schnitzer, Matthew Schwarz, Yuri Shirman and Andy Strominger for
 useful comments. He is grateful to Werner Israel, Hideo Kodama and Lev
 Kofman for their continuing encouragement and Alan Coley for his warm
 hospitality at Dalhousie University, where the first version of this
 paper was completed. The author's research was supported in part by
 JSPS and NSF grant PHY-0201124.
\end{acknowledgments}


\appendix

\section{Derivation of equations of motion}
\label{app:derivation}

In this appendix we consider a more general action of the form
%
\begin{equation}
 {\bf I}[g_{\mu\nu},\phi] 
  = \int d^Dx \sqrt{-g}
  \left[\frac{R}{2\kappa^2}+{\bf L}(R,X,K,\phi)\right], 
  \label{eqn:more-general-action}
\end{equation}
where $X=R^{\mu\nu}R_{\mu\nu}$,
$K=-\kappa^4\partial^{\mu}\phi\partial_{\mu}\phi$, and $R$ and
$R_{\mu\nu}$ are the Ricci scalar and the Ricci tensor of the metric
$g_{\mu\nu}$. (See Appendix \ref{app:equivalence} for an alternative
description.) Our sign convention for the metric is $(-+++)$.

\subsection{Variational formulas}

In this subsection we derive some general formulas for variations of
geometrical objects in $D$-dimension. 

Let us decompose the metric $g_{\mu\nu}$ into the background
$g_{\mu\nu}^{(0)}$ and perturbation $\delta g_{\mu\nu}$: 
%
\begin{equation}
 g_{\mu\nu} =g_{\mu\nu}^{(0)}+\delta g_{\mu\nu}. 
\end{equation}
The indices of any linear-order quantities are lowered by
$g_{\mu\nu}^{(0)}$ and raised by the inverse $g^{(0)\mu\nu}$ of
$g_{\mu\nu}^{(0)}$. For example, 
%
\begin{equation}
 \delta g^{\mu\nu} \equiv
  g^{(0)\mu\rho}g^{(0)\nu\sigma}\delta g_{\rho\sigma}. 
\end{equation}
Hence, 
%
\begin{equation}
 g^{\mu\nu} =g^{(0)\mu\nu}-\delta g^{\mu\nu}
\end{equation}
up to the linear order.

Next, we can easily expand $\sqrt{-g}$ and the Christoffel symbol
$\Gamma^{\rho}_{\mu\nu}$ as follows, where $g$ is the determinant of
$g_{\mu\nu}$. 
%
\begin{eqnarray}
 \sqrt{-g} & = & 
        \sqrt{-g^{(0)}}\left(1+\frac{1}{2}\delta g\right), \nonumber\\ 
 \Gamma^{\rho}_{\mu\nu} & = & 
        {\Gamma^{(0)}}^{\rho}_{\mu\nu}+\delta\Gamma^{\rho}_{\mu\nu}
\end{eqnarray}
up to the linear order, where $\delta g=g^{(0)\mu\nu}\delta g_{\mu\nu}$,
${\Gamma^{(0)}}^{\rho}_{\mu\nu}$ is the Christoffel symbol for the
background metric $g_{\mu\nu}^{(0)}$, and 
%
\begin{equation}
 \delta\Gamma^{\rho}_{\mu\nu} \equiv 
  \frac{1}{2}g^{(0)\rho\sigma}(\delta g_{\sigma\mu;\nu}
  +\delta g_{\sigma\nu;\mu}-\delta g_{\mu\nu;\sigma}).
\end{equation}
Here, a semicolon denotes the covariant derivative compatible with the
background metric $g_{\mu\nu}^{(0)}$.

Thirdly, the Ricci tensor is expanded as follows.
%
\begin{equation}
 R_{\mu\nu} = R^{\rho}_{\mu\rho\nu}=\Gamma^{\rho}_{\mu\nu,\rho}
        -\Gamma^{\rho}_{\mu\rho,\nu}+\Gamma^{\rho}_{\sigma\rho}
        \Gamma^{\sigma}_{\mu\nu}-\Gamma^{\rho}_{\sigma\nu}
        \Gamma^{\sigma}_{\mu\rho}
 = R^{(0)}_{\mu\nu}+\delta R_{\mu\nu}
\end{equation}
up to the linear order, where $R^{(0)}_{\mu\nu}$ is the Ricci tensor for
the background metric $g_{\mu\nu}^{(0)}$, and
%
\begin{equation}
 \delta R_{\mu\nu} = 
  \frac{1}{2}(\delta g^{\rho}_{\mu;\nu\rho}+\delta g^{\rho}_{\nu;\mu\rho}
  -\delta g_{\mu\nu;\rho}^{\quad;\rho}-\delta g_{;\mu\nu}). 
\end{equation}
Therefore, we obtain
%
\begin{eqnarray}
 R & = & R^{(0)} + \delta R, \nonumber\\
 X & = & R^{(0)\mu\nu}R^{(0)}_{\mu\nu} + \delta X,
\end{eqnarray}
where
%
\begin{eqnarray}
 \delta R & = & -R^{(0)\mu\nu}\delta g_{\mu\nu} 
  + (\delta g^{\mu\nu}_{\quad;\nu}-\delta g^{;\mu})_{;\mu},
  \nonumber\\
 \delta X & = & 2R^{(0)\mu\nu}\delta R_{\mu\nu} 
  -2R^{(0)\mu\rho}R_{\rho}^{(0)\nu}\delta g_{\mu\nu}. 
\end{eqnarray}

\subsection{Equations of motion}

By using the variational formulas presented in the previous subsection,
the variation of the action $\delta {\bf I}$ is calculated as
%
\begin{equation}
 \delta {\bf I} = \int d^Dx\sqrt{-g}
  \left[-\frac{1}{2\kappa^2}\left(G^{\mu\nu}
   -\kappa^2T_{\phi}^{\mu\nu}\right)\delta g_{\mu\nu} 
   + E_{\phi}\delta\phi\right],
\end{equation}
where
%
\begin{eqnarray}
 T_{\phi}^{\mu\nu} & \equiv & 
  2\kappa^4{\bf L}_{,K}\partial^{\mu}\phi\partial^{\nu}\phi
  + {\bf L}g^{\mu\nu}
  - 2{\bf L}_{,R}R^{\mu\nu} + 2({\bf L}_{,R})^{;\mu\nu}
  - 2({\bf L}_{,R})^{;\rho}_{;\rho}g^{\mu\nu}\nonumber\\
 & &  + 2({\bf L}_{,X}R^{\mu\rho})^{;\nu}_{;\rho}
  + 2({\bf L}_{,X}R^{\rho\nu})^{;\mu}_{;\rho}
  - 2({\bf L}_{,X}R^{\mu\nu})^{;\rho}_{;\rho}
  - 2({\bf L}_{,X}R^{\rho\sigma})_{;\rho\sigma}g^{\mu\nu}
  - 4{\bf L}_{,X}R^{\mu\rho}R_{\rho}^{\nu}, \nonumber\\
 E_{\phi} & \equiv & 
  2\kappa^4({\bf L}_{,K}\partial_{\mu}\phi)^{;\mu} + {\bf L}_{,\phi}.
\end{eqnarray}
Hence we
obtain the following set of equations of motion. 
%
\begin{eqnarray}
 G^{\mu\nu} & = & \kappa^2 
  \left(T_{\phi}^{\mu\nu}+T_{other}^{\mu\nu}\right), \nonumber\\
 E_{\phi} & = & 0,
\end{eqnarray}
where $T_{other}^{\mu\nu}$ is the stress energy tensor of other fields
whose action should be added to the action
(\ref{eqn:more-general-action}).

Let us consider a homogeneous $\phi=\phi(t)$ in the $D=4$ flat FRW
background (\ref{eqn:FRW-metric}). With this ansatz, the stress energy
tensor is 
%
\begin{equation}
 T^{\mu}_{\phi\nu} = \left(\begin{array}{cccc}
	-\rho_{\phi} & 0 & 0 & 0 \\
	0 & p_{\phi} & 0 & 0 \\
	0 & 0 & p_{\phi} & 0 \\
	0 & 0 & 0 & p_{\phi}
	\end{array}\right),
\end{equation}
where
%
\begin{eqnarray}
 \rho_{\phi} & = & 
  2\kappa^4{\bf L}_{,K}\dot{\phi}^2-{\bf L}
  -6H\left[{\bf L}_{,R}+2(3H^2+2\dot{H}){\bf L}_{,X}\right]^{\cdot}
  +6\left[(H^2+\dot{H}){\bf L}_{,R}
     +2(3H^4+3H^2\dot{H}+2\dot{H}^2){\bf L}_{,X}\right], \nonumber\\
 \rho_{\phi}+p_{\phi} & = & 
  2\kappa^4{\bf L}_{,K}\dot{\phi}^2 
  +2\left[{\bf L}_{,R}+2(3H^2+2\dot{H}){\bf L}_{,X}\right]^{\cdot\cdot}
  -2H({\bf L}_{,R}+6H^2{\bf L}_{,X})^{\cdot}
  +4\dot{H}\left[{\bf L}_{,R}+6(H^2+\dot{H}){\bf L}_{,X}\right],
\end{eqnarray}
and
%
\begin{eqnarray}
 R & = & 6(\dot{H}+2H^2),     \nonumber\\
 X & = & 12(\dot{H}^2+3H^2\dot{H}+3H^4). 
\end{eqnarray}
Here, a dot represents the derivative with respect to the proper time
$t$. The equations of motion are 
%
\begin{eqnarray}
 H^2 & = & \frac{\kappa^2}{3}\left(\rho_{\phi}+\rho_{other}\right),
  \label{eqn:Friedmann-eq-general}\\
 \dot{H} & = & -\frac{\kappa^2}{2}
  \left[\left(\rho_{\phi}+p_{\phi}\right)
   +\left(\rho_{other}+p_{other}\right)\right], 
   \label{eqn:dynamical-eq-general}\\
 0 & = &  \dot{\pi} +3H\pi  -{\bf L}_{,\phi}
  \label{eqn:scalar-eq-general}
\end{eqnarray}
and the equation of motion for other fields, where 
%
\begin{equation}
 \pi \equiv 2\kappa^4{\bf L}_{,K}\dot{\phi},
\end{equation}
and $\rho_{other}(t)$ and $p_{other}(t)$ are energy density and pressure
of other fields: 
%
\begin{equation}
 T^{\quad\mu}_{other\ \nu} = \left(\begin{array}{cccc}
        -\rho_{other} & 0 & 0 & 0 \\
        0 & p_{other} & 0 & 0 \\
        0 & 0 & p_{other} & 0 \\
        0 & 0 & 0 & p_{other}
        \end{array}\right).
\end{equation}
The conservation of $T^{\quad\mu}_{other\ \nu}$ is expressed as
%
\begin{equation}
 \dot{\rho}_{other} +3H(\rho_{other}+p_{other}) = 0.
  \label{eqn:eom-matter}
\end{equation}

Note that (\ref{eqn:Friedmann-eq-general}),
(\ref{eqn:dynamical-eq-general}) and (\ref{eqn:scalar-eq-general}) are
not independent. Actually, there is an identity 
%
\begin{equation}
 \dot{\rho}_{\phi} + 3H(\rho_{\phi}+p_{\phi})+\dot{\phi}E_{\phi}=0,
\end{equation}
which can be checked explicitly and corresponds to the conservation
equation of $T_{\phi\mu\nu}$.

\subsection{First order equations}

Hereafter in this appendix, we assume that ${\bf L}$ is of the form 
%
\begin{equation}
 {\bf L} = \frac{\kappa^{-4}}{2q}F(R,X)K^q - V(\phi) + G(R,X),
  \label{eqn:special-L}
\end{equation}
where $F$ and $G$ are functions of $R$ and $X$, and $q$ is a
constant. In this case, $\rho_{\phi}$ and $p_{\phi}$ are expressed as 
%
\begin{eqnarray}
 \kappa^4\rho_{\phi} & = & \frac{2q-1}{2q}
  F^{-\frac{1}{2q-1}}(\kappa^4\pi^2)^{\frac{q}{2q-1}}
  + \kappa^4(V - G)
  + \frac{3H^2\Omega}{q}
  \left[F_{,R} + 2H^2(2\Omega+3)F_{,X}\right]
  \left(\frac{\kappa^4\pi^2}{F}\right)^{\frac{q}{2q-1}} \nonumber\\
 & & + 6\kappa^4H^2\Omega
  \left[G_{,R}+2H^2(2\Omega+3)G_{,X}\right]
  -3\kappa^4 H\varphi, \nonumber\\
 \kappa^4(\rho_{\phi}+p_{\phi}) & = & 
  F^{-\frac{1}{2q-1}}(\kappa^4\pi^2)^{\frac{q}{2q-1}}
  + \frac{3H^2\Omega}{q}
  \left[F_{,R} + 2H^2(2\Omega+3)F_{,X}\right]
  \left(\frac{\kappa^4\pi^2}{F}\right)^{\frac{q}{2q-1}} \nonumber\\
 & & + 6\kappa^4H^2\Omega
  \left[G_{,R}+2H^2(2\Omega+3)G_{,X}\right]
  + \kappa^4\dot{\varphi},
\end{eqnarray}
where
%
\begin{eqnarray}
 \pi & \equiv & FK^{q-1}\dot{\phi}, \nonumber\\
 \Omega & \equiv & \frac{\dot{H}}{H^2}, \nonumber\\
 \varphi & \equiv & 2[{\bf L}_{,R}+2(3H^2+2\dot{H}){\bf L}_{,X}]^{\cdot}
  - 2H({\bf L}_{,R}+6H^2{\bf L}_{,X}).
\end{eqnarray}

Hence, the equations of motion (\ref{eqn:dynamical-eq-general}),
(\ref{eqn:scalar-eq-general}) and (\ref{eqn:eom-matter}) are rewritten
as the following set of first-order equations:
%
\begin{eqnarray}
 \dot{\phi} & = & f\pi\cdot
  (\kappa^4\pi^2)^{\frac{1}{2(2q-1)}-\frac{1}{2}}, \nonumber\\
 \dot{\pi} & = & -3H\pi-V'(\phi), \nonumber\\
 \dot{H} & = & H^2\Omega, \nonumber\\
 \epsilon\dot{\Omega} & = & H{\cal F},  \nonumber\\
 \kappa^2\dot{\varphi} & = & -H^2({\cal G}_0 +\Delta{\cal G}), \nonumber\\ 
 \dot{\rho}_{other} & = & -3H(\rho_{other}+p_{other}), 
  \label{eqn:dphi_i-general}
\end{eqnarray}
where
%
\begin{eqnarray}
 f & \equiv & F^{-\frac{1}{2q-1}}, \nonumber\\
 \epsilon & \equiv & 
  12\kappa^2H^2
  \left[3G_{,RR}+12H^2(2\Omega+3)G_{,RX}+12H^4(2\Omega+3)^2G_{,XX}
     +2G_{,X}\right] \nonumber\\
 & &  -\frac{6(2q-1)H^2}{q\kappa^2}
  \left[3f_{,RR}+12H^2(2\Omega+3)f_{,RX}+12H^4(2\Omega+3)^2f_{,XX}
     +2f_{,X}\right](\kappa^4\pi)^{\frac{q}{2q-1}}, \nonumber\\
 {\cal F} & \equiv & \frac{3\kappa^2\varphi}{H}
  - \frac{6(3H\pi+V')}{H}
  \left[f_{,R}+2H^2(2\Omega+3)f_{,X}\right]\kappa^2\pi\cdot
  (\kappa^4\pi^2)^{-\frac{q-1}{2q-1}}\nonumber\\
 & &  + \frac{3(2q-1)}{q\kappa^2}
  \left\{-f_{,R}+2(4\Omega^2+6\Omega-3)H^2f_{,X}\right.\nonumber\\
 & & \left.  + 12\Omega\left[(\Omega+2)H^2f_{,RR}
	      +2(4\Omega^2+13\Omega+12)H^4f_{,RX}
	      +8(2\Omega+3)(\Omega^2+3\Omega+3)H^6f_{,XX}\right]
  \right\}(\kappa^4\pi^2)^{\frac{q}{2q-1}} \nonumber\\
 & &  -6\kappa^2
  \left\{-G_{,R}+2(4\Omega^2+6\Omega-3)H^2G_{,X}\right.\nonumber\\
 & & \left.  + 12\Omega\left[(\Omega+2)H^2G_{,RR}
	      +2(4\Omega^2+13\Omega+12)H^4G_{,RX}
	      +8(2\Omega+3)(\Omega^2+3\Omega+3)H^6G_{,XX}\right]
  \right\}, \nonumber\\
 {\cal G}_0 & \equiv & 2\Omega 
  + \frac{1}{\kappa^2H^2}\left[\kappa^4(\rho_{other}+p_{other})
  + f\cdot(\kappa^4\pi^2)^{\frac{q}{2q-1}}\right], \nonumber\\
 \Delta{\cal G} & \equiv & 
  -\frac{3(2q-1)\Omega}{q\kappa^2}
  \left[f_{,R} + 2H^2(2\Omega+3)f_{,X}\right]
  (\kappa^4\pi^2)^{\frac{q}{2q-1}}
  + 6\kappa^2\Omega\left[G_{,R} + 2H^2(2\Omega+3)G_{,X}\right],
  \nonumber\\
 R & = & 6H^2(\Omega+2), \nonumber\\
 X & = & 12H^4(\Omega^2+3\Omega+3).
\end{eqnarray}
The constraint equation (\ref{eqn:Friedmann-eq-general}) is now written
as 
%
\begin{equation}
 \frac{3\kappa^2\varphi}{H} = {\cal F}_0+\Delta{\cal G}
  -\frac{\kappa^2G}{H^2},
  \label{eqn:constraint-eq-general}
\end{equation}
where
%
\begin{eqnarray}
 {\cal F}_0 & \equiv &
  \frac{\kappa^2(\rho_{other} + \rho_{\phi,0}) -3H^2}{H^2},
  \nonumber\\
 \rho_{\phi,0} & \equiv & 2KL_{kin,K}-L_{kin}+V 
  = \frac{2q-1}{2q\kappa^4}f\cdot(\kappa^4\pi^2)^{\frac{q}{2q-1}}
  + V. 
\end{eqnarray}
When the system is analyzed numerically, we can use the constraint
equation to set the initial condition for $\varphi$ and also to check
numerical accuracy.

It is also possible to use the constraint equation to eliminate
$\varphi$ from the set of first-order equations: ${\cal F}$ in the forth
equation in (\ref{eqn:dphi_i-general}) is rewritten as 
%
\begin{equation}
 {\cal F} = {\cal F}_0 + \Delta{\cal F},
\end{equation}
where
%
\begin{eqnarray}
 \Delta{\cal F} & = & 
  - \frac{6(3H\pi+V')}{H}
  \left[f_{,R}+2H^2(2\Omega+3)f_{,X}\right]\kappa^2\pi\cdot
  (\kappa^4\pi^2)^{-\frac{q-1}{2q-1}}\nonumber\\
 & & + \frac{3(2q-1)}{q\kappa^2}
  \left\{-(\Omega+1)f_{,R}+2(2\Omega^2+3\Omega-3)H^2f_{,X}
  \right.\nonumber\\
 & & \left. + 12\Omega\left[(\Omega+2)H^2f_{,RR}
	      +2(4\Omega^2+13\Omega+12)H^4f_{,RX}
	      +8(2\Omega+3)(\Omega^2+3\Omega+3)H^6f_{,XX}\right]
  \right\}(\kappa^4\pi^2)^{\frac{q}{2q-1}} \nonumber\\
 & &  -6\kappa^2
  \left\{-(\Omega+1)G_{,R}+2(2\Omega^2+3\Omega-3)H^2G_{,X}
  \right.\nonumber\\
 & & \left.  + 12\Omega\left[(\Omega+2)H^2G_{,RR}
	      +2(4\Omega^2+13\Omega+12)H^4G_{,RX}
	      +8(2\Omega+3)(\Omega^2+3\Omega+3)H^6G_{,XX}\right]
  \right\} -\frac{\kappa^2G}{H^2}.
\end{eqnarray}


\section{Equivalent action}
\label{app:equivalence}

In this appendix let us consider an action of the form
%
\begin{equation}
 {\bf I}[g_{\mu\nu},\phi] = \int d^Dx \sqrt{-g}{\bf F}(R,K,\phi),
  \label{eqn:general-action} 
\end{equation}
where $R$ is the Ricci scalar of the metric $g_{\mu\nu}$, $\phi$ is a
scalar field and
$K=-\kappa^4g^{\mu\nu}\partial_{\mu}\phi\partial_{\nu}\phi$. We 
shall show that this action is equivalent to the following action. 
%
\begin{eqnarray}
 \tilde{\bf I}[\tilde{g}_{\mu\nu}, \psi,\phi] & = &
  \int d^Dx\sqrt{-\tilde{g}}
  \left[ \frac{1}{2\kappa^2}\tilde{R} 
   -\frac{1}{2}\tilde{g}^{\mu\nu}\partial_{\mu}\psi\partial_{\nu}\psi
   \right.
   \nonumber\\
   & & \left.
        +e^{-\alpha\kappa\psi}
        {\bf F}\left(R(\psi,e^{\beta\kappa\psi}\tilde{K},\phi),
        e^{\beta\kappa\psi}\tilde{K},\phi\right)
   -\frac{1}{2\kappa^2}e^{-\beta\kappa\psi}
   R(\psi,e^{\beta\kappa\psi}\tilde{K},\phi) \right],
   \label{eqn:general-equiv-action} 
\end{eqnarray}
where the new field $\psi$ and the function $R(\psi,K,\phi)$ are defined 
by 
%
\begin{equation}
  e^{\gamma\kappa\psi} = 
   2\kappa^2{\bf F}_R(R,K,\phi) \leftrightarrow R=R(\psi,K,\phi),
   \label{eqn:def-psi}
\end{equation}
the metric $\tilde{g}_{\mu\nu}$ is defined by 
%
\begin{equation}
 \tilde{g}_{\mu\nu} = e^{\beta\kappa\psi}g_{\mu\nu},
\end{equation}
and $\tilde{R}$ is the Ricci scalar for the new metric
$\tilde{g}_{\mu\nu}$. Here, $\kappa$ is an arbitrary positive
constant, ${\bf F}_R\equiv(\partial {\bf F}/\partial R)_{K,\phi}$,
$\tilde{K}\equiv -\kappa^4
\tilde{g}^{\mu\nu}\partial_{\mu}\phi\partial_{\nu}\phi$, 
$\tilde{g}^{\mu\nu}\equiv(\tilde{g}^{-1})^{\mu\nu}$, 
and constants $\alpha$, $\beta$ and $\gamma$ are 
%
\begin{eqnarray}
 \alpha & = & \frac{D}{\sqrt{(D-1)(D-2)}}, \nonumber\\
 \beta & = & \frac{2}{\sqrt{(D-1)(D-2)}}, \nonumber\\
 \gamma & = & \sqrt{\frac{D-2}{D-1}}.
  \label{eqn:alpha-beta-gamma}
\end{eqnarray}
In (\ref{eqn:def-psi}) the function $R(\psi,K,\phi)$ is defined by
solving the left equation with respect to $R$ and, thus, we have
implicitly assumed that $\partial^2{\bf F}/\partial R^2\ne 0$. 
The action (\ref{eqn:general-equiv-action}) actually
describes the Einstein gravity plus two scalar fields $\phi$ and
$\psi$. Equations of motion derived from the two actions are the same
and are 
%
\begin{eqnarray}
 \tilde{G}_{\mu\nu} - \kappa^2
  \left(\partial_{\mu}\psi\partial_{\nu}\psi
   -\frac{1}{2}\tilde{g}^{\rho\sigma}
   \partial_{\rho}\psi\partial_{\sigma}\psi\tilde{g}_{\mu\nu}\right)
  - \kappa^2 
  \left(e^{-\alpha\kappa\psi}{\bf F}\tilde{g}_{\mu\nu}  
   +2\kappa^4e^{-\gamma\kappa\psi}{\bf F}_K
   \partial_{\mu}\phi\partial_{\nu}\phi \right)
  + \frac{1}{2}e^{-\beta\kappa\psi}R\tilde{g}_{\mu\nu} & = & 0,
  \nonumber\\
 \frac{\kappa^4}{\sqrt{-\tilde{g}}}\partial_{\mu}
  \left[\sqrt{-\tilde{g}}\tilde{g}^{\mu\nu}e^{-\gamma\kappa\psi}
   {\bf F}_K\partial_{\nu}\phi\right] 
  +\frac{1}{2}e^{-\alpha\kappa\psi}{\bf F}_{\phi}
  & = & 0,\label{eqn:eom}
\end{eqnarray}
where it is understood that 
$R=R(\psi,e^{\beta\kappa\psi}\tilde{K},\phi)$ and
$K=e^{\beta\kappa\psi}\tilde{K}$ are substituted into $R$, ${\bf F}$,
${\bf F}_K\equiv(\partial {\bf F}/\partial K)_{R,\phi}$ and 
${\bf F}_{\phi}\equiv(\partial {\bf F}/\partial \phi)_{R,K}$.
The equation of motion for $\psi$ can be derived from the second
equation and the divergence of the first equation, and is 
%
\begin{equation}
\frac{1}{\sqrt{-\tilde{g}}}\partial_{\mu}
  \left[\sqrt{-\tilde{g}}\tilde{g}^{\mu\nu}\partial_{\nu}\psi\right]
  -\alpha\kappa e^{-\alpha\kappa\psi}{\bf F} 
  -\beta\kappa^5 e^{-\gamma\kappa\psi}{\bf F}_K\tilde{g}^{\mu\nu}
  \partial_{\mu}\phi\partial_{\nu}\phi
  +\frac{\beta}{2\kappa}e^{-\beta\kappa\psi}R = 0.
  \label{eqn:eom-psi}
\end{equation}
Again, it is understood that
$R=R(\psi,e^{\beta\kappa\psi}\tilde{K},\phi)$ and
$K=e^{\beta\kappa\psi}\tilde{K}$ are substituted into $R$, ${\bf F}$ and 
${\bf F}_K$.

Now let us show that the actions (\ref{eqn:general-action}) and
(\ref{eqn:general-equiv-action}) are classically equivalent. It is an
extension of ref.~\cite{Maeda}, in which the scalar field $\phi$ has a
usual canonical kinetic term.

First, it is easy to show that 
%
\begin{eqnarray}
 \frac{1}{\sqrt{-g}}\delta\left(\sqrt{-g}{\bf F}(R,K,\phi)\right)
  & = & \frac{1}{2}{\bf F}g^{\mu\nu}\delta g_{\mu\nu} 
  + {\bf F}_R\delta R
  + {\bf F}_K\delta K + {\bf F}_{\phi}\delta\phi \nonumber\\
  & = & E^{\mu\nu}\delta g_{\mu\nu} + E\delta\phi + X^{\mu}_{\ ;\mu},
\end{eqnarray}
where
%
\begin{eqnarray}
 E^{\mu\nu} & = & 
  \frac{1}{2}{\bf F}g^{\mu\nu} 
  - {\bf F}_RR^{\mu\nu}+{\bf F}_R^{\ ;\mu\nu}
  -{\bf F}_{R\ ;\rho}^{\ ;\rho}g^{\mu\nu} 
  +\kappa^4{\bf F}_K\phi^{;\mu}\phi^{;\nu},
  \nonumber\\
 E & = & 
  2\kappa^4({\bf F}_K\partial_{\mu}\phi)^{;\mu}+{\bf F}_{\phi}, \nonumber\\
 X^{\mu} & = & 
  {\bf F}_R(\delta g^{\mu\nu}_{\quad ;\nu}-\delta g^{\nu\ ;\mu}_{\ \nu})
  -{\bf F}_{R;\nu}\delta g^{\mu\nu}
  +{\bf F}_R^{\ ;\mu}\delta g^{\nu}_{\ \nu}
  -2\kappa^4{\bf F}_K\phi^{;\mu}\delta\phi,
\end{eqnarray}
and a semicolon represents the covariant derivative compatible with the
metric $g_{\mu\nu}$. Equations of motion derived from $\delta I=0$ are
$E^{\mu\nu}=0$ and $E=0$. These equations, supplemented by suitable
initial conditions, govern dynamics of the system. Evidently, unless
${\bf F}_R$ is independent of $R$, $E^{\mu\nu}$ includes up to the forth order 
derivatives of the metric $g_{\mu\nu}$. In the following, we shall
introduce an auxiliary field so that the resulting action includes only
up to the second derivatives of fields. Moreover, it will be shown that 
after a conformal transformation, the gravitational part of the
resulting action is of the form of the Einstein-Hilbert action.

Second, let us perform a conformal transformation 
%
\begin{equation}
 \tilde{g}_{\mu\nu} = e^{2\omega}g_{\mu\nu}, 
\end{equation}
where $\omega$ is a function to be determined below. The relation
between the Ricci tensor $R_{\mu\nu}$ of the original metric
$g_{\mu\nu}$ and the Ricci tensor $\tilde{R}_{\mu\nu}$ of the
conformally transformed metric $\tilde{g}_{\mu\nu}$ can be found in
textbooks in general relativity~\cite{Wald}. 
%
\begin{equation}
 R_{\mu\nu} = \tilde{R}_{\mu\nu} 
  -(D-2)\omega_{;\mu}\omega_{;\nu}
  +(D-2)\omega^{;\rho}\omega_{;\rho}g_{\mu\nu}
  +(D-2)\omega_{;\mu\nu}+\omega^{;\rho}_{\ ;\rho}g_{\mu\nu}. 
\end{equation}
Accordingly, the relation between Einstein tensors $G_{\mu\nu}$ and
$\tilde{G}_{\mu\nu}$ for $g_{\mu\nu}$ and $\tilde{g}_{\mu\nu}$,
respectively, is 
%
\begin{equation}
 G_{\mu\nu} =  \tilde{G}_{\mu\nu} -(D-2)
  \left[\omega_{;\mu}\omega_{;\nu}
   +\frac{1}{2}(D-3)\omega^{;\rho}\omega_{;\rho}g_{\mu\nu}\right]
   +(D-2)(\omega_{;\mu\nu}-\omega^{;\rho}_{\ ;\rho}g_{\mu\nu}). 
\end{equation}
Thus, we obtain the following expression of $E_{\mu\nu}$.
%
\begin{eqnarray}
 -E_{\mu\nu} & = & {\bf F}_R\tilde{G}_{\mu\nu} 
  - \frac{1}{2}({\bf F}-{\bf F}_RR)g_{\mu\nu}
  -(D-2){\bf F}_R
  \left[\omega_{;\mu}\omega_{;\nu}
   +\frac{1}{2}(D-3)\omega^{;\rho}\omega_{;\rho}g_{\mu\nu}\right]
  -\kappa^4{\bf F}_K\phi_{;\mu}\phi_{;\nu}. \nonumber\\
  & & +(D-2){\bf F}_R(\omega_{;\mu\nu}-\omega^{;\rho}_{\ ;\rho}g_{\mu\nu})
  -({\bf F}_{R;\mu\nu}-{\bf F}_{R\ ;\rho}^{\ ;\rho}g_{\mu\nu})
\end{eqnarray}
The higher derivative terms in the last line can be canceled up to lower
derivative terms if we chose
%
\begin{equation}
 \omega = \frac{1}{D-2}\ln(2\kappa^2{\bf F}_R),
\end{equation}
where $\kappa$ is an arbitrary positive constant. With this choice of
the conformal factor, $E_{\mu\nu}$ and $E$ are reduced to 
%
\begin{eqnarray}
 -E_{\mu\nu} & = &{\bf F}_R\tilde{G}_{\mu\nu} 
  - \frac{1}{2}({\bf F}-{\bf F}_RR)e^{-2\omega}\tilde{g}_{\mu\nu}
  -(D-1)(D-2){\bf F}_R
  \left(\omega_{,\mu}\omega_{,\nu}
   -\frac{1}{2}\tilde{g}^{\rho\sigma}
   \omega_{,\rho}\omega_{,\sigma}\tilde{g}_{\mu\nu}\right)
  -\kappa^4{\bf F}_K\phi_{,\mu}\phi_{,\nu}. \nonumber\\
 E & = & 2\kappa^4\frac{e^{D\omega}}{\sqrt{-\tilde{g}}}
  \partial_{\mu}
  \left(\sqrt{-\tilde{g}}\tilde{g}^{\mu\nu}e^{-(D-2)\omega}{\bf F}_K
   \partial_{\nu}\phi\right) +{\bf F}_{\phi}. 
\end{eqnarray}
These expressions include only up to the second derivatives. However,
they include not only the conformally transformed metric
$\tilde{g}_{\mu\nu}$ and the original scalar field $\phi$ but also the
new field $\omega$.

Thirdly, the new field $\omega$ can be canonically normalized by
defining the normalized field $\psi$ by
%
\begin{equation}
 \kappa\psi = \sqrt{(D-1)(D-2)}\omega 
  = \sqrt{\frac{D-1}{D-2}}\ln(2\kappa^2{\bf F}_R). 
\end{equation}
With this normalization, 
%
\begin{eqnarray}
 E_{\mu\nu} & = & -\frac{e^{\gamma\kappa\psi}}{2\kappa^2}
  \left[\tilde{G}_{\mu\nu} 
  - \frac{1}{2}(2\kappa^2e^{-\gamma\kappa\psi}{\bf F}-R)
  e^{-\beta\kappa\psi}\tilde{g}_{\mu\nu}
  -\kappa^2
  \left(\psi_{,\mu}\psi_{,\nu}
   -\frac{1}{2}\tilde{g}^{\rho\sigma}
   \psi_{,\rho}\psi_{,\sigma}\tilde{g}_{\mu\nu}\right)
  -2\kappa^6e^{-\gamma\kappa\psi}
  {\bf F}_K\phi_{,\mu}\phi_{,\nu}\right]. \nonumber\\
 E & = & 2\kappa^4\frac{e^{\alpha\kappa\psi}}{\sqrt{-\tilde{g}}}
  \partial_{\mu}
  \left(\sqrt{-\tilde{g}}\tilde{g}^{\mu\nu}e^{-\gamma\kappa\psi}{\bf F}_K
   \partial_{\nu}\phi\right) + {\bf F}_{\phi},
\end{eqnarray}
where constants $\alpha$, $\beta$ and $\gamma$ are defined by
(\ref{eqn:alpha-beta-gamma}). Therefore, the equations of motion
$E_{\mu\nu}=0$ and $E=0$ are given by (\ref{eqn:eom}).

Fourthly, it is easy to show the following equality.
%
\begin{equation}
 \tilde{\nabla}^{\mu}\left( e^{-\gamma\kappa\psi}E_{\mu\nu} \right) 
  - \frac{1}{2} e^{-\alpha\kappa\psi}E\phi_{,\nu}
  = \frac{1}{2}E_{\psi}\psi_{,\nu},
\end{equation}
where
%
\begin{equation}
 E_{\psi} = 
  \tilde{\nabla}^{\mu}\tilde{\nabla}_{\mu}\psi
  -\alpha\kappa e^{-\alpha\kappa\psi}{\bf F} 
  -\beta\kappa^5 e^{-\gamma\kappa\psi}{\bf F}_K\tilde{g}^{\mu\nu}
  \partial_{\mu}\phi\partial_{\nu}\phi
  +\frac{\beta}{2\kappa}e^{-\beta\kappa\psi}R,
\end{equation}
and $\tilde{\nabla}$ is the covariant derivative compatible with the 
conformally transformed metric $\tilde{g}_{\mu\nu}$. This means that
(\ref{eqn:eom-psi}) is derived from (\ref{eqn:eom}).

Fifthly, the variation of the action $\tilde{I}$ defined by
(\ref{eqn:general-equiv-action}) can be calculated as 
%
\begin{equation}
 \delta\tilde{\bf I} = \int d^Dx\sqrt{-\tilde{g}}
  \left[ e^{-\gamma\kappa\psi}E_{\rho\sigma}
   \tilde{g}^{\rho\mu}\tilde{g}^{\sigma\nu}\delta\tilde{g}_{\mu\nu}
   +e^{-\alpha\kappa\psi}E\delta\phi
   +E_{\psi}\delta\psi\right].
\end{equation}
Thus, the set of equations of motion derived from the action $\tilde{I}$
is the same as that from $I$.

Finally, by using the relations
%
\begin{eqnarray}
 R & = & e^{\beta\kappa\psi}\tilde{R} 
  + 2\gamma^{-1}\kappa\psi^{;\mu}_{\ ;\mu}
  + \kappa^2\psi^{;\mu}\psi_{\ ;\mu}, \nonumber\\
 \psi^{;\mu}_{\ ;\mu} & = & 
  e^{\beta\kappa\psi}\tilde{\nabla}^{\mu}\tilde{\nabla}_{\mu}\psi
  -\gamma\kappa e^{\beta\kappa\psi}\tilde{g}^{\mu\nu}
  \psi_{,\mu}\psi_{,\nu},
\end{eqnarray}
it is shown that 
$\tilde{\bf I}[\tilde{g}_{\mu\nu}, \psi,\phi]={\bf I}[g_{\mu\nu},\phi]$.


\section{Harmonics on a plane}
        \label{app:harmonics}

In this Appendix we give definitions of scalar, vector and tensor
harmonics on an $n$-dimensional plane. For definitions and properties of
more general harmonics, see Appendix B of ref.~\cite{Mukohyama2000b} and
Appendix A of ref.~\cite{Mukohyama-Kofman}.

\subsection{Scalar harmonics}

The scalar harmonics are given by 
%
\begin{equation}
 Y = exp(-ik_jx^j),
\end{equation}
by which any function $f$ can be expanded as 
%
\begin{equation}
 f = \int d^nk\ c Y,
\end{equation}
where $c$ is a constant depending on $k_i$. Hereafter, we write $k_i$ as
$k$ in most cases, and sometimes we omit it.

\subsection{Vector harmonics}

In general, any vector field $v_i$ can be decomposed as
%
\begin{equation}
 v_i=v_{(T)i}+\partial_{i} f ,
\end{equation}
where $f$ is a function and $v_{(T)i}$ is a transverse vector field:
%
\begin{equation}
 \delta^{ij}\partial_{i}v_{(T)j}=0 .
\end{equation}

Thus, the vector field $v_i$ can be expanded by using the scalar 
harmonics $Y$ and transverse vector harmonics $V_{(T)i}$ as 
%
\begin{equation}
 v_i = \int d^nk
  \left[c_{(T)}V_{(T)i}+c_{(L)}\partial_i Y\right].
        \label{eqn:dY+V}
\end{equation}
Here, $c_{(T)}$ and $c_{(L)}$ are constants depending on $k$, and the
transverse vector harmonics $V_{(T)i}$ is given by 
%
\begin{equation}
 V_{(T)i} = u_i\exp(-ik_jx^j),
\end{equation}
where the constant vector $u_i$ satisfies the condition
%
\begin{equation}
 \delta^{ij}k_iu_j=0.
\end{equation}

Because of the expansion (\ref{eqn:dY+V}), it is convenient to define
longitudinal vector harmonics $V_{(L)i}$ by
%
\begin{equation}
 V_{(L)i} \equiv \partial_i Y = -ik_iY. 
\end{equation}

\subsection{Tensor harmonics}

In general, a symmetric second-rank tensor field $t_{ij}$ can be
decomposed as
%
\begin{equation}
 t_{ij}=t_{(T)ij} + \partial_iv_j+\partial_jv_i + f\delta_{ij},
\end{equation}
where $f$ is a function, $v_i$ is a vector field and $t_{(T)ij}$
is a transverse traceless symmetric tensor field:
%
\begin{eqnarray}
 \delta^{ij}t_{(T)ij} & = & 0,\nonumber\\
 \delta^{ii'}\partial_i t_{(T)i'j} & = &0.
        \label{eqn:trasverse-traceless}
\end{eqnarray}

Thus, the tensor field $t_{ij}$ can be expanded by using the scalar
harmonics $Y$, the vector harmonics $V_{(T)i}$ and $V_{(L)i}$, and
transverse traceless tensor harmonics $T_{(T)ij}$ as 
%
\begin{eqnarray}
 t_{ij} & = & \int d^nk\left[
        c_{(T)}T_{(T)ij}+c_{(LT)}
        (\partial_iV_{(T)j}+\partial_jV_{(T)i})\right.
        \nonumber\\
 & &    \left.
         + c_{(LL)}(\partial_iV_{(L)j}+\partial_jV_{(L)i})
        + \tilde{c}_{(Y)}Y\delta_{ij}\right].
        \label{eqn:dV+T}
\end{eqnarray}
Here, $c_{(T)}$, $c_{(LT)}$, $c_{(LL)}$, and $\tilde{c}_{(Y)}$ are
constants depending on $k$, and the transverse traceless tensor
harmonics $T_{(T)ij}$ is given by 
%
\begin{equation}
 T_{(T)ij} = s_{ij}\exp(-ik_{i'}x^{i'}),
\end{equation}
where the constant symmetric second-rank tensor $s_{ij}$ satisfies the
condition
%
\begin{eqnarray}
 \delta^{ii'}k_is_{i'j} & = & 0, \nonumber\\
 \delta^{ij}s_{ij} & = & 0.
\end{eqnarray}

Because of the expansion (\ref{eqn:dV+T}), it is convenient to define
tensor harmonics $T_{(LT)ij}$, $T_{(LL)ij}$, and $T_{(Y)ij}$ by 
%
\begin{eqnarray}
 T_{(LT)ij} & \equiv & \partial_iV_{(T)j}
  +\partial_jV_{(T)i}, \nonumber\\
 & = & -i(u_ik_j+u_jk_i)Y, \nonumber\\
 T_{(LL)ij} & \equiv & \partial_iV_{(L)j}
  +\partial_jV_{(L)i}
        -\frac{2}{n}\delta_{ij}\delta^{i'j'}\partial_{i'}V_{(L)j'} 
	\nonumber\\
 & = & \left(-2k_ik_j+\frac{2}{n}\delta^{i'j'}k_{i'}k_{j'}
        \delta_{ij}\right)Y,  \nonumber\\ 
 T_{(Y)ij} & \equiv & \delta_{ij}Y. 
\end{eqnarray}


\section{Detailed analysis of linearized gravity}
        \label{app:linear-gravity}

In this appendix we show that linearized Einstein gravity in Minkowski
background is recovered at distances longer than the length scale
$l_*=\sqrt{\alpha}\kappa$ and at energies lower than $l_*^{-1}$. For
simplicity we consider the $q=1$ case, but extension to a general $q$
($>1/2$) should be straightforward.

For this purpose we investigate in detail perturbations around the
flat FRW background in the longitudinal gauge and take the 
$\kappa H\to +0$ limit in the end of the calculation. This strategy
makes it possible to analyze linear perturbations around Minkowski
background, on which the action of the scalar field perturbation is not
apparently well-defined. For simplicity, we assume that the FRW
background is driven by the scalar field $\phi$. Of course, we include a
general matter stress-energy tensor into perturbations and carefully
investigate how they couple to gravity. A critical point for this
analysis is that constraint equations (($0i$)-components and the
traceless part of ($ij$)-components of Einstein equation) prevent $\phi$
from fluctuating freely, and  it turns out that the $\phi$ perturbation
is of sufficiently high order in the $\kappa H$ expansion that the 
perturbative analysis is under control.

As stated in Sec.~\ref{sec:gravity} the result of the analysis is
simple. The linearized gravity in our model is similar to that in the
theory $R/2\kappa^2+\alpha R^2$ in the sense that the only difference is
the existence of an extra scalar-type massless mode in our model. The
extra massless mode is decoupled from the matter stress-energy tensor at
the linearized level. Hence, as far as we are concerned with classical,
linear perturbations generated by matter sources, these two theories
give exactly the same prediction. Therefore, in our model the linearized
Einstein gravity in Minkowski background is recovered at distances
longer than the length scale $l_*=\sqrt{\alpha}\kappa$ and at energies
lower than $l_*^{-1}$.

Let us begin by defining the metric perturbation $\delta g_{\mu\nu}$ by
%
\begin{equation}
 g_{\mu\nu} = g^{(0)}_{\mu\nu} + \delta g_{\mu\nu}, 
\end{equation}
where
%
\begin{equation}
 g^{(0)}_{\mu\nu}dx^{\mu}dx^{\nu} = -dt^2 + a(t)^2\delta_{ij}dx^idx^j
\end{equation}
($i=1,2,3$). In order to investigate linearized gravity in Minkowski
background, in the end of calculations, we shall take the limit  
%
\begin{equation}
 \kappa H\to +0,
\end{equation}
keeping $\partial_t^nH/H^{n+1}$ ($n=1,2,3$) finite, where
$H=\partial_ta/a$. In the following, we shall denote $\partial_t$ by an 
overdot.

\subsection{Gauge choice and three types of perturbations}
\label{subsec:gauge-choice}

In order to take the full advantage of the symmetry of the background
spacetime, we expand the metric perturbation by harmonics on the
$3$-plane: 
%
\begin{eqnarray}
 \delta g_{\mu\nu}dx^{\mu}dx^{\nu} & = & 
        h_{00}Ydt^2
        + 2(h_{(T)0}V_{(T)i}+h_{(L)0}V_{(L)i})dtdx^i
        \nonumber\\
 & & + (h_{(T)}T_{(T)ij}+h_{(LT)}T_{(LT)ij}+
        h_{(LL)}T_{(LL)ij}+h_{(Y)}T_{(Y)ij})dx^idx^j,\nonumber\\
	\label{eqn:expansion-g}
\end{eqnarray}
where $Y$, $V_{(T,L)}$ and $T_{(T,LT,LL,Y)}$ are scalar, vector and
tensor harmonics, respectively, and the coefficients $h_{00}$,
$h_{(T,L)0}$, $h_{(T,LT,LL,Y)}$ depend only on the time $t$. In this
expression and hereafter, we omit the $3$-momentum $k_i$ and the
integration over $3$-momenta. See Appendix~\ref{app:harmonics} for
definitions of the harmonics.

We also expand the perturbation of the scalar field $\phi$ by harmonics
as 
%
\begin{equation}
 \phi = \phi^{(0)}(t) + \delta\phi Y,
  \label{eqn:expansion-phi}
\end{equation}
where $\phi^{(0)}(t)$ is the background depending only on the time $t$
and the coefficient $\delta\phi$ of the perturbation also depends only
on the time $t$. In the following we shall derive the equation of motion
linearized with respect to  $h$'s and $\delta\phi$.

Now, it is easily shown from the orthogonality among different kinds of
harmonics that each of the following sets of variables form a closed set
of equations: (S) $h_{00}$, $h_{(L0)}$, $h_{(LL)}$, $h_{Y}$ and
$\delta\phi$; (V) $h_{(T)0}$, $h_{(LT)}$; and (T) $h_{T}$. Perturbations
in different sets are completely decoupled from each other at the
linearized level. The perturbations in the category (S) form a 
spin-$0$ representation and are called {\it scalar perturbations}. The
perturbations in the category (V) form a spin-$1$ representation and are
called {\it vector perturbations}.  Finally, the perturbations in the
category (T) form a spin-$2$ representation and are called  
{\it tensor perturbations}.

The metric perturbation $\delta g_{\mu\nu}$ includes not only physical 
degrees of freedom but also gauge freedom. An infinitesimal gauge
transformation is given by
%
\begin{eqnarray}
 \delta g_{\mu\nu} & \to & \delta g_{\mu\nu} 
        - \bar{\xi}_{\mu;\nu} - \bar{\xi}_{\nu;\mu},\nonumber\\
 \delta\phi Y & \to & \delta\phi Y 
  - \bar{\xi}^{\mu}\partial_{\mu}\phi^{(0)},
\end{eqnarray}
where $\bar{\xi}_{\mu}$ is an arbitrary vector field and a semicolon
denotes the covariant derivative compatible with the background metric 
$g^{(0)}_{\mu\nu}$. Hence, by expanding the vector $\bar{\xi}_{\mu}$ in 
terms of harmonics as 
%
\begin{equation}
 \bar{\xi}_{\mu}dx^{\mu} = \xi_0Ydt + 
        (\xi_{(T)}V_{(T)i}+\xi_{(L)}V_{(L)i})dx^i, 
\end{equation}
we obtain the following infinitesimal gauge transformation for the 
expansion coefficients in (\ref{eqn:expansion-g}) and
(\ref{eqn:expansion-phi}) with $k_i\ne 0$. 
%
\begin{eqnarray}
 h_{00} & \to & h_{00} - 2\dot{\xi}_0, \nonumber\\
 h_{(T)0} & \to & h_{(T)0} 
        - a^2(a^{-2}\xi_{(T)})^{\cdot}, \nonumber\\
 h_{(L)0} & \to & h_{(L)0} - \xi_0
        - a^2(a^{-2}\xi_{(L)})^{\cdot}, \nonumber\\
 h_{(T)} & \to & h_{(T)}, \nonumber\\
 h_{(LT)} & \to & h_{(LT)} - \xi_{(T)}, \nonumber\\
 h_{(LL)} & \to & h_{(LL)} - \xi_{(L)}, \nonumber\\
 h_{(Y)} & \to & h_{(Y)} + 2a^2H\xi_0
       + \frac{2}{3}k^2\xi_{(L)},\nonumber\\
 \delta\phi & \to &\delta\phi + \xi_0\dot{\phi}^{(0)}.
        \label{eqn:gauge-tr}
\end{eqnarray}
From the gauge transformation (\ref{eqn:gauge-tr}) it is evident that
for modes with $k_i\ne 0$, we can choose the gauge so that 
%
\begin{equation}
 h_{(L)0} = h_{(LT)} = h_{(LL)} = 0. \label{eqn:gauge-choice}
\end{equation}
With this condition, the gauge degrees of freedom represented by
$\xi_0$, $\xi_{(T)}$ and $\xi_{(L)}$ are completely fixed for modes with
$k_i\ne 0$. On the other hand, we do not need to consider modes with
$k_i=0$ since they can be absorbed into the background without loss of
generality.

We also take into account the stress-energy tensor $T_{\mu\nu}$ of other
fields, consider itself as a perturbation around $T_{\mu\nu}=0$ and
expand it by harmonics as
%
\begin{eqnarray}
 T_{\mu\nu}dx^{\mu}dx^{\nu} & = & 
        \tau_{00}Ydt^2
        + 2(\tau_{(T)0}V_{(T)i}+\tau_{(L)0}V_{(L)i})dtdx^i
        \nonumber\\
 & & + (\tau_{(T)}T_{(T)ij}+\tau_{(LT)}T_{(LT)ij}+
        \tau_{(LL)}T_{(LL)ij}+\tau_{(Y)}T_{(Y)ij})dx^idx^j. 
\end{eqnarray}
Since gauge freedom is already completely fixed, all the coefficients 
$\tau$'s represent gauge-inequivalent degrees of freedom.

Hence, in the gauge (\ref{eqn:gauge-choice}), all gauge-inequivalent
degrees of freedom are represented by the following variables: (S)
$h_{00}$, $h_{(Y)}$, $\delta\phi$, $\tau_{00}$, $\tau_{(L)0}$,
$\tau_{(LL)}$ and $\tau_{(Y)}$ for scalar perturbations; (V) $h_{(T)0}$,
$\tau_{(T)0}$ and $\tau_{(LT)}$ for vector perturbations; and (T)
$h_{(T)}$ and $\tau_{(T)}$ for tensor perturbations. As already
mentioned, scalar-, vector- and tensor-type perturbations are completely
decoupled from each other at the linearized level. Hence, in the
following, we shall analyze each type separately.

For vector- and tensor-type perturbations, as shown in
Sec.~\ref{sec:gravity}, there is no difference between weak gravity in
our model and that in Einstein theory. Hence, in the following we
consider scalar-type perturbations only.

\subsection{Scalar perturbations}
\label{subsec:lineargravity-scalar}

For scalar-type metric perturbations, we have two gauge-inequivalent
degrees $h_{00}$ and $h_{(Y)}$. By introducing $\Phi(t)$ and $\Psi(t)$ 
by $\Phi\equiv -h_{00}/2$ and $\Psi\equiv -a^{-2}h_{(Y)}/2$, the
perturbed metric is written as 
%
\begin{equation}
 g_{\mu\nu}dx^{\mu}dx^{\nu} = 
  -(1+2\Phi Y)dt^2 + (1-2\Psi Y)a^2\delta_{ij}dx^idx^j,
  \label{eqn:longitudinal-gauge-appendix}
\end{equation}
For this metric, the perturbed Ricci tensor is 
%
\begin{equation}
 R_{\mu\nu}dx^{\mu}dx^{\nu} = R^{(0)}_{\mu\nu}dx^{\mu}dx^{\nu}
  + \delta R_{00}Ydt^2 + 2\delta R_{(L)0}V_{(L)i}dtdx^i
  + (\delta R_{(LL)}T_{(LL)ij}+\delta R_{(Y)}T_{(Y)ij})dx^idx^j. 
\end{equation}
up to the linear order in $\Phi$ and $\Psi$, where $R^{(0)}_{\mu\nu}$ is
the background Ricci tensor
%
\begin{eqnarray}
 R^{(0)}_{00} & = & -3(\dot{H}+H^2), \nonumber\\
 R^{(0)}_{ij} & = & (\dot{H}+3H^2)a^2\delta_{ij},
\end{eqnarray}
and 
%
\begin{eqnarray}
 \delta R_{00} & = & 
  3\ddot{\Psi}+6H\dot{\Psi} + 3H\dot{\Phi} 
  -\frac{k^2}{a^2}\Phi, \nonumber\\
 \delta R_{(L)0} & = & 2\dot{\Psi} + 2H\Phi, \nonumber\\
 \delta R_{(LL)} & = & \frac{1}{2}\left(\Psi-\Phi\right), \nonumber\\ 
 \delta R_{(Y)} & = &  -a^2
  \left[ \ddot{\Psi}+6H\dot{\Psi}+H\dot{\Phi} 
   + 2(\dot{H}+3H^2)(\Psi+\Phi) + \frac{4}{3}\frac{k^2}{a^2}\Psi 
   -\frac{1}{3}\frac{k^2}{a^2}\Phi
  \right]. 
\end{eqnarray}
Hence, the perturbed Ricci scalar and the perturbed Einstein tensor are
%
\begin{eqnarray}
 R & = & R^{(0)} + \delta R Y, \nonumber\\
 G_{\mu\nu}dx^{\mu}dx^{\nu} & = &G^{(0)}_{\mu\nu}dx^{\mu}dx^{\nu} 
  + \delta G_{00}Ydt^2 + 2\delta G_{(L)0}V_{(L)i}dtdx^i
  + (\delta G_{(LL)}T_{(LL)ij}+\delta G_{(Y)}T_{(Y)ij})dx^idx^j 
\end{eqnarray}
up to the linear order, where $R^{(0)}$ and $G^{(0)}_{\mu\nu}$ are the
background Ricci scalar and the background Einstein tensor
%
\begin{eqnarray}
 R^{(0)} & = & 6(\dot{H}+2H^2), \nonumber\\
 G^{(0)}_{00} & = & 3H^2, \nonumber\\
 G^{(0)}_{ij} & = & -(2\dot{H}+3H^2)a^2\delta_{ij},
\end{eqnarray}
and 
%
\begin{eqnarray}
 \delta R & = & 
  -2\left[3\ddot{\Psi}+12H\dot{\Psi}+3H\dot{\Phi}+6(\dot{H}+2H^2)\Phi
    +2\frac{k^2}{a^2}\Psi-\frac{k^2}{a^2}\Phi\right],\nonumber\\
 \delta G_{00} & = & 
  -2\left[3H\dot{\Psi}+\frac{k^2}{a^2}\Psi\right], \nonumber\\ 
 \delta G_{(L)0} & = & \delta R_{(L)0}, \nonumber\\ 
 \delta G_{(LL)} & = & \delta R_{(LL)}, \nonumber\\ 
 \delta G_{(Y)} & = & 
  2a^2\left[ \ddot{\Psi} + 3H\dot{\Psi} + H\dot{\Phi} 
      + (2\dot{H}+3H^2)(\Psi+\Phi)  
      + \frac{1}{3}\frac{k^2}{a^2}(\Psi-\Phi) \right].
  \label{eqn:deltaG-components}
\end{eqnarray}
The Bianchi identity 
%
\begin{equation}
 \nabla^{\mu}G_{\mu\nu}=0,
\end{equation}
where $\nabla$ is the covariant derivative compatible with the perturbed
metric $g_{\mu\nu}$, is reduced to 
%
\begin{eqnarray}
 \dot{\delta G}_{00} +3H\delta G_{00} +3H\frac{\delta G_{(Y)}}{a^2} 
  +\frac{k^2}{a^2}\delta G_{(L)0} & = & 
  \left(3G^{(0)}_{00}+a^{-2}\delta^{ij}G^{(0)}_{ij}\right)
  \dot{\Psi} + 2G^{(0)}_{00}\dot{\Phi}
  -2a^{-2}\delta^{ij}G^{(0)}_{ij}H(\Psi+\Phi), \nonumber\\
 \dot{\delta G}_{(L)0}+3H\delta G_{(L)0} 
  +\frac{4}{3}\frac{k^2}{a^2}G_{(LL)}-\frac{\delta G_{(Y)}}{a^2} & = & 
  \frac{1}{3}\left(3G^{(0)}_{00}+a^{-2}\delta^{ij}G^{(0)}_{ij}\right)\Phi 
  +\frac{2}{3}a^{-2}\delta^{ij}G^{(0)}_{ij}\Psi.
  \label{eqn:Bianchi}
\end{eqnarray}
These two equations are, of course, satisfied by the components shown in
(\ref{eqn:deltaG-components}).

As for fields other than $\phi$, we have the stress energy tensor 
%
\begin{equation}
 T^{other}_{\mu\nu}dx^{\mu}dx^{\nu} = \tau_{00}Ydt^2
        + 2\tau_{(L)0}V_{(L)i}dtdx^i
	+ (\tau_{(LL)}T_{(LL)ij}+\tau_{(Y)}T_{(Y)ij})dx^idx^j. 
	\label{eqn:T-harmonics}
\end{equation}
The conservation equation 
%
\begin{equation}
 \nabla^{\mu}T^{other}_{\mu\nu}=0,
\end{equation}
where $\nabla$ is the covariant derivative compatible with the perturbed
metric $g_{\mu\nu}$, is reduced to 
%
\begin{eqnarray}
 \dot{\tau}_{00} +3H\tau_{00} +3H\frac{\tau_{(Y)}}{a^2} 
  +\frac{k^2}{a^2}\tau_{(L)0} & = & 0, \nonumber\\
 \dot{\tau}_{(L)0}+3H\tau_{(L)0} 
  +\frac{4}{3}\frac{k^2}{a^2}\tau_{(LL)}-\frac{\tau_{(Y)}}{a^2} & = & 0.
  \label{eqn:conservation}
\end{eqnarray}

On the other hand, the stress energy tensor of $\phi$ (with $q=1$) is
given by 
%
\begin{equation}
 T_{\phi}^{\mu\nu} \equiv 
  \frac{1}{f}
  \partial^{\mu}\phi\partial^{\nu}\phi + L g^{\mu\nu}
  -2L_RR^{\mu\nu}+2L_R^{\ ;\mu\nu}-2L_{R;\rho}^{\ ;\rho}g^{\mu\nu},
\end{equation}
where 
%
\begin{eqnarray}
 L & \equiv & \alpha R^2+L_{kin}-V, \nonumber\\
 L_R & \equiv & 2\alpha R
  - \frac{\kappa^{-4}f'K}{2f^2},
\end{eqnarray}
and a prime applied to $f$ denotes derivative with respect to $R$. 
We now expand $T_{\phi\mu\nu}$ up to the linear order:
%
\begin{equation}
 T_{\phi\mu\nu}dx^{\mu}dx^{\nu} = T^{(0)}_{\phi\mu\nu}dx^{\mu}dx^{\nu} 
  + \tau_{\phi 00}Ydt^2 + 2\tau_{\phi(L)0}V_{(L)i}dtdx^i
  + (\tau_{\phi(LL)}T_{(LL)ij}+\tau_{\phi(Y)}T_{(Y)ij})dx^idx^j. 
\end{equation}
The background $T^{(0)}_{\phi\mu\nu}$ is given by
%
\begin{equation}
 T^{\mu}_{\phi\nu} = \left(\begin{array}{cccc}
        -\rho_{\phi} & 0 & 0 & 0 \\
        0 & p_{\phi} & 0 & 0 \\
        0 & 0 & p_{\phi} & 0 \\
        0 & 0 & 0 & p_{\phi}
        \end{array}\right),
 \label{eqn:Tmunu-FRW}
\end{equation}
where
%
\begin{eqnarray}
 \rho_{\phi} & = &
  \pi\dot{\phi}^{(0)} -L
  -6H(L_R)^{\cdot} +6(H^2+\dot{H})L_R, \nonumber\\
 \rho_{\phi}+p_{\phi} & = &
  \pi\dot{\phi}^{(0)}
  +2(L_R)^{\cdot\cdot} -2H(L_R)^{\cdot}
  +4\dot{H}L_R, \label{eqn:rho-p-for-phi}
\end{eqnarray}
%
\begin{eqnarray}
 \pi & \equiv & \frac{\dot{\phi}}{f}, \nonumber\\
 L & = & \alpha R^2
  + \frac{\dot{\phi}^{(0)2}}{2f} - V, \nonumber\\
 L_R & = & 2\alpha R
  - \frac{f'\dot{\phi}^{(0)2}}{2f^2},  \nonumber\\
 R & = & 6(\dot{H}+2H^2).
\end{eqnarray}
and a dot represents the derivative with respect to the proper time
$t$. The linear part is 
%
\begin{eqnarray}
 \tau_{\phi 00} & = & F^{(0)}\dot{\phi}^{(0)}\dot{\delta\phi} 
  + \frac{1}{2}F^{(0)}_{,R}\dot{\phi}^{(0)2}\delta R 
  + 2(V^{(0)}-\alpha R^{(0)2})\Phi 
  + {V^{(0)}}'\delta\phi -2\alpha\kappa^2R^{(0)}\delta R \nonumber\\
 & & -3(\dot{H}+H^2)\delta W 
  + W^{(0)}\delta R_{00} + 3H\dot{\delta W} + \frac{k^2}{a^2}\delta W 
  - 3\dot{W}^{(0)}\dot{\Psi}, \nonumber\\
 \tau_{\phi(L)0} & = & F^{(0)}\dot{\phi}^{(0)}\delta\phi 
  + W^{(0)}\delta R_{(L)0} 
  - \dot{\delta W} + H\delta W + \dot{W}^{(0)}\Phi, \nonumber\\ 
 \tau_{\phi(LL)} & = & -\frac{1}{2}\delta W + W^{(0)}\delta R_{(LL)}, 
  \nonumber\\
 \tau_{\phi(Y)} & = & a^2\left[F^{(0)}\dot{\phi}^{(0)}\dot{\delta\phi} 
  + \frac{1}{2}F^{(0)}_{,R}\dot{\phi}^{(0)2}\delta R 
  -F^{(0)}\dot{\phi}^{(0)2}(\Psi+\Phi)+ 2(V^{(0)}-\alpha R^{(0)2})\Psi 
  - {V^{(0)}}'\delta\phi\right. \nonumber\\ 
 & & +2\alpha R^{(0)}\delta R 
  +(\dot{H}+3H^2)\delta W + W^{(0)}a^{-2}\delta R_{(Y)}
  -\ddot{\delta W}-2H\dot{\delta W} -\frac{2}{3}\frac{k^2}{a^2}\delta W
  \nonumber\\
 & & \left. +2\ddot{W}^{(0)}(\Psi+\Phi) + 4H\dot{W}^{(0)}(\Psi+\Phi)
      + \dot{W}^{(0)}\left(2\dot{\Psi}+\dot{\Phi}\right)\right],
\end{eqnarray}
where
%
\begin{eqnarray}
 V^{(0)} & \equiv & V(\phi^{(0)}), \nonumber\\
 V^{(0)'} & \equiv & V'(\phi^{(0)}), \nonumber\\
 F^{(0)} & \equiv & \frac{1}{f(R^{(0)})}, \nonumber\\
 F^{(0)}_{,R} & \equiv & -\frac{f'(R^{(0)})}{[f(R^{(0)})]^2},
  \nonumber\\
 F^{(0)}_{,RR} & \equiv & 
  \frac{2[f'(R^{(0)})]^2-f(R^{(0)})f''(R^{(0)})}{[f(R^{(0)})]^3},
  \nonumber\\
 W^{(0)} & = & -F^{(0)}_{,R}\dot{\phi}^{(0)2} -4\alpha R^{(0)},
  \nonumber\\ 
 \delta W & = & -F^{(0)}_{,RR}\dot{\phi}^{(0)2}\delta R 
  + 2F^{(0)}_{,R}\dot{\phi}^{(0)2}\Phi
  -2F^{(0)}_{,R}\dot{\phi}^{(0)}\dot{\delta\phi}
  -4\alpha\delta R. 
\end{eqnarray}

Note that there is an identity 
%
\begin{equation}
 \dot{\tau}_{\phi(L)0}+3H\tau_{\phi(L)0} 
  +\frac{4}{3}\frac{k^2}{a^2}\tau_{\phi(LL)}-\frac{\tau_{\phi(Y)}}{a^2} 
  =  
  \frac{1}{3}\left(3T^{(0)}_{\phi 00}
	      +a^{-2}\delta^{ij}T^{(0)}_{\phi ij}\right)\Phi 
  +\frac{2}{3a^2}\delta^{ij}T^{(0)}_{\phi ij}\Psi
  -E^{(0)}_{\phi}\delta\phi,
  \label{eqn:conservation-phi}
\end{equation}
where $E^{(0)}_{\phi}$ is given by 
%
\begin{equation}
 E^{(0)}_{\phi} = -\left[F^{(0)}\dot{\phi}^{(0)}\right]^{\cdot}
  -3HF^{(0}\dot{\phi}^{(0)}-V'(\phi),
\end{equation}
and gives the field equation $E^{(0)}_{\phi}=0$ for the background
$\phi$. The identity (\ref{eqn:conservation-phi}) is equivalent to the 
$(L)$-component of the linearized conservation equation of
$T_{\phi\mu\nu}$. Hence, combining this identity with the Bianchi
identity (\ref{eqn:Bianchi}) and the conservation equation
(\ref{eqn:conservation}) for other fields, it is concluded that the
$(Y)$-component $\delta G_{(Y)}=\kappa^2(\tau_{\phi(Y)}+\tau_{(Y)})$ of
the linearized equation of motion follows from other components. On the
other hand, the $0$-component of the linearized conservation equation of
$T_{\phi\mu\nu}$ is not an identity but is equivalent to the linearized 
field equation of $\phi$. Therefore, the $00$-, $(L)0$- and
$(LL)$-components give all independent equations.

Actually, the $(L)0$- and $(LL)$-components of the equation of motion
can be, in a sense, considered as constraint equations since we have
imposed the gauge condition $h_{(L)0}=h_{(LL)}=0$. The $(LL)$-component 
$\delta G_{(LL)}=\kappa^2(\tau_{\phi(LL)}+\tau_{(LL)})$ is 
%
\begin{equation}
 \kappa^2\delta W =
  -\left(1-\kappa^2W^{(0)}\right)\Psi_- + 2\kappa^2\tau_{(LL)}. 
  \label{eqn:LL-component}
\end{equation}
The $(L)0$-component 
$\delta G_{(L)0}=\kappa^2(\tau_{\phi(L)0}+\tau_{(L)0})$ is, by using
the $(LL)$-component, reduced to 
%
\begin{equation}
 \kappa^2F^{(0)}\dot{\phi}^{(0)}\delta\phi = 
  \left(1-\kappa^2W^{(0)}\right)\left(\dot{\Psi}_+ +H\Psi_+\right)
  + \frac{\kappa^2}{2}\dot{W}^{(0)}(-\Psi_+ +3\Psi_-)
  + \kappa^2(2\dot{\tau}_{(LL)}-2H\tau_{(LL)}-\tau_{(L)0}). 
  \label{eqn:L0-component}
\end{equation}
To be precise, this equation is 
$\delta G_{(L)0}-2a(\delta G_{(LL)}/a)^{\cdot}
=\kappa^2(\tau_{\phi(L)0}+\tau_{(L)0})
-2\kappa^2a[(\tau_{\phi(LL)}+\tau_{(LL)})/a]^{\cdot}$. 
Here, we have defined 
%
\begin{equation}
 \Psi_{\pm} \equiv \Psi\pm\Phi. 
\end{equation}

The remaining $00$-component 
$\delta G_{00}=\kappa^2(\tau_{\phi 00}+\tau_{00})$ is much more
complicated than the above two components since it is a dynamical
equation while the above two are constraint equations. By using the
$(LL)$- and $(L)0$-components, the $00$-component is reduced to
%
\begin{equation}
 \left(
 C_{2+}\kappa^2\ddot{\Psi}_+ + C_{1+}\kappa\dot{\Psi}_+ 
 + C_{0+}\Psi_+ \right)
+ \left(
   C_{2-}\kappa^2\ddot{\Psi}_- + C_{1-}\kappa\dot{\Psi}_-
   + C_{0-}\Psi_- \right)+ S_{00} = 0, \label{eqn:00-component-mod}
\end{equation}
where
%
\begin{eqnarray}
 C_{2+} & = & -1 + 2\kappa^2F_{,R}^{(0)}\dot{\phi}^{(0)2} 
  - 4\alpha\kappa^2R^{(0)}, \nonumber\\
 C_{1+} & = & 
  3\kappa^3\dot{W}^{(0)}
  +\left(\frac{\kappa\dot{F}^{(0)}}{F^{(0)}}
  +\frac{\kappa\ddot{\phi}^{(0)}}{\dot{\phi}^{(0)}}
  -\frac{\kappa{V^{(0)}}'}{F^{(0)}\dot{\phi}^{(0)}}
  \right) \left(1-\kappa^2W^{(0)}\right)
  -\kappa H\left(4+\frac{7}{2}\kappa^2W^{(0)}
	    -\frac{15}{2}\kappa^2F^{(0)}_{,R}\dot{\phi}^{(0)2}
	    +30\alpha\kappa^2R^{(0)} \right), \nonumber\\
 C_{0+} & = &
  -\kappa^4\left(V^{(0)}-\alpha R^{(0)2}\right)
  -\kappa^2\left[H(1-\kappa^2W^{(0)})\right]^{\cdot}
  - \frac{\kappa^2k^2}{a^2}\left(1-\frac{1}{2}\kappa^2W^{(0)}
   -\frac{1}{2}\kappa^2F^{(0)}_{,R}\dot{\phi}^{(0)2}
   +2\alpha\kappa^2R^{(0)}\right)
  +\frac{1}{2}\kappa^4\ddot{W}^{(0)}
  \nonumber\\
 & &
  +\left(\frac{\kappa\dot{F}^{(0)}}{F^{(0)}}
  +\frac{\kappa\ddot{\phi}^{(0)}}{\dot{\phi}^{(0)}}
  -\frac{\kappa{V^{(0)}}'}{F^{(0)}\dot{\phi}^{(0)}} \right) 
  \left[\kappa H(1-\kappa^2W^{(0)})-\frac{1}{2}\kappa^3\dot{W}^{(0)}\right]
  +3\kappa^2(\dot{H}+2H^2)
  \left(\kappa^2F^{(0)}_{,R}\dot{\phi}^{(0)2}
   -4\alpha\kappa^2R^{(0)}\right),  \nonumber\\
 C_{2-} & = & 3\kappa^2F^{(0)}_{,R}\dot{\phi}^{(0)2},
  \nonumber\\
 C_{1-} & = & 9\kappa^3HF_{,R}^{(0)}\dot{\phi}^{(0)2},
  \nonumber\\
 C_{0-} & = & 
  \kappa^4\left(V^{(0)}-\alpha R^{(0)2}\right)
  -3\kappa^2(\dot{H}+H^2)\left(1-\kappa^2W^{(0)}\right)
  +3\kappa^2F^{(0)}_{,R}\dot{\phi}^{(0)2}\frac{\kappa^2k^2}{a^2}
  -\frac{3}{2}\kappa^4\ddot{W}^{(0)}\nonumber\\
  & & +\frac{3}{2}
   \left(-2\kappa H + \frac{\kappa\dot{F}^{(0)}}{F^{(0)}}
    +\frac{\kappa\ddot{\phi}^{(0)}}{\dot{\phi}^{(0)}}
    -\frac{\kappa{V^{(0)}}'}{F^{(0)}\dot{\phi}^{(0)}}
   \right)\kappa^3\dot{W}^{(0)}
  -3\kappa^2(\dot{H}+2H^2)
  \left(\kappa^2F^{(0)}_{,R}\dot{\phi}^{(0)2}
   -4\alpha\kappa^2R^{(0)}\right),
  \nonumber\\
 S_{00} & = & -\kappa^4\tau_{00}+\kappa^4\dot{\tau}_{(L)0}
  -2\kappa^4\left[\ddot{\tau}_{(LL)}+2H\dot{\tau}_{(LL)}
	     -(4\dot{H}+3H^2)\tau_{(LL)}+\frac{k^2}{a^2}\tau_{(LL)}\right]
  \nonumber\\
 & & 
  -\left(\frac{\kappa\dot{F}^{(0)}}{F^{(0)}}
    +\frac{\kappa\ddot{\phi}^{(0)}}{\dot{\phi}^{(0)}}
    -\frac{\kappa{V^{(0)}}'}{F^{(0)}\dot{\phi}^{(0)}} \right)
  (\kappa^3\tau_{(L)0}-2\kappa^3\dot{\tau}_{(LL)}+2\kappa^3H\tau_{(LL)}). 
\end{eqnarray}

The $(LL)$-component (\ref{eqn:LL-component}) can also be rewritten as a
differential equation for $\Psi_{\pm}$ by using the $(L)0$-component
(\ref{eqn:L0-component}) as follows. 
%
\begin{equation}
 \left(
  D_{2+}\kappa^2\ddot{\Psi}_+ + D_{1+}\kappa\dot{\Psi}_+ 
  + D_{0+}\Psi_+ \right)
 + \left(
    D_{2-}\kappa^2\ddot{\Psi}_- + D_{1-}\kappa\dot{\Psi}_-
    + D_{0-}\Psi_- \right)+ S_{(LL)} = 0,
  \label{eqn:LL-component-mod} 
\end{equation}
where
%
\begin{eqnarray}
 D_{2+} & = & -(1-\kappa^2W^{(0)})
  +\frac{3}{2}\frac{\kappa^2F^{(0)}}{F^{(0)}_{,R}}
  (F^{(0)}_{,RR}\dot{\phi}^{(0)2} +4\alpha), 
  \nonumber\\
 D_{1+} & = & 
  \left(\frac{\kappa\dot{F}^{(0)}}{F^{(0)}} 
  +\frac{\kappa\ddot{\phi}^{(0)}}{\dot{\phi}^{(0)}}
  -\kappa H\right)(1-\kappa^2W^{(0)})
  +\frac{3}{2}\kappa^3\dot{W}^{(0)}
  +\frac{15}{2}\frac{\kappa^2F^{(0)}}{F^{(0)}_{,R}}
  (F^{(0)}_{,RR}\dot{\phi}^{(0)2}+4\alpha)\kappa H,
  \nonumber\\
 D_{0+} & = & \frac{1}{2}\kappa^4F^{(0)}\dot{\phi}^{(0)2}
  +  \left(\frac{\kappa\dot{F}^{(0)}}{F^{(0)}}
  +\frac{\kappa\ddot{\phi}^{(0)}}{\dot{\phi}^{(0)}}\right)
  \left[\kappa H\left(1-\kappa^2W^{(0)}\right)
   -\frac{1}{2}\kappa^3\dot{W}^{(0)}\right]
  -\kappa\left[\kappa H\left(1-\kappa^2W^{(0)}\right)\right]^{\cdot}
  \nonumber\\
 & & 
  +\frac{1}{2}\kappa^4\ddot{W}^{(0)}
  +\frac{1}{2}\frac{\kappa^2F^{(0)}}{F^{(0)}_{,R}}
  (F^{(0)}_{,RR}\dot{\phi}^{(0)2}+4\alpha)
  \left[\frac{\kappa^2k^2}{a^2 }+6\kappa^2(\dot{H}+2H^2)\right], 
  \nonumber\\
 D_{2-} & = & \frac{3}{2}\frac{\kappa^2F^{(0)}}{F^{(0)}_{,R}}
  (F^{(0)}_{,RR}\dot{\phi}^{(0)2}+4\alpha),
  \nonumber\\
 D_{1-} & = & -\frac{3}{2}\kappa^3\dot{W}^{(0)}
  +\frac{9}{2}\frac{\kappa^2F^{(0)}}{F^{(0)}_{,R}}
  (F^{(0)}_{,RR}\dot{\phi}^{(0)2}+4\alpha)\kappa H,
  \nonumber\\
 D_{0-} & = & \frac{\kappa^2F^{(0)}}{F^{(0)}_{,R}}
  \left\{\frac{1}{2} + 2\alpha\kappa^2R^{(0)}
  +\frac{3}{2}(F^{(0)}_{,RR}\dot{\phi}^{(0)2}+4\alpha)
  \left[\frac{\kappa^2k^2}{a^2}
   -2\kappa^2(\dot{H}+2H^2)\right]  \right\}\nonumber\\
  & & +\frac{3}{2}
  \left(\frac{\kappa\dot{F}^{(0)}}{F^{(0)}}
  +\frac{\kappa\ddot{\phi}^{(0)}}{\dot{\phi}^{(0)}}\right)
  \kappa^3\dot{W}^{(0)}
  -\frac{3}{2}\kappa^4\ddot{W}^{(0)},
  \nonumber\\
 S_{(LL)} & = & 
  -\kappa^4(2\ddot{\tau}_{(LL)}-\dot{\tau}_{(L)0})
  + \left(\frac{\kappa\dot{F}^{(0)}}{F^{(0)}}
  +\frac{\kappa\ddot{\phi}^{(0)}}{\dot{\phi}^{(0)}}\right)
  \kappa^3(2\dot{\tau}_{(LL)}-2H\tau_{(LL)}-\tau_{(L)0})
  +2\kappa^4(H\tau_{(LL)})^{\cdot}
  - \frac{\kappa^4F^{(0)}}{F^{(0)}_{,R}}\tau_{(LL)}.
\end{eqnarray}

Now let us solve the equations (\ref{eqn:00-component-mod}) and
(\ref{eqn:LL-component-mod}) with respect to $\Psi_{\pm}$ up to the
zeroth order in the $\kappa H$ expansion. For this purpose, as we shall 
see, we actually need to keep terms up to the second order in $\kappa H$
in intermediate steps.

First, let us introduce a small dimensionless parameter $\epsilon$,
assume that 
%
\begin{eqnarray}
 \kappa H & = & O(\epsilon), \nonumber\\
 \frac{\kappa^2k^2}{a^2} & = & O(\epsilon^0),
\end{eqnarray}
and estimate orders of other quantities as
%
\begin{eqnarray}
 F^{(0)} & = & O(\epsilon^{-4m}), \nonumber\\
 \kappa^{-2}F^{(0)}_{,R} & = & O(\epsilon^{-4m-2}), \nonumber\\
 \kappa^{-4}F^{(0)}_{,RR} & = & O(\epsilon^{-4m-4}), \nonumber\\
 \kappa^2R^{(0)} & = & O(\epsilon^2), \nonumber\\
 W^{(0)} & = & O(\epsilon^2), \nonumber\\
 \kappa^3{V^{(0)}}' & = & c = O(\epsilon^0), \nonumber\\
 \kappa^2F^{(0)}\dot{\phi}^{(0)} & = & O(\epsilon^{-1}). 
  \label{eqn:order-other-quantities}
\end{eqnarray}

Next, $\kappa^2\dot{H}$ is shown to be of order $\epsilon^4$ by using
the equation of motion. 
%
\begin{equation}
 \kappa^2\dot{H} = 
  \frac{-\kappa^4F^{(0)}\dot{\phi}^{(0)2}+\kappa^4\ddot{W}^{(0)}
  -\kappa^4H\dot{W}^{(0)}}{2(1-\kappa^2W^{(0)})}
  = O(\epsilon^4). \label{eqn:dH}
\end{equation}
This means that $\kappa H$ can be considered as a constant up to the
order $O(\epsilon^2)$, which is sufficient for our purpose. As a
consequence, we can regard $F^{(0)}$, $F^{(0)}_{,R}$, $F^{(0)}_{,RR}$
and $R^{(0)}$ as constants up to the order $O(\epsilon^2)$. The
estimate (\ref{eqn:dH}) leads to the expansion of $1/a^2$ as 
%
\begin{equation}
 \frac{1}{a^2} = \frac{1}{a_0^2}
  \left[ 1-2H(t-t_0)+2H^2(t-t_0)^2+O(\epsilon^2)\right],
\end{equation}
around the time $t=t_0$ when an experiment is performed, where
$a_0=a(t_0)$ is a constant. Here, we have assumed that the duration
$t-t_0$ of the experiment is much shorter than the cosmological time
scale $H^{-1}$. Of course, this assumption is justified since we shall
take the $\kappa H\to +0$ limit in the end of calculation.

Thirdly, from (\ref{eqn:eom-B}) and the statement after that, we obtain
the estimate that
%
\begin{equation}
 \frac{\kappa^3HF^{(0)}\dot{\phi}^{(0)}}{c}+\frac{1}{3} 
  = {\cal B}+\frac{1}{3} = O(\dot{H}/H^2) = O(\epsilon^2),
\end{equation}
where we have used (\ref{eqn:dH}). With this estimation and the
background equation of motion, we obtain 
%
\begin{equation}
 \frac{\kappa\dot{F}^{(0)}}{F^{(0)}} 
  +\frac{\kappa\ddot{\phi}^{(0)}}{\dot{\phi}^{(0)}}
  = -3\kappa H - \frac{c}{\kappa^2F^{(0)}\dot{\phi}^{(0)}}
  = O(\epsilon^3). \nonumber\\
\end{equation}
By using the background equations of motion, we also obtain 
%
\begin{eqnarray}
 \kappa^4\left(V^{(0)}-\alpha R^{(0)2}\right) & = & 
  3\kappa^2H^2\left(1+\kappa^2W^{(0)}\right) 
  + \kappa^2\dot{H}\left(1+2\kappa^2W^{(0)}\right)
  -\frac{1}{2}\kappa^4\ddot{W}^{(0)} 
  -\frac{5}{2}\kappa^4H\dot{W}^{(0)}
  = 3\kappa^2H^2+O(\epsilon^4). 
\end{eqnarray}

Fourthly, let us expand all other relevant quantities as
%
\begin{eqnarray}
 \Psi_{\pm} & = & \sum_{i=0}^{\infty}\Psi_{\pm[i]},\nonumber\\
 \tau_{00} & = & 
  \sum_{i=0}^{\infty}\tau_{00[i]},\nonumber\\
 \tau_{(L)0} & = & 
  \sum_{i=0}^{\infty}\tau_{(L)0[i]},\nonumber\\
 \tau_{(Y)} & = & 
  \sum_{i=0}^{\infty}\tau_{(Y)[i]},\nonumber\\
 \tau_{(LL)} & = & 
  \sum_{i=0}^{\infty}\tau_{(LL)[i]},
\end{eqnarray}
where terms with the subscript $[i]$ is of order
$O(\epsilon^i)$. According to the expansion of $\tau$'s, we obtain the
following expansion of the conservation equation
(\ref{eqn:conservation}): 
%
\begin{eqnarray}
 \dot{\tau}_{00[0]}+\frac{k^2}{a_0^2}\tau_{(L)0[0]} & = & 0,\nonumber\\
 \dot{\tau}_{(L)0[0]} + \frac{4}{3}\frac{k^2}{a_0^2}\tau_{(LL)[0]}
  - \frac{\tau_{(Y)[0]}}{a_0^2} & = & 0
  \label{eqn:conservation-zeroth}
\end{eqnarray}
in the order $O(\epsilon^0)$,
%
\begin{eqnarray}
 \dot{\tau}_{00[1]}+\frac{k^2}{a_0^2}\tau_{(L)0[1]} 
  + 3H\tau_{00[0]} + 3H\frac{\tau_{(Y)[0]}}{a_0^2} 
  - 2\frac{k^2}{a_0^2}\tau_{(L)0[0]}H(t-t_0) & = & 0,\nonumber\\
 \dot{\tau}_{(L)0[1]} + \frac{4}{3}\frac{k^2}{a_0^2}\tau_{(LL)[1]}
  - \frac{\tau_{(Y)[1]}}{a_0^2}  + 3H\tau_{(L)0[0]} 
  -\frac{8}{3}\frac{k^2}{a_0^2}\tau_{(LL)[0]}H(t-t_0)
  +2\frac{\tau_{(Y)[0]}}{a_0^2}H(t-t_0) & = & 0.
  \label{eqn:conservation-first}
\end{eqnarray}
in the order $O(\epsilon^1)$, and 
%
\begin{eqnarray}
 \dot{\tau}_{00[2]} + \frac{k^2}{a^2}\tau_{(L)0[2]} 
  + 3H\tau_{00[1]} + 3H\frac{\tau_{(Y)[1]}}{a_0^2} 
  -2\frac{k^2}{a_0^2}\tau_{(L)0[1]}H(t-t_0) & & \nonumber\\
 - 6\frac{\tau_{(Y)[0]}}{a_0^2}H^2(t-t_0)
  + 2\frac{k^2}{a_0^2}\tau_{(L)0[0]}H^2(t-t_0)^2
  & = & 0, \nonumber\\
 \dot{\tau}_{(L)0[2]} +\frac{4}{3}\frac{k^2}{a_0^2}\tau_{(LL)[2]}
  -\frac{\tau_{(Y)[2]}}{a_0^2} +3H\tau_{(L)0[1]} 
  -\frac{8}{3}\frac{k^2}{a_0^2}\tau_{(LL)[1]}H(t-t_0)
  +2\frac{\tau_{(Y)[1]}}{a_0^2}H(t-t_0) & & \nonumber\\
  +\frac{8}{3}\frac{k^2}{a_0^2}\tau_{(LL)[0]}H^2(t-t_0)^2
  -2\frac{\tau_{(Y)[0]}}{a_0^2}H^2(t-t_0)^2
  & = & 0.
  \label{eqn:conservation-second}
\end{eqnarray}
in the order $O(\epsilon^2)$. These conservation equations are used
throughout the forthcoming calculations.

Fifthly, the coefficients of the equations (\ref{eqn:00-component-mod})
and (\ref{eqn:LL-component-mod}) are correspondingly expanded as
%
\begin{eqnarray}
 C_{I\pm} = \sum_{i=0}^{\infty}C_{I\pm[i]}, \nonumber\\
 S_{00} = \sum_{i=0}^{\infty}S_{00[i]}, \nonumber\\
 D_{I\pm} = \sum_{i=0}^{\infty}D_{I\pm[i]}, \nonumber\\
 S_{(LL)} = \sum_{i=0}^{\infty}S_{(LL)[i]},
\end{eqnarray}
where 
%
\begin{eqnarray}
 C_{2+[0]} & = & -1, \quad
 C_{2+[1]} =  0, \quad
 C_{2+[2]} = -4\alpha\kappa^2R^{(0)}, \nonumber\\
 C_{1+[0]} & = & 0, \quad
 C_{1+[1]} = -\kappa H, \quad
 C_{1+[2]} = 0,   \nonumber\\
 C_{0+[0]} & = & -\frac{\kappa^2k^2}{a_0^2}, \quad
 C_{0+[1]} = 2\frac{\kappa^2k^2}{a_0^2}H(t-t_0), \quad
 C_{0+[2]} = -2\frac{\kappa^2k^2}{a_0^2}H^2(t-t_0)^2
 - 4\alpha\frac{\kappa^2k^2}{a_0^2}\kappa^2R^{(0)},  \nonumber\\
 C_{2-[0]} & = & C_{2-[1]} = C_{2-[2]} = 
 C_{1-[0]} = C_{1-[1]} = C_{1-[2]} = C_{0-[0]} = 
 C_{0-[1]} = C_{0-[2]} = 0, \nonumber\\
 S_{00[0]} & = & -\kappa^4\tau_{00[0]} + \kappa^4\dot{\tau}_{(L)0[0]}
  -2\kappa^4\ddot{\tau}_{(LL)[0]} 
  -2\frac{\kappa^2k^2}{a_0^2}\kappa^2\tau_{(LL)[0]}, \nonumber\\
 S_{00[1]} & = & -\kappa^4\tau_{00[1]} + \kappa^4\dot{\tau}_{(L)0[1]}
  -2\kappa^4\ddot{\tau}_{(LL)[1]} 
  -2\frac{\kappa^2k^2}{a_0^2}\kappa^2\tau_{(LL)[1]} \nonumber\\
 & & 
  +2\kappa^4H\dot{\tau}_{(LL)0[0]}-3\kappa^4H\tau_{(L)0[0]}
  +4\frac{\kappa^2k^2}{a_0^2}\kappa^2\tau_{(LL)[0]} H (t-t_0), 
  \nonumber\\
 S_{00[2]} & = & -\kappa^4\tau_{00[2]} + \kappa^4\dot{\tau}_{(L)0[2]}
  -2\kappa^4\ddot{\tau}_{(LL)[2]} 
  -2\frac{\kappa^2k^2}{a_0^2}\kappa^2\tau_{(LL)[2]}\nonumber\\
 & & 
  +2\kappa^4H\dot{\tau}_{(LL)0[1]}-3\kappa^4H\tau_{(L)0[1]}
  +4\frac{\kappa^2k^2}{a_0^2}\kappa^2\tau_{(LL)[1]} H (t-t_0)
  -4\frac{\kappa^2k^2}{a_0^2}\kappa^2\tau_{(LL)[0]} H^2(t-t_0)^2, 
  \nonumber\\
 D_{2+[0]} & = & -1, \quad
 D_{2+[1]} = 0, \quad
 D_{2+[2]} = -4\alpha\kappa^2R^{(0)}
 + 6\alpha\frac{\kappa^2F^{(0)}}{F^{(0)}_{,R}},  \nonumber\\
 D_{1+[0]} & = & 0, \quad
 D_{1+[1]} = -\kappa H, \quad
 D_{1+[2]} = 0, \nonumber\\
 D_{0+[0]} & = & D_{0+[1]}  = 0, \quad
 D_{0+[2]} = 2\alpha\frac{\kappa^2k^2}{a_0^2}
 \frac{\kappa^2F^{(0)}}{F^{(0)}_{,R}},  \nonumber\\
 D_{2-[0]} & = &  D_{2-[1]} = 0, \quad
 D_{2-[2]} = 6\alpha\frac{\kappa^2F^{(0)}}{F^{(0)}_{,R}},
  \nonumber\\
 D_{1-[0]} & = & D_{1-[1]} = D_{1-[2]} = 0, \nonumber\\
 D_{0-[0]} & = & D_{0-[1]}  = 0, \quad
 D_{0-[2]} = \frac{\kappa^2F^{(0)}}{F^{(0)}_{,R}}
 \left(\frac{1}{2}+6\alpha\frac{\kappa^2k^2}{a_0^2}\right),  \nonumber\\
 S_{(LL)[0]} & = & 
  \kappa^4\dot{\tau}_{(L)0[0]}-2\kappa^4\ddot{\tau}_{(LL)[0]}, 
  \nonumber\\
 S_{(LL)[1]} & = & 
  \kappa^4\dot{\tau}_{(L)0[1]}-2\kappa^4\ddot{\tau}_{(LL)[1]}
  +2\kappa^4H\dot{\tau}_{(LL)[0]},   \nonumber\\
 S_{(LL)[2]} & = & 
  \kappa^4\dot{\tau}_{(L)0[2]}-2\kappa^4\ddot{\tau}_{(LL)[2]}
  +2\kappa^4H\dot{\tau}_{(LL)[1]} 
  -\frac{\kappa^2F^{(0)}}{F^{(0)}_{,R}}\kappa^2\tau_{(LL)[0]}.
\end{eqnarray}

Now it is time to expand the equations (\ref{eqn:00-component-mod}) and 
(\ref{eqn:LL-component-mod}). For this purpose we shall use the
conservation equations (\ref{eqn:conservation-zeroth}),
(\ref{eqn:conservation-first}) and (\ref{eqn:conservation-second}). In
the order $O(\epsilon^0)$, by subtracting (\ref{eqn:LL-component-mod})
from  (\ref{eqn:00-component-mod}) we obtain 
%
\begin{equation}
 \frac{\kappa^2k^2}{a_0^2}\Psi_{+[0]} = 
  -\kappa^4\left(\tau_{00[0]}
	    +2\frac{k^2}{a_0^2}\tau_{(LL)[0]}\right).
  \label{eqn:sol-zeroth}
\end{equation}
It is easily shown by using the conservation equations that both 
(\ref{eqn:00-component-mod}) and (\ref{eqn:LL-component-mod}) are
simultaneously satisfied by this solution in the order $O(\epsilon^0)$.

In the order $O(\epsilon^1)$, by subtracting
(\ref{eqn:LL-component-mod}) from (\ref{eqn:00-component-mod}),
substituting the solution (\ref{eqn:sol-zeroth}), and using the
conservation equations, we obtain 
%
\begin{equation}
 \frac{\kappa^2k^2}{a_0^2}\Psi_{+[1]} = 
  -\kappa^4\left(\tau_{00[1]}
	    +2\frac{k^2}{a_0^2}\tau_{(LL)[1]}
	    +3H\tau_{(L)0[0]}+2\tau_{00[0]}H(t-t_0)
	   \right).\label{eqn:sol-first}
\end{equation}
It is easily shown by using the conservation equations that both 
(\ref{eqn:00-component-mod}) and (\ref{eqn:LL-component-mod}) are
simultaneously satisfied by this solution in the order $O(\epsilon^1)$.

In the order $O(\epsilon^2)$, by subtracting
(\ref{eqn:LL-component-mod}) from (\ref{eqn:00-component-mod}),
substituting the solutions (\ref{eqn:sol-zeroth}) and
(\ref{eqn:sol-first}), and using the conservation equations, we obtain 
%
\begin{equation}
 \frac{\kappa^2k^2}{a_0^2}\Psi_{+[2]} = 
  6\alpha\frac{\kappa^2F^{(0)}}{F^{(0)}_R}
  \left[
   \kappa^2\Box\left(\Psi_{-[0]}-2\kappa^2\tau_{(LL)[0]}\right) 
   -\frac{1}{12\alpha}
   \left(\Psi_{-[0]}-2\kappa^2\tau_{(LL)[0]}\right) \right]
   + R_{-[2]},  \label{eqn:sol-second}
\end{equation}
where
%
\begin{eqnarray}
 \Box & = & -\partial_t^2 -\frac{k^2}{a_0^2}, \nonumber\\
 R_{-[2]} & = & 
  -\kappa^4\tau_{00[2]} -2\frac{\kappa^2k^2}{a_0^2}\kappa^2\tau_{(LL)[2]} 
  -2\kappa^4\tau_{00[1]}H(t-t_0) -3\kappa^4H\tau_{(L)0[1]}
  -6\alpha\frac{\kappa^2F^{(0)}}{F^{(0)}_{,R}}
  \frac{\kappa^4\tau_{(Y)[0]}}{a_0^2}\nonumber\\
 & &   +\left[2\alpha\frac{\kappa^2F^{(0)}}{F_R}
  +4\alpha\kappa^2R^{(0)}-2H^2(t-t_0)^2\right]
  \kappa^4\tau_{00[0]}
  -6\kappa^4\tau_{(L)0[0]}H^2(t-t_0) 
  +8\alpha\frac{\kappa^2k^2}{a_0^2}\kappa^4R^{(0)}\tau_{(LL)[0]}. 
\end{eqnarray}
By substituting this back to (\ref{eqn:00-component-mod}) and using
conservation equations, we obtain 
%
\begin{equation}
 \Box\left[\left(1-12\alpha\kappa^2\Box\right)
      \left(\Psi_- -2\kappa^2\tau_{(LL)[0]}\right)
      + 4\alpha\kappa^4
      \left(-\tau_{00[0]}+3\frac{\tau_{(Y)[0]}}{a_0^2}\right)
     \right] = 0.
  \label{eqn:eq-for-psi-m-pre}
\end{equation}
Note that $\Psi_{-[0]}$ appeared only in the equations of order
$O(\epsilon^2)$ (and higher). This is the reason why we had to keep
terms of up to order $O(\epsilon^2)$.

By using the conservation equation, it is easy to show that the
expression $\kappa^4(-\tau_{00[0]}+3\tau_{(Y)[0]}/a_0^2)$
in (\ref{eqn:eq-for-psi-m-pre}) is rewritten as 
%
\begin{equation}
\kappa^4\left(-\tau_{00[0]}+3\frac{\tau_{(Y)[0]}}{a_0^2}\right)
  = \kappa^2\left(3\ddot{\Psi}_{+[0]} 
	     + \frac{k^2}{a_0^2}\Psi_{+[0]}\right)
  -6\kappa^4\Box\tau_{(LL)[0]}. 
\end{equation}
Hence, (\ref{eqn:eq-for-psi-m-pre}) is rewritten as 
%
\begin{equation}
 \Box\left[\left(1-12\alpha\kappa^2\Box\right)\Psi_{-[0]}
      + 4\alpha\kappa^2
      \left(3\ddot{\Psi}_{+[0]} + \frac{k^2}{a_0^2}\Psi_{+[0]}\right)
      -2\kappa^2\tau_{(LL)[0]}
     \right] = 0.
  \label{eqn:eq-for-psi-m}
\end{equation}

Finally, by taking the $\kappa H\to 0$ limit in (\ref{eqn:sol-zeroth})
and (\ref{eqn:eq-for-psi-m}), we obtain the following equations for
linearized gravity in Minkowski background. 
%
\begin{eqnarray}
 \kappa^2k^2\Psi_+  +\kappa^4\left(\tau_{00}
	    +2k^2\tau_{(LL)}\right) & = & 0, \nonumber\\ 
 \Box\left[\left(1-12\alpha\kappa^2\Box\right)\Psi_- 
      + 4\alpha\kappa^2
      \left(3\ddot{\Psi}_+ + k^2\Psi_+\right)
      -2\kappa^2\tau_{(LL)}
     \right] & = & 0,
\end{eqnarray}
where we have redefined the spatial coordinates as $a_0x^i\to x^i$ and,
thus, 
%
\begin{equation}
 \Box \equiv -\partial_t^2 -k^2
\end{equation}
The perturbed metric, after the $\kappa H\to +0$ limit and the above
redefinition of the spatial coordinates, is 
%
\begin{equation}
 ds^2 = 
  -(1+2\Phi Y)dt^2 + (1-2\Psi Y)\delta_{ij}dx^idx^j,
\end{equation}
and $\Psi_{\pm}=\Psi\pm\Phi$. The conservation equation is reduced to 
%
\begin{eqnarray}
 \dot{\tau}_{00}+k^2\tau_{(L)0} & = & 0,\nonumber\\
 \dot{\tau}_{(L)0} + \frac{4}{3}k^2\tau_{(LL)}
  - \tau_{(Y)} & = & 0,
\end{eqnarray}
where $\tau$'s are defined by (\ref{eqn:T-harmonics}).



\begin{references}
 \bibitem{paper1}
 S.~Mukohyama and L.~Randall, hep-th/0306108.
\bibitem{reviews}
 S.~Weinberg, Rev. Mod. Phys. {\bf 61}, 1 (1989);
 P.~J.~E.~Peebles and B.~Ratra, Rev. Mod. Phys. {\bf75}, 559 (2003);
 T.~Padmanabhan, Phys. Rept. {\bf 380}, 235 (2003).
 \bibitem{Witten}
 E.~Witten, hep-ph/0002297.
 \bibitem{Perlmutter}
 S.~Perlmutter, et al., Astrophys. J. {\bf 517}, 565 (1999).
 \bibitem{Schmidt}
 B.~P.~Schmidt, et al., Astrophys. J. {\bf 507}, 46 (1998).
 \bibitem{Riess}
 A.~G.~Riess et al., Astron. J. {\bf 116}, 1009 (1998).
 \bibitem{quintessence}
 R.~R.~Caldwell, R.~Dave, P.~J.~Steinhardt, Phys. Rev. Lett. {\bf 80},
 1582 (1998); P.~J.~Steinhardt, L.~M.~Wang, I.~Zlatev, Phys. Rev. 
{\bf D59}, 123504 (1999). 
 \bibitem{k-essence}
 T.~Chiba, T~Okabe, M.~Yamaguchi, Phys. Rev. {\bf D62}, 023511 (2000);
 C.~Armendariz-Picon, V.~Mukhanov, P.~J.~Steinhardt,
 Phys. Rev. Lett. {\bf 85}, 4438 (2000). 
 \bibitem{AHKM}
 N.~Arkani-Hamed, L.~J.~Hall, C.~F.~Kolda, H.~Murayama,
 Phys. Rev. Lett. {\bf 85}, 4434 (2000). 
 \bibitem{Yokoyama}
 J.~Yokoyama, Phys. Rev. Lett. {\bf 88}, 151302 (2002). 
 \bibitem{Baum}
 E.~Baum, Phys. Lett. {\bf B133}, 185 (1983). 
 \bibitem{Hawking}
 S.~W.~Hawking, Phys. Lett. {\bf B134}, 403 (1984). 
 \bibitem{Coleman}
 S.~R.~Coleman, Nucl. Phys. {\bf B310}, 643 (1988). 
 \bibitem{Brown-Teitelboim}
 J.~D.~Brown and C.~Teitelboim, Phys. Lett. {\bf B195}, 177 (1987); 
 Nucl. Phys. {\bf B297}, 787 (1988). 
 \bibitem{Bousso-Polchinski}
 R.~Bousso and J.~Polchinski, JHEP {\bf 0006}, 006 (2000)
 [hep-th/0004134]. 
 \bibitem{Banks-Dine-Motl}
 T.~Banks, M.~Dine and L.~Motl, JHEP {\bf 0101}, 031 (2001)
 [hep-th/0007206]. 
\bibitem{FMSW}
J.~L.~Feng, J.~March-Russell, S.~Sethi and F.~Wilczek,
Nucl.~Phys. {\bf B602}, 307 (2001)
[arXiv:hep-th/0005276].
 \bibitem{Garriga-Vilenkin}
 J.~Garriga and A.~Vilenkin, Phys. Rev. {\bf D61}, 083502 (2000)
 [astro-ph/9908115]. 
 \bibitem{Weinberg2000}
 S.~Weinberg, Phys. Rev. {\bf D61}, 103505 (2000) [astro-ph/0002387]. 
 \bibitem{Dolgov} A.~D.~Dolgov, in {\it The Very Early Universe},
 G.~W.~Gibbons, S.~W.~Hawking, and S.~T.~C.~Siklos, eds. (Cambridge
 University Press, 1983).
 \bibitem{Ford}
 L.~H.~Ford, Phys. Rev. D {\bf 35}, 2339 (1987);
 gr-qc/0210096.
 \bibitem{eternal-inflation}
 A.~Vilenkin, Phys. Rev. {\bf D27}, 2848 (1983); A.~D.~Linde,
 Phys. Lett. {\bf B175}, 395 (1986). 
 \bibitem{extra-dim}
 S.~Kachru, M.~B.~Schulz, E.~Silverstein, Phys. Rev. {\bf D62}, 045021
 (2000); N.~Arkani-Hamed, S.~Dimopoulos, N.~Kaloper and R.~Sundrum,
 Phys. Lett. {\bf B480}, 193 (2000); 
 C.~Deffayet, G.~R.~Dvali, G.~Gabadadze and A.~Lue, Phys. Rev. {\bf
 D64}, 104002 (2001). 
 \bibitem{Rubakov}
 V.~A.~Rubakov, Phys.Rev. {\bf D61}, 061501 (2000). 
 \bibitem{Hebecker-Wetterich}
 A.~Hebecker and C.~Wetterich, Phys. Rev. Lett. {\bf 85}, 3339 (2000). 
 \bibitem{FKPS}
 W.~Fischler, I.~R.~Klebanov, J.~Polchinski and L.~Susskind,
 Nucl. Phys. {\bf B327}, 157 (1989).  
 \bibitem{Duff}
 M.~J.~Duff, Phys. Lett. {\bf B226}, 36 (1989). 
 \bibitem{Hartle-Hawking}
 J.~B.~Hartle and S.~W.~Hawking, Phys. Rev. {\bf D28}, 2960 (1983). 
 \bibitem{Linde}
 A.~D.~Linde, Sov. Phys. JETP {\bf 60}, 211 (1984); 
 Zh. Eksp. Teor. Fiz. {\bf 87}, 369 (1984).  
 \bibitem{Vilenkin1986}
 A.~Vilenkin, Phys. Rev. {\bf D33}, 3560 (1986).  
 \bibitem{Vilenkin1995}
 A.~Vilenkin, Phys. Rev. Lett. {\bf 74}, 846 (1995). 
 \bibitem{Anthropic}
 For example, J.~Garriga and A.~Vilenkin, Phys. Rev. {\bf D67},
 043503 (2003); R.~Kallosh and A.~Linde, Phys. Rev. {\bf D67}, 023510
 (2003); and references therein.
  \bibitem{Garriga-Mukhanov}
 J.~Garriga and V.~F.~Mukhanov, Phys. Lett. {\bf B458}, 219 (1999).
 \bibitem{Muller-Schmidt}
 V.~M\"uller and H.-J.~Schmidt, Gen. Rel. and Grav. {\bf 17}, 769
 (1985).
 \bibitem{Lindebook}
 A.~Linde, {\it Particle physics and inflationary cosmology},
 Harwood (1990).
 \bibitem{Stelle}
 K.~S.~Stelle, Gen. Rel. and Grav. {\bf 9}, 353 (1978).
 \bibitem{Mukhanov}
 V.~F.~Mukhanov, Sov. Phys. JETP {\bf 67}, 1297 (1988).
 \bibitem{Maeda}
 K.~Maeda, Phys. Rev. {\bf D39}, 3159 (1989). 
 \bibitem{Wald}
 For example, R.~M.~Wald, {\it General Relativity}, The University of
 Chicago Press (1984). 
 \bibitem{Mukohyama2000b}
 S.~Mukohyama, Phys. Rev. {\bf D62}, 084015 (2000) [hep-th/0004067].
 \bibitem{Mukohyama-Kofman} 
 S.~Mukohyama and L.~Kofman, Phys. Rev. {\bf D65}, 124025 (2002)
 [hep-th/0112115].
\end{references}
\end{document}